\documentclass[aps,prd,twocolumn,superscriptaddress,groupedaddress,nofootinbib]{revtex4}

\usepackage[utf8]{inputenc}

\usepackage{hyperref}
\usepackage{tikz}
\usepackage{pgfplots}
\pgfplotsset{compat=newest}

\def\d{\partial}
\newcommand{\bd}[1]{{\bf #1}}

\newcommand{\lmfp}{\ensuremath{\ell_{\mathrm{MFP}}^\gamma}}
\newcommand{\lyamfp}{\ensuremath{\ell_{\mathrm{MFP}}^{\mathrm{Ly}\alpha}}}
\newcommand{\zfs}{\ensuremath{z_{\mathrm{FS}}}}
\newcommand{\kcons}{\ensuremath{k_{\mathrm{cons}}}}
\newcommand{\feq}{\ensuremath{f_{\mathrm{eq}}}}

\newcommand{\eq}[1]{\begin{equation}
    \begin{split}
        #1
    \end{split}
\end{equation}}

\usepackage{amsmath,amsfonts}
\usepackage{amssymb}
\usepackage{amsthm}
\usepackage{xcolor}
\usepackage{tikz}

\usepackage{array}
\usepackage{cancel}
\usepackage{comment}
\usepackage{physics}
\usepackage{bm}

\usepackage{graphicx}
\usepackage{caption}
\usepackage{subcaption}
\usepackage{orcidlink}

\usetikzlibrary{arrows}
\usetikzlibrary{calc}

\usepackage[normalem]{ulem} % in the preamble

\usepackage{xr}

\makeatletter
\newcommand*{\addFileDependency}[1]{% argument=file name and extension
\typeout{(#1)}% latexmk will find this if $recorder=0
\@addtofilelist{#1}
\IfFileExists{#1}{}{\typeout{No file #1.}}
}\makeatother

\makeatletter
\renewcommand*\env@matrix[1][\arraystretch]{%
  \edef\arraystretch{#1}%
  \hskip -\arraycolsep
  \let\@ifnextchar\new@ifnextchar
  \array{*\c@MaxMatrixCols c}}
\makeatother

\begin{document}

\title{Primordial magnetic fields and modified recombination histories}
\author{Jonathan Schiff\,\orcidlink{0000-0001-7802-0798}}
\affiliation{Department of Physics, University of California, Santa Barbara, CA 93106, USA}
\author{Tejaswi Venumadhav\,\orcidlink{0000-0002-1661-2138}}
\affiliation{Department of Physics, University of California, Santa Barbara, CA 93106, USA}
\affiliation{International Centre for Theoretical Sciences, Tata Institute of Fundamental Research, Bangalore 560089, India}
\date{\today}

\begin{abstract}
    Recent cosmological data and astrophysical observations, such as the Hubble tension and the increasing preference from galaxy surveys for dynamical dark energy, have begun to challenge the standard $\Lambda$-cold dark matter cosmological model. 
    Primordial magnetic fields (PMFs) offer a mechanism to alleviate these tensions within the framework of the standard model. 
    These fields source excess small-scale baryon clumping, which can speed up recombination and shrink the comoving sound horizon at the surface of last scattering. 
    Computing the modified recombination history requires coupling the radiative transport of Lyman-$\alpha$ photons to compressible magnetohydronamic simulations.
    Since doing so is generically computationally intractable, we have developed a linearized treatment which self-consistently computes the modified recombination history in the presence of PMF induced baryon clumping for fields with red-tilted spectra. 
    The clumping factors we find are too small to alleviate outstanding cosmological tensions, but our general framework can be applied to other PMF spectra, and provides a significant theoretical step towards a complete account of recombination in the presence of small-scale baryon clumping. 
\end{abstract}

\maketitle

{
  \hypersetup{linktocpage=true}
  \tableofcontents
}

\section{Introduction }

The $\Lambda$-cold dark matter (LCDM) model of cosmology has proven to be an incredibly successful theory, capable of describing a wide-range of phenomena and observations from the early to the late Universe. 
Despite its many successes, incompatibilities between LCDM and cosmological observations have begun to emerge, perhaps none more significant than the so-called Hubble tension. The Hubble tension is the discrepancy between direct measurements of the present Hubble expansion rate $H_0$ from the local universe and the inference of $H_0$ from the statistics of the cosmic microwave background (CMB) fluctuations within the LCDM model. 
The current state of the art local measurements from the $\text{SH}_0\text{ES}$ collaboration \cite{Riess_2022, breuval2024small} places the tension at more than $5\sigma$ with the most recent CMB results from Planck \cite{2020}. 
In the absence of yet unaccounted for systematic errors, a resolution to the Hubble tension requires a modification to the LCDM paradigm to bring the CMB inference of $H_0$ into agreement with the direct, late-universe measurements. 
An intriguing recent proposal to alleviate the tension is that primordial magnetic fields (PMFs), should they be present in the early Universe, would introduce small-scale baryon clumping that would substantially alter the recombination history \cite{jedamzik2011weak}. 
A modified recombination history that pushes back the surface of last scattering has also been shown to provide a better fit than LCDM when jointly fitting to the baryon acoustic oscillations (BAO) observed both in the CMB and in the Dark Energy Spectroscopic Instrument's (DESI) galaxy survey \cite{jedamzik2023primordialmagneticfieldshubble,lynch2024reconstructing,lynch2024desihubbletensionlight,jedamzik2025hints}. 
There is therefore considerable interest in developing a detailed formalism for modified recombination histories in the presence of inhomogeneities beyond those strictly sourced in LCDM.

PMFs find theoretical motivation from extensions to the standard model with first order phase transitions, such as those considered in the context of electroweak baryogenesis, which can excite the gauge potential and source PMFs \cite{cornwall1997speculations, vachaspati2001estimate}. 
There is also indirect observational evidence from the non-detection of secondary GeV emission from blazars, which are most naturally explained by the presence of magnetic fields permeating the cosmic voids \cite{neronov2010evidence, Tavecchio_2010, Tavecchio_2011, Taylor_2011, Vovk_2012, Dolag_2010, Podlesnyi_2022, Acciari_2023, Xia_2024, Dzhatdoev_2023, Aharonian_2023, Huang_2023, vovk2023constraintintergalacticmagneticfield}. 
The large coherence scale needed to explain the blazar observations are difficult to generate astrophysically, and therefore a primordial origination remains a distinct and compelling possibility. 
If it can be shown that PMFs also provide a viable resolution to the Hubble tension, this would provide further evidence for their existence and allow for a straightforward alleviation of the Hubble tension without the introduction of otherwise unmotivated extensions to the LCDM paradigm. 

$H_0$ is inferred from the CMB by using the precisely measured angular scale of the BAO in the CMB's power spectra, $\theta_s \sim r_s/D_A$, where $r_s$ is the comoving sound horizon at the time of the CMB's release and $D_A$ is the angular diameter distance to the CMB, which depends on $H_0$. 
The time of the CMB's release is set by the peak of the visibility function, $g(z) = \dot{\tau}e^{-\tau}$, where $\dot{\tau} = d\tau/d\eta = an_e\sigma_T$ is the differential Thomson optical depth. 
$a$ is the scale factor, $n_e$ is the free electron number density, $\sigma_T$ is the Thomson cross section, and $\eta$ is conformal time.
The visibility function gives the probability that a CMB photon last scattered at a certain redshift. 

Resolutions to the Hubble tension largely fall into two camps, early- and late-universe. 
Early-Universe solutions seek to shrink $r_s$ by introducing new physical effects before the CMB's release. 
These include introducing novel energy contributions, such as early dark energy \cite{Karwal_2016,Poulin_2019,poulin2023upsdownsearlydark}, which alter the expansion history prior to the CMB's release, or speeding up the recombination history, as we consider here. 
Modified recombination histories have also been considered in the context of varying fundamental constants, such as the electron mass, and have been shown to have some success in reducing the Hubble tension \cite{Hart_2020, Hart_2017, Knox_2020, Sekiguchi_2021, Sch_neberg_2022, Lee_2023, khalife2024reviewhubbletensionsolutions, calabrese2025atacamacosmologytelescopedr6}.
Late-Universe solutions on the other hand seek to modify $D_A$ from its LCDM value by introducing new physics in the post-recombination Universe, such as introducing dynamical, phantom dark energy models \cite{Di_Valentino_2016,Di_Valentino_2017,Joudaki_2018}. 
For a recent review of different efforts to resolve the Hubble tension, see Ref.~\cite{Sch_neberg_2022}.

The Lorentz force from PMFs is capable of sourcing fluctuations below the scale of diffusive damping, which erases primordially sourced adiabatic fluctuations on small scales \cite{jedamzik2011weak,jedamzik2013small,Jedamzik_2019}. 
The excess inhomogeneities lead to an overall faster recombination rate and as a result shrink the comoving sound horizon. 
Initial assessments of the viability of PMFs in alleviating the Hubble tension relied on a simplified ``three-zone model" of baryon clumping and an assumption that inhomogeneous recombination can be treated locally \cite{Jedamzik_2020,Thiele_2021,Rashkovetskyi_2021,Galli_2022}.
However, on the small scales where PMFs source compressible power prior to the CMB's release, recombination can no longer be treated as a local process, as the non-local transport of Lyman-$\alpha$ photons must be taken into account \cite{jedamzik2024cosmicrecombinationpresenceprimordial}. 
A fully self-consistent perturbed recombination code would therefore require coupling compressible magnetohydrodynamic (MHD) simulations that follow the fluid's motion in the presence of evolving magnetic fields to a radiative transfer code implementing the transport of line photons, which is computationally infeasible. In order to avoid this, recent work has used monte carlo simulations to characterize the mixing length of Lyman-$\alpha$ photons in the early Universe along with a semi-analytical treatment of radiative transfer in which a modified recombination framework was derived when Lyman-$\alpha$ photons are fully mixed \cite{jedamzik2024cosmicrecombinationpresenceprimordial}.

Despite this recent progress, there is still much more to be understood about the rich microphysics that governs perturbed recombination in the presence of small-scale inhomogeneities. 
To this end, we have devised a linearized scheme to self-consistently treat perturbed recombination in the presence of small-scale baryon clumping -- when the clumping is sourced by magnetic fields, the linearization can only be justified for certain choices of the fields' spectra. 
Fully non-linear density fluctuations or magnetic field configurations with general spectra fall outside the purview of this work, but there may nevertheless be a large region of PMF parameter space that only sources linear fluctuations and for which our framework can be directly applied. 
For the purposes of a proof of principle computation, we only consider PMF spectra which admit a self-consistent linearization, but comment throughout on the applicability of this work beyond the narrow scope of linearizable spectra that we consider here.
For a comprehensive review of commonly considered magnetogenesis scenarios and PMF spectra, see Refs.~\cite{durrer2013cosmological, subramanian2016origin, Vachaspati_2021}.

This paper can largely be considered as two distinct and independent sections. 
The first is on linear MHD (LMHD) in the early Universe and how to evolve LMHD modes and compute the statistics of LMHD fluctuations. 
The second is on perturbed recombination in the presence of small-scale fluctuations. 
The only interdependence of the two sections is that the recombination history is important for accurately evolving LMHD modes near the time of recombination, and that we apply the formalism for perturbed recombination to the specific case of LMHD sourced fluctuations.

The rest of the paper is organized as follows. 
In Section \ref{sec:MHD_derivation}, we introduce the linear MHD equations and the characteristic changes that the equations undergo as modes transition from the tight coupling regime, in which baryons and photons can be treated as one tightly coupled fluid, to the photon free-streaming regime, in which the baryon fluid is free to slip with respect to the free-streaming photons. 
We then evolve these equations to solve for transfer functions for each Fourier mode from the time of their horizon reentry to and through recombination. 
We compute linear transfer functions for the fluid density and velocity fields, the small-scale PMF perturbations, and, near recombination, for the linearly perturbed ionization fraction as well. Computing transfer functions for fluid variables and the perturbed ionization fraction near the time of recombination requires solving the linearized Boltzmann equation for Lyman-$\alpha$ photons, which mediate recombination. 
In Section \ref{sec:BE_all}, we solve the Boltzmann equation for Lyman-$\alpha$ photons and derive the perturbed recombination equations to first order, extending the results of Ref.~\cite{Venumadhav_Hirata_15} to account for the vortical modes that PMFs source. 
In Section \ref{sec:mod_rec}, we present results for the lowest order shift to the background ionization fraction, which are sourced by the linear fluid perturbations.
We then show results for the modification to the recombination history for the linearizable PMF spectra that we consider. 
In Section \ref{sec:Discussion}, we conclude and provide a discussion on the consequences of this work. 

\section{Linear MHD in the early Universe}
\label{sec:MHD_derivation}

\subsection{Linear MHD: background and formalism}
\label{sec:LMHD_background}
We begin by presenting the equations of magnetohydrodynamics, which we evolve in a cosmological background to and through recombination. 
The equations have been derived in the literature before \cite{Brandenburg_1996,SB98,JKO98}, but there are various subtleties relating to the setup, our linearization scheme, and different regimes that are realized over cosmic time, that are worth noting explicitly. 
We then qualitatively discuss the evolution of modes through the relevant regimes in the early Universe, and the different damping terms that the fluid experiences in these regimes.
Finally, we present the full LMHD equations and establish useful conventions which we use to numerically evolve the equations in the next section. 

% \vspace{15pt}
\subsubsection{Overview and our approach}

At a high level, we're studying a coupled system of fluids and magnetic fields evolving in a homogeneous, flat  Friedmann-Lemaître-Robertson-Walker (FLRW) spacetime -- we assume metric perturbations themselves are described by a scalar Newtonian-like potential, so our analysis applies within the cosmological horizon. 
In this background, imposing the conservation of the stress-energy tensor along with Maxwell's equations gives rise to the continuity and Euler equations for the fluid, and the induction equation for the magnetic field. 
These need to be supplemented with the Poisson equation for the scalar potential $\varphi$ and the generalized Ohm's law to close the system. 
A rough derivation can be found in Appendix~\ref{app:MHD_deriv}, which closely follows the derivations of Refs.~\cite{Brandenburg_1996,SB98,JKO98}. 

Under these assumptions, the full non-linear ideal MHD equations for a single fluid with energy density $\rho$, pressure $p$, and velocity $\bd{v}$, in the presence of a magnetic field with comoving strength $\bd{B}$ are
\eq{\label{eq:MHD_full}
\frac{\d \rho}{\d t} + \frac{1}{a}\nabla \cdot &[(\rho + p)\bd{v}] + 3H(\rho + p) = 0\\
\frac{1}{a^4}\frac{\d}{\d t}[a^4(\rho + p)\bd{v}] + &\frac{(\bd{v}\cdot\nabla)[(\rho+p)\bd{v}]}{a} + \frac{ \nabla \cdot [(\rho + p )\bd{v}]}{a}\bd{v}\\
+ \frac{\nabla p}{a} + &\rho\frac{\nabla \varphi}{a} + \frac{\bd{B} \times (\nabla \times \bd{B})}{4\pi a^5} = \bd{F}_d\\
\frac{\d \bd{B}}{\d t} - &\frac{1}{a}\nabla \times (\bd{v}\times \bd{B}) = 0,
}
% \begin{widetext}
% \eq{\label{eq:MHD_full}
% \frac{\d \rho}{\d t} + \frac{1}{a}\nabla \cdot [(\rho + p)\bd{v}] + 3H(\rho + p) &= 0\\
% \frac{1}{a^4}\frac{\d}{\d t}[a^4(\rho + p)\bd{v}] + \frac{(\bd{v}\cdot\nabla)[(\rho+p)\bd{v}]}{a} + \frac{ \nabla \cdot [(\rho + p )\bd{v}]}{a}\bd{v} + \frac{\nabla p}{a} + \rho\frac{\nabla \varphi}{a} + \frac{\bd{B} \times (\nabla \times \bd{B})}{4\pi a^5} &= \bd{F}_d\\
% \frac{\d \bd{B}}{\d t} - \frac{1}{a}\nabla \times (\bd{v}\times \bd{B}) &= 0,
% }
% \end{widetext}
where the symbol $\bd{F}_d$ is a generic damping term. 
The microscopic character of the fluid, and consequently the pressure $p$ and the form of the damping term change over cosmic time; we will discuss the various regimes of interest in Section \ref{sec:diff_damp}. 

The goal is to evolve the system of equations above through the epochs of interest with physically-motivated initial conditions. 
In a cosmological setting, the initial conditions are typically adiabatic fluctuations for fluid and metric variables. 
We also need initial conditions for the magnetic fields, which are supplied by the magnetogenesis model being considered.
We will consider only non-helical PMFs in this work, for which the spectrum can generically be written as~\cite{durrer2013cosmological}
\eq{\label{eq:B_spec}
\langle B_i(\bd{k},t)B_j^*(\bd{k}',t) \rangle &= (2\pi)^3 \delta^{(3)}(\bd{k}-\bd{k}')P_{ij} P_B(k),
}
where the the projection tensor $P_{ij} = \delta_{ij} - \hat{{k}}_i\hat{{k}}_j$ enforces the divergence free condition on the PMF, and $P_B(k)$ is the magnetic field's power spectrum, which we parameterize by
\eq{\label{eq:P_B(k)} 
P_B(k) = A\bigg( \frac{k}{k_0} \bigg)^{n_B}.
}
% \tv{Where's the cutoff scale in the parameterization?} \js{Addressed in following paragraph.}
This parameterization is not unique, and does not capture all generation models. For all stochastic variables, we denote the dimensionless power per log wavenumber by $\Delta_i^2(k) = k^3 P_i(k)/(2\pi^2)$. 

The system in Eq.~\eqref{eq:MHD_full} is inherently non-linear, so in general, we need to specify initial conditions for and evolve a range of scales to capture the relevant behavior. 
It is hard to use (semi-)analytical approaches for the general problem. 
In this paper, we will specialize to a particular scenario for which we can heuristically argue that an analytical approach might work -- a Universe with primordial magnetic fields with a red-tilted power spectrum, modeled as a set of small-scale modes independently evolving in the background of a large-scale `mean-field'. 
First, we decompose the PMF into a large-scale field $\mathbf{B_0}$ and a small-scale perturbation $\bd{b}$, such that $\bd{B} = \mathbf{B_0} + \bd{b}$ \cite{SB98,JKO98}. 
Second, we take the comoving mean field $\mathbf{B_0}$ to be spatially homogeneous over the length scales we consider and conserved in time ($\nabla \mathbf{B_0} = 0, \; \dot{\bd{B}}_0 = 0$), and assume the small-scale perturbations ($\bd{b}$ and associated fluid perturbations) evolve according to linear equations in this background (these are the LMHD equations, which are linearized versions of Eqs.~\eqref{eq:MHD_full}).  
We then evolve the LMHD equations to solve for linear transfer functions for each small-scale Fourier mode for all fluid and magnetic-field variables, which makes the problem very computationally tractable. 

In Appendix~\ref{appendix:LMHD_validity}, we argue this approach can work for inflation-generated PMFs with a red-tilted spectrum. We therefore write the spectral index as $n_B = \epsilon - 3$, where $\epsilon = 0$ would correspond to a perfectly scale invariant field, and $\epsilon < 0$ leads to a tilt redward from scale invariance. 
Red-tilted spectra have an infrared divergence as $k \to 0$, so we require a cutoff scale $k_{\Lambda}$ such that $P_B(k< k_{\Lambda}) = 0$. 
We require the mean field to be constant in space and time, so we conservatively set the cutoff scale to $\Lambda = 2\pi/k = 1 \text{ Gpc}$ so that it is super-horizon at the time of recombination. 
Further details and conventions for the normalization of the power spectrum, which we write in terms of the cutoff scale and the mean field, are provided in Appendix~\ref{app:PMF_conventions}.

As long as we are in the linear regime, transfer functions sourced by $\bd{b}$ and driven by $\mathbf{B_0}$ can be computed independently from standard LCDM adiabatic fluctuations \cite{Ralegankar_2024,ralegankar2024matterpowerspectruminduced}. The small scales on which PMFs naturally source baryon clumping is well below the CMB's diffusive damping tail. Primordially sourced adiabatic fluctuations on these scales are rapidly damped away and we therefore disregard them. 
We discuss this in more detail in Appendix~\ref{appendix:IC_PCR}. 

In later sections, we use transfer functions for various fluid variables to compute the lowest-order shift to the \emph{background} cosmological recombination rate in the presence of PMFs. 
This is related to the framework presented in Ref.~\cite{Lee_2021}, but generalized to account for cosmological recombination's non-local nature.

\begin{figure}[!t]
\centering
    \includegraphics[width=1\linewidth]{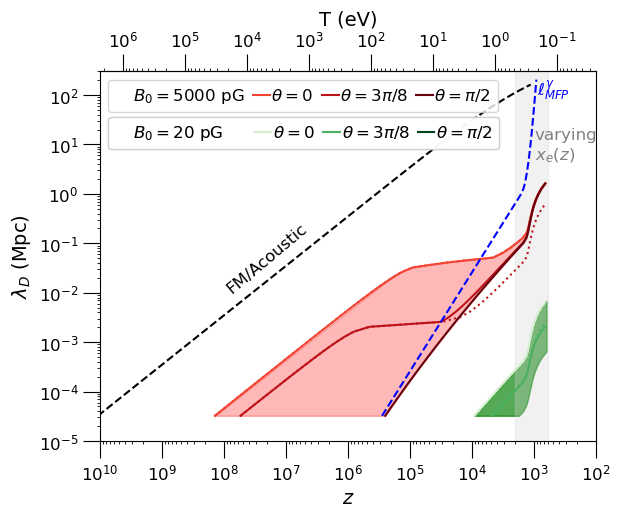}
    \caption{Damping scales for LMHD waves. 
    Fast magnetosonic (FM), slow magnetosonic (SM), and Alfvén modes are shown with dashed, solid, and dotted lines. FM modes damp like acoustic waves in LCDM, while the damping of SM and Alfvén modes depends on the background field strength $B_0$ (indicated by color), and angle $\theta$ w.r.t $\mathbf{B_0}$ (shown by a band encompassing $\theta \in [0,\pi/2]$ for SM modes). 
    SM and Alfvén modes become overdamped and their $\lambda_D$ flattens out (discussed in detail in Section \ref{sec:TCR}). 
    The blue dashed line shows a photon's comoving mean free path.
    The gray band shows the approximate region where recombination occurs; details are in Sections \ref{sec:BE_all} and \ref{sec:mod_rec}.}
    \vspace{-10pt}
    \label{fig:FM_SM_damping}
\end{figure}

\subsubsection{Fluid regimes in the early Universe} \label{sec:diff_damp}

For the inflationary generated PMFs, modes should be initialized at the time of their horizon reentry, i.e. when $\lambda \sim (aH)^{-1}$. For simplicity, we focus on the evolution of the fluid after neutrino decoupling $(T\approx 1 \text{ MeV})$. 
We therefore initialize modes which enter the horizon prior to neutrino decoupling at $T = 1 \text{ MeV}$ (the approximate temperature at which neutrinos decouple), so that we can skip over the periods of neutrino diffusion and free-streaming damping. Neutrino damping can be incorporated in a similar fashion to that due to photons following the prescriptions of Refs.~\cite{JKO98, banerjee2004evolution}.

After neutrino decoupling, the relevant length scale which determines whether or not the baryons and photons are tightly coupled is the photon's comoving mean free path $\lmfp = (an_e\sigma_T)^{-1}$.
As long as a mode's wavelength is larger than this comoving mean free path, i.e., $\lambda > \lmfp$ (or equivalently $k < (\lmfp)^{-1}$), Thomson scattering keeps the baryons tightly coupled to the photons, and the two behave as one fluid; we refer to this regime as the tight-coupling regime (TCR).
In LCDM, diffusion of CMB photons gives rise to a characteristic damping scale for scalar perturbations during the TCR, which is approximately
\eq{\label{eq:diff_damp_WKB}
k_D^{-2} = \frac{1}{6} \int \frac{dt}{a} \lmfp \frac{R^2 + 16(1+R)/15}{(1+R)^2},
} 
where we have introduced $R = 3 \rho_b/4\rho_\gamma$. A mode's amplitude is then damped by $\exp[-(k/k_D)^2]$ \cite{Hu_1997}. In the absence of PMFs, modes which satisfy $k > k_D$ are exponentially suppressed, erasing velocity and density perturbations at these scales. Figure \ref{fig:FM_SM_damping} shows both the photon mean free path $\lmfp$ and the acoustic damping scale.

Once a mode's wavelength becomes comparable to $\lmfp$, the photons begin to decouple from the baryon fluid and the tight-coupling approximation breaks down. 
At this point, the mode transitions to the free-streaming regime (FSR) and is now described by a single baryon fluid that experiences a viscous drag from the free-streaming photons. 
We denote the redshift at which this occurs for a particular mode by $\zfs$.

This transition scheme is itself a simplification of a more complicated and gradual process, which we discuss in Appendix~\ref{app:regimes_damping}. 
The transition epoch offers another mechanism by which a mode can be damped even if $k < k_D$, in principle. 
At the onset of the FSR, the fluid description for photons begins to break down, and the full Boltzmann hierarchy must be solved. 
At this point, all photon density and velocity perturbations are erased as power is transferred to higher moments, which are then exponentially suppressed \cite{Hu_1997}. Prior to recombination, Thomson scattering drags the baryon perturbations to zero along with the photon perturbations if a mode retains power when it satisfies $k \sim (\lmfp)^{-1}$. 

In Appendix~\ref{app:regimes_damping}, we find the damping term $\bd{F}_d$ of Eq.~\eqref{eq:MHD_full} to be given by
\eq{\label{eq:euler_F_d}
\bd{F}_d = 
\begin{cases}
     \frac{\eta}{a^2}[\nabla^2 \bd{v} + \frac{1}{3}\nabla (\nabla \cdot \bd{v})] \hspace{0.3cm} \text{TCR $(k < (\lmfp)^{-1})$}\\
     -\alpha\rho_b \bd{v} \hspace{2.4cm} \text{ FSR $(k > (\lmfp)^{-1})$},
\end{cases}
}
where $\eta$, the diffusive damping coefficient, and $\alpha$, the free-streaming damping coefficient, are given by~\cite{SB98,JKO98}
\eq{\label{eq:damp_coeffs}
\eta &= \frac{4}{15}\rho_\gamma a \lmfp\\
\alpha &= \frac{n_e\sigma_T}{R}.
}
In LCDM, if either damping criterion is met ($k> k_D$ or $k>(\lmfp)^{-1}$), no additional power can be sourced at this scale. Before recombination, acoustic sound waves always reach $k_D$ prior to $(\lmfp)^{-1}$, so that in LCDM no mode ever reaches the FSR prior to recombination without being significantly damped. 

In Section \ref{sec:MHD_evo}, we discuss how PMFs alter diffusive damping in the TCR in detail. 
Here, we will just emphasize a few key differences between LCDM and LMHD.  
In Fig.~\ref{fig:FM_SM_damping}, we plot the damping scales for LMHD modes. 
Compressional LMHD modes damp almost identically to acoustic sound waves in LCDM and therefore always reach $k_D$ prior to $(\lmfp)^{-1}$. 
In contrast, non-compressional modes can become overdamped during the TCR due to the driving provided by the Lorentz force, which causes the damping scale to flatten out and power to be retained on small scales~\cite{JKO98}. 
Fluid perturbations will still be erased as the overdamped modes transition to the FSR, but the remaining power in the PMF survives this transition. 
The baryonic fluid then evolves in a highly viscous state, with the Lorentz force driving the fluid and free-streaming photons damping the fluid's motion.

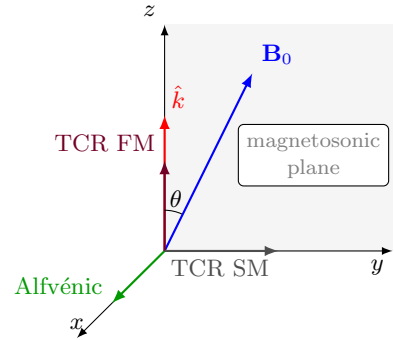
\begin{figure}[!t]
    \centering
    \begin{tikzpicture}[scale=3, >=latex]

    % Shade the y-z plane
    \fill[gray!15, opacity=0.5] (0,0,0) -- (0,1,0) -- (1,1,0) -- (1,0,0) -- cycle;

    % Axes
    \draw[->] (0,0,0) -- (0,0,1) node[anchor=south] {$x$};
    \draw[->] (0,0,0) -- (1,0,0) node[anchor=north east] {$y$};
    \draw[->] (0,0,0) -- (0,1,0) node[anchor=south east] {$z$};

    % Red k-hat vector along z
    \draw[->, red, thick] (0,0,0) -- (0,0.6,0);
    \node[red] at (0.08,0.7,0.05) {$\hat{k}$};  % shifted label

    % Vector B_0 in y-z plane
    \draw[->, thick, blue] (0,0,0) -- (0,0.4,-1) node[anchor=south west] {$\mathbf{B}_0$};

    % Alfvénic vector along x-axis
    \draw[->, thick, green!60!black] (0,0,0) -- (0,0,0.6) node[anchor=south east] {Alfvénic};

    \draw[->, thick, purple!60!black] (0,0,0) -- (0,0.4,0) node[anchor=south east] {TCR FM};
    \draw[->, thick, gray!60!black] (0,0,0) -- (0.5,0,0) node[anchor=north east] {TCR SM};

    % Draw theta arc in the y-z plane
    \draw (0,0.18,0) arc[start angle=90,end angle=65,radius=0.18];
    \node at (0.05,0.23,0) {$\theta$};

    \node[anchor=south east, font=\footnotesize, fill=white, draw=black, rounded corners=2pt, text = gray] at (0.6,-0.1,-1) {\shortstack{magnetosonic\\plane}};

    \end{tikzpicture}
    \caption{Coordinate conventions. We align the wavevector such that $\hat{\bd{k}} = \hat{\bd{z}}$, and place the mean field in the $\bd{y-z}$ plane. We define $\hat{\bd{k}} \cdot \hat{\bd{B}}_{\bd{0}} = \cos\theta$. The Alfvénic modes are always along $\hat{\bd{x}}$. During the TCR, the SM modes are along $\hat{\bd{y}}$ and the FM modes are compressible, i.e., aligned along $\hat{\bd{z}}$.}
    \label{fig:coord_conventions}
\end{figure}

Before moving to the LMHD equations in each regime and their numerical evolution, we outline a few simplifying assumptions that we make throughout the fluid's evolution:
\begin{itemize}
    \item \textbf{$e^+ e^-$ annihilation:} The diffusion coefficient is proportional to the photon's mean free path, which is itself inversely proportional to the free electron number density $n_e$. 
    %so that the greater the number of free electrons, the weaker the effect of diffusive damping. 
    Once the temperature falls below the electron's mass, $(T \lesssim 0.5 \text{ MeV})$, $e^+ e^-$ annihilation proceeds and concludes around $T = 20 \text{ keV}$. 
    We ignore this effect and set the electron fraction $x_e = n_e/n_H$ to unity at all temperatures above $T = 20 \text{ keV}$. 
    Therefore, we underestimate the free electron number density and as a result overestimate the amount of diffusive damping that fluid perturbations undergo at high temperatures ($T \gtrsim 20 \text{ keV}$). This is only relevant for modes which enter the horizon prior, and we explicitly checked for several PMF spectra that this 
%erroneous overestimation of diffusive damping when $20 \text{ keV} \lesssim T \lesssim 1 \text{ MeV}$ 
    does not appreciably alter the evolution of our modes and our final result for the modified recombination history. 
    \item \textbf{Helium recombination:} We ignore Helium recombination, which introduces extra free electrons and modifies the speed of sound in the FSR.
    \item \textbf{Single-fluid approximation:} Once a non-trivial amount of the Universe has recombined, we really should consider two fluids, one of ions which experience the Lorentz force and one of neutral atoms which do not. 
    We nevertheless treat the neutrals and ions as a single fluid and do not include ambipolar diffusion~\cite{banerjee2004evolution}.
    \item \textbf{Baryon heating and energy dissipation:} We always set the baryon temperature equal to the photon temperature: $T_b = T_\gamma \approx 2.725 \text{ K}(1+z)$. 
    This is well-justified prior to recombination when the baryons remain thermally coupled to the CMB photons \cite{Lee_2021}. 
    However, after recombination the baryon fluid can become turbulent and small-scale damping, as well as ambipolar diffusion, can heat the baryon fluid, which need not thermally equilibrate with the CMB \cite{sethi2005primordial,chluba2015effectprimordialmagneticfields,Trivedi_2018}. 
    Baryon heating ultimately slows down the recombination rate and increases the ionization fraction freeze-out value \cite{jedamzik2024cosmicrecombinationpresenceprimordial,chluba2015effectprimordialmagneticfields}.
    % After recombination, the Thomson scattering rate falls dramatically as $x_e$ plummets 
    % As a result, the linear drag $\alpha$ falls as well, which can lead to the development of turbulence in the full non-linear evolution, and the inefficiency of Thomson scattering also allows the baryons to thermally decouple from the photons.
    % The non-linear mode couplings in the post-recombination trubulent evolution causes kinetic and magnetic power to transfer to small scales where it is subsequently dissipated, heating the baryons \cite{Trivedi_2018}. 
    Since the LMHD scheme we employ is inherently incapable of capturing turbulent evolution and we ignore ambipolar diffusion, we cannot accurately account for baryon heating.
    % In the presence of PMFs, there is still uncertainty about how to accurately treat baryon heating and spectral distortions due to non-ideal fluid dissipation. 
    In Appendix~\ref{app:bar_heating}, we incorporate some relevant baryon heating effects in the linear context, but further investigation is needed. 
\end{itemize}

\subsubsection{The linearized MHD system of equations}
\label{sec:LMHD_system}

Having described the various damping terms that must be incorporated in each regime, we can use our results from Appendix~\ref{app:MHD_deriv} to write down the complete set of LMHD equations. Using dots to denote derivatives with respect to comoving time, the LMHD equations in the TCR are
\eq{
\dot{\delta}_\gamma &= \frac{1}{3}\dot{\delta}_b = -\frac{1}{3a}(\nabla \cdot \bd{v})\\
(1 + R)\dot{\bd{v}} &+ HR\bd{v} + \frac{\nabla \delta_\gamma}{a} + \frac{\bd{{B}}_0 \times (\nabla \times \bd{{b}})}{4\pi a^5 (\rho_\gamma + p_\gamma)}\\
&= \frac{\eta}{a^2(\rho_\gamma + p_\gamma)}\bigg(\nabla^2\bd{v} + \frac{1}{3}\nabla(\nabla \cdot \bd{v})\bigg)\\
\dot{{\bd{b}}} &= \frac{1}{a}\nabla \times (\bd{v} \times \mathbf{B_0}),
}
while during the FSR, the equations are
\eq{\label{eq:FSR_LMHD}
\dot{\delta_b} &= -\frac{1}{a}(\nabla \cdot \bd{v})\\
\dot{\bd{v}} &+ H\bd{v} + \frac{k_BT\nabla n_b^1}{a \rho_b } + \frac{\nabla \varphi}{a} + \frac{\bd{{B}}_0 \times (\nabla \times \bd{b})}{4\pi a^5 \rho_b} = -\alpha\bd{v}\\
\dot{\bd{b}} &= \frac{1}{a}\nabla \times (\bd{v} \times {\bd{B}}_0).
}
Moving forward, we will work with the LMHD equations in Fourier space. In the rest of this paper, we use the following Fourier conventions for all fields
\eq{
f(\bd{x}) &= \int \frac{d^3 k}{(2\pi)^{3}} f(\bd{k}) e^{i\bd{k}\cdot\bd{x}}\\
f(\bd{k}) &= \int d^3x f(\bd{x})e^{-i\bd{k}\cdot\bd{x}}.
}
We align $\hat{\bd{k}} = \hat{\bd{z}}$ and use the additional degree of freedom to put $\bd{B_0} = {B}_y\hat{\bd{y}} + {B}_z\hat{\bd{z}}$. 
We also define $\hat{\bd{k}} \cdot \hat{{\bd{B}}}_\bd{0} =\cos\theta $, so that $\bd{B_0} = {B}_0\sin\theta\hat{\bd{y}} + {B}_0\cos\theta\hat{\bd{z}}$. With these conventions, the divergence free condition on the magnetic field gives $\bd{{b}} = {b}_x\hat{\bd{x}} + {b}_y\hat{\bd{y}}$. 
We also define $\Theta = i\bd{k}\cdot\bd{v}, \Phi_x = ikv_x, \Phi_y = ikv_y$, since these quantities will naturally show up in our computation of the modified recombination history. 

\begin{table*}[t]
\large
\renewcommand{\arraystretch}{2}
\centering
\begin{tabular}{|c|c|c|c|c|}
    \hline
      & $\gamma_x,\gamma_y$ & $\gamma_z$ & $c_s^2$ & $v_A^2 = \frac{B_0^2}{4\pi a^4(\rho_0 + p_0)}$\\
     \hline
     Tight-coupling & $\frac{H + 2HR + 3\eta'\frac{k^2}{a^2}}{1+R}$ & $\frac{H + 2HR + 4\eta'\frac{k^2}{a^2}}{1+R}$ & $\frac{1}{3(1+R)}$ & $\frac{B_0^2}{4\pi a^4(\rho_\gamma + p_\gamma)(1+R)}$ \\
     \hline
     Free-streaming & $2H + \alpha$ & $2H + \alpha$ & $\frac{k_BT n_b}{\rho_b}$ & $ \frac{B_0^2}{4\pi a^4\rho_b}$ \\
     \hline
\end{tabular}%
\caption{Damping, sound speed, and Alfvén speed definitions, as they appear in Eqs.~\eqref{eq: LMHD_xi_TC} and \eqref{eq: LMHD_xi_FS}}
\label{tab:Fluid_vars_table2}
\end{table*} 

For ideal MHD, it is quite useful to define the comoving displacement field of the fluid $\bm{\xi}$, which is related to the peculiar velocity field by $\bd{v} = a\dot{\bm{\xi}}$~\cite{SB98}. Doing so allows us to integrate the induction and continuity equations to find
\eq{\label{eq:integ_induc_cont}
\delta_b &= -ik\xi_z\\
{b}_x &= ik{B}_0\xi_x\cos\theta\\
{b}_y &= ik{B}_0(\xi_y\cos\theta - \xi_z\sin\theta).
}
We can then rewrite our system of equations in terms of the displacement fields, a damping term $\gamma_i$, the speed of sound $c_s$, and the Alfvén velocity $v_A$ (we provide the expressions for each of these parameters in the TCR and FSR in Table \ref{tab:Fluid_vars_table2}). 
In the TCR, the equations are
\eq{\label{eq: LMHD_xi_TC}
\ddot{\xi}_x &+ \gamma_x \dot{\xi}_x + \frac{k^2}{a^2} v_A^2\cos^2\theta\xi_x = 0 \\
\ddot{\xi}_y &+ \gamma_y \dot{\xi}_y + \frac{k^2}{a^2} v_A^2\cos^2\theta\xi_y - \frac{k^2}{a^2} v_A^2\sin\theta\cos\theta\xi_z = 0 \\
\ddot{\xi}_z &+ \gamma_z \dot{\xi}_z  + \frac{k^2}{a^2} [ c_s^2 + v_A^2\sin^2\theta] \xi_z\\
&\hspace{2cm}- \frac{k^2}{a^2} v_A^2\sin\theta\cos\theta\xi_y = 0. 
}
The equations take a very similar form in the FSR, except for two additional phenomena: 
first, the gravitational potential becomes important for modes above the Jeans scale, so it has to be restored in the Euler equation. 
Second, the baryon thermal pressure couples the fluid equations to the evolution of the free electron fraction $x_e$.

The total baryon number density is the sum of that of Hydrogen and Helium atoms and free electrons, i.e., $n_b = n_H + n_{He} + n_e = n_H(1 + f_{He} + x_e)$, where $f_{He} = n_{He}/n_{H}$ is the Helium to Hydrogen abundance ratio, related to the Hydrogen mass fraction $X_H = 0.7546$ through $f_{He} = (1-X_H)/4X_H$. 
We also introduce the linearly perturbed free electron fraction $\delta x_e$. 
To close our free-streaming equations, we need to supplement them with an equation for the evolution of $\delta x_e$. 
This is worked out in detail in Section \ref{sec:BE_all} which closely follows the derivation from Ref.~\cite{Venumadhav_Hirata_15}. 
We introduce a rescaled quantity $ik\delta \tilde{x}_e = \delta x_e$ so that $\delta \tilde{x}_e$ has the same units as $\bm{\xi}$; with this definition, the FSR LMHD equations are
\eq{\label{eq: LMHD_xi_FS}
\ddot{\xi}_x &+ \gamma_x \dot{\xi}_x + \frac{k^2}{a^2} v_A^2\cos^2\theta\xi_x  = 0 \\
\ddot{\xi}_y &+ \gamma_y \dot{\xi}_y + \frac{k^2}{a^2} v_A^2\cos^2\theta\xi_y - \frac{k^2}{a^2} v_A^2\sin\theta\cos\theta\xi_z = 0 \\
\ddot{\xi}_z &+ \gamma_z \dot{\xi}_z  + \bigg(\frac{k^2}{a^2} [ c_s^2 + v_A^2\sin^2\theta] - 4\pi G \rho_b\bigg) \xi_z\\
&- \frac{k^2}{a^2} v_A^2\sin\theta\cos\theta\xi_y- \frac{k^2}{a^2} X_H \frac{k_B T}{m_H}\delta \tilde{x}_e = 0. 
}

We see that in both regimes the equations reduce to a coupled set of damped harmonic oscillators in the $\bd{y-z}$ plane and another equation for a damped harmonic oscillator in $\hat{\bd{x}}$ direction. These equations give rise to the linear waves of MHD now generalized to an expanding spacetime. Under our conventions, slow and fast magnetosonic waves (SM and FM) are in the $\bd{y-z}$ plane and Alfvén waves are in the $\hat{\bd{x}}$ direction, see Fig.~\ref{fig:coord_conventions}.

\subsubsection{WKB approximation and linear MHD waves}

% \js{cut down this section}

% \js{figure showing axes, with k-hat, x,y, and then the mean field and the direction that it points in or move it up to where i establisht he conventions in previous section }

Although we will proceed by directly numerically evolving all of our modes in the next section, we can first work under the WKB approximation as in Ref.~\cite{JKO98} and assume a time dependence for each of our displacement fields given by $\xi_i(k,t) = \xi_i(k) \exp[i\int_0^t \omega(t')dt']$ to develop better insight into the evolution and behavior of the LMHD waves. The WKB approximation in the TCR gives rise to a matrix equation in terms of our displacement fields $M^i_j[\omega(t)] \xi_i = 0$.

The dispersion relations for the linear waves are then the values of $\omega$ for which ${\rm Det}(M[\omega(t)]) = 0$, and the displacement vectors associated with each linear wave are the null space generators for $M[\omega(t)]$. Alfvén waves' dispersion relation is quadratic in $\omega$ and the corresponding displacement field is $\bm{\xi}_A = \xi_x$. The magnetosonic plane's dispersion relation is a quartic equation in $\omega$. This yields two pairs of solutions which approximately differ only by a phase. One pair corresponds to FM waves and one pair to SM waves. For the modes and magnetic field strengths we consider, we solve for the null space generators, to find that the FM modes are compressible, and the SM modes are rotational, i.e. $\bm{\xi}_{FM} \approx \xi_z, \bm{\xi}_{SM} \approx \xi_y$. Figure \ref{fig:coord_conventions} pictorially depicts the modes. Therefore, during the TCR, to a good approximation, each component of the displacement field is associated with one of the three LMHD modes and evolves independently. From the imaginary part of the dispersion relation, we can define a characteristic damping scale $k_D$ and overdamping scale for each mode in each regime. This is worked out in detail in Ref.~\cite{JKO98}.

In the next section, we further discuss the nature of these equations and describe how we numerically evolve them from horizon crossing to and through recombination.

\subsection{Linear MHD: evolution}
\label{sec:MHD_evo}

Equipped with the complete set of LMHD equations, we are ready to evolve our fluid variables. We start by establishing initial conditions for our modes and setting conventions for our transfer functions. We then turn to the evolution of the linear modes in each regime. For the mean field, we consider present day magnetic field strengths in the range $5 \text{ pG} \leq B_{0} \leq 5 \text{ nG}$. We evolve modes in the range $2\times 10^5 \text{ Mpc}^{-1} \lesssim  k \lesssim 3 \times 10^{-2} \text{ Mpc}^{-1}$. The largest mode was chosen to be comparable to the size of the horizon at the time of recombination, and the smallest mode was chosen based on initial simulations which established the smallest scales that clump around the time of recombination. Since we are in the regime in which the Alfvén velocity is much less than the tight-coupling speed of sound $v_A^2 \ll 1/(3[1+R])$, we do not expect the LMHD modes to disrupt the primary CMB. 

\begin{figure*}[!t]
    \includegraphics[width=1\textwidth]{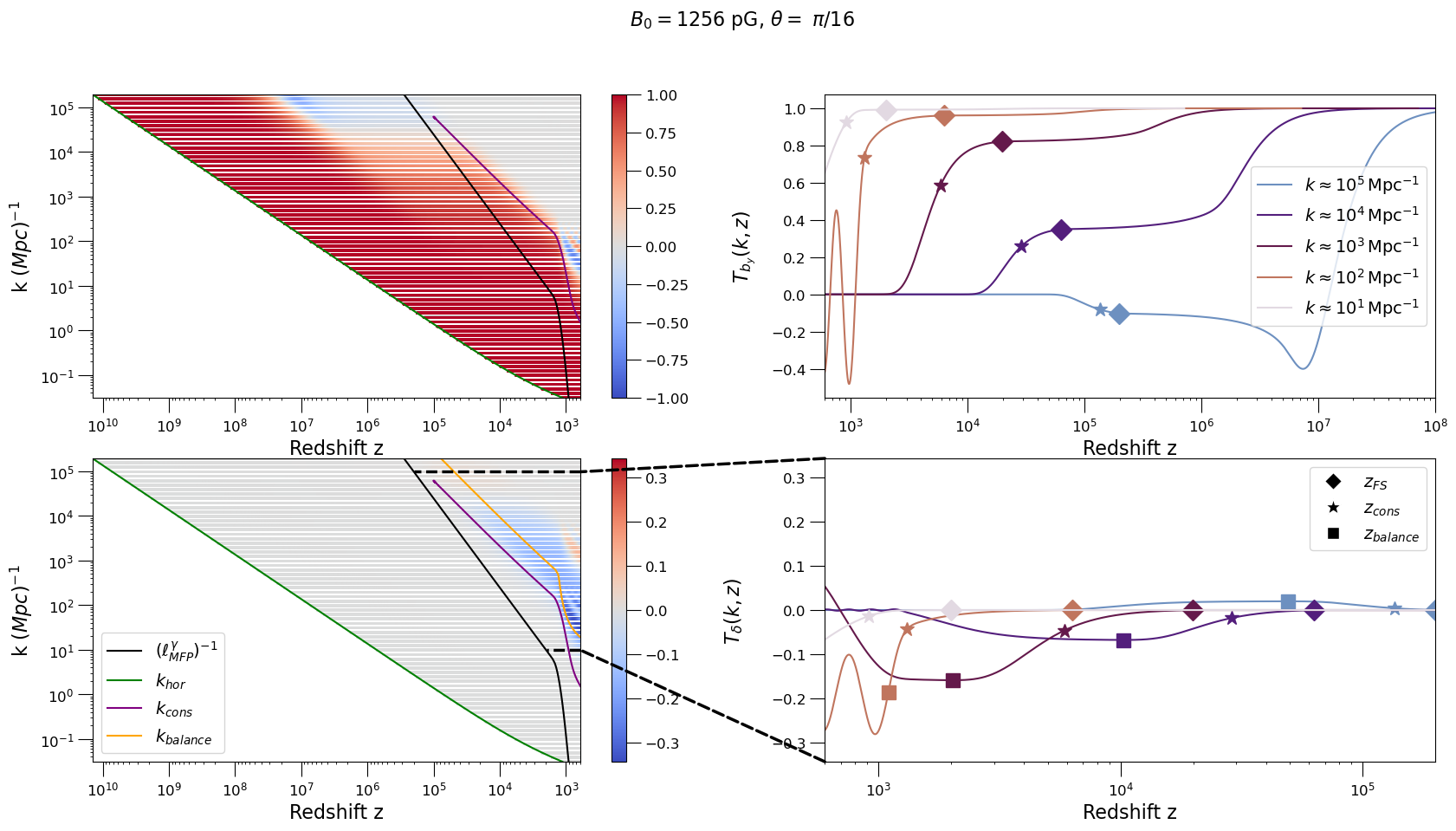}  
    \captionsetup{width=1\linewidth}
    \caption{Magnetosonic transfer functions for $b_y$ and $\delta$. (Top/Bottom row) $T_{b_y}(k,\theta,z)/T_{\delta}(k,\theta,z)$ shown across all regimes. (Left Column)  All modes are initialized at horizon reentry $k \sim k_{\text{hor}} \approx aH$, shown in green. Modes transition to free-streaming when $k \sim  (\lmfp)^{-1}$, shown in black. The purple line is a numerical calculation of the conservation scale for the PMF as given by Eq.~\eqref{eq:k_cons}. The orange line is a numerical calculation of when the various pressure terms first balance the magnetic tension in the Euler equation during the FSR, which approximates when a mode reaches its maximum compression. (Right) A few characteristic transfer functions. The redshifts at which modes transition to free-streaming, when a mode satisfies $k = \kcons$, and when a mode achieves balance between the pressure and tension terms are shown with diamonds, stars, and squares, respectively.   }
    \label{fig:Tfs_master}
\end{figure*}

\subsubsection{Initial conditions, transfer functions, and cross-correlations}
\label{sec:IC_Tfs_CCs}

We now provide some details about how we initialize various physical variables. As discussed earlier, we follow modes from either the time of their horizon reentry ($k = aH$), or after neutrino decoupling at $T \approx 1 \text{ MeV}$, whichever happens later.

For inflationary generated PMFs, perturbations are blown up to super-horizon scales ($k \ll aH$) over the course of inflation. 
The super-horizon limit of the LMHD equations leads to the conservation of the comoving magnetic field $\dot{\bd{b}} = 0$, so at horizon crossing, $\bd{b}$ is drawn from its primordial spectrum. 
In principle, adiabatic fluctuations also lead to fluctuations in the energy densities $\delta_b, \delta_r$, which could ostensibly influence the evolution of modes in the magnetosonic plane.
However, as we argue in Appendix~\ref{appendix:IC_PCR}, this turns out to be negligible for the PMF strengths and spectra we consider. 
We therefore initialize all modes with the adiabatic scalar perturbations set to zero, zero velocity field, and only an initial PMF. 

At later times, fluctuations in all fluid variables are related to the initial magnetic field fluctuations by transfer functions. 
To calculate these transfer functions, we initialize our equations with $b_x = 1$ (for Alfvénic modes) and $b_{y} = 1$ (for magnetosonic modes). 
The statistical fluctuations in all quantities are related to those in $\bd{b}$ by
\eq{\label{eq:Tfs_b}
A(\bd{k},z) = T_A(\bd{k},z)b_i(\bd{k},z_{HC}),
}
where if $A$ is any of the fluid variables in the magnetosonic plane ($\delta, \Phi_y, \Theta, b_y$) then $b_i = b_y$ and if $A$ is either $b_x$ or $\Phi_x$ then $b_i = b_x$.

With these definitions, we can also compute power spectra for and cross-correlations between fluid variables, which are needed to compute the modified recombination history and characterize the statistics of fluid perturbations. We first note that our transfer functions are not functions of the entire wavevector $\bd{k}$, but only of the magnitude $k$, and the angle $\theta$ between the large scale field and the wavevector. To compute auto- and cross-correlations, such as the clumping factor $\langle \delta^2_b \rangle$, we first define
\eq{\label{eq:cross_corr_ang_avg}
\overline{T}_{AB}(k,z) = \int_0^{\pi} d\theta \sin\theta T_A(k,\theta,z)T^*_B(k,\theta,z),
}
where $A(\bd{k},z), B(\bd{k},z)$ can be $\delta,\Phi_x,\Phi_y$, or $\Theta$ and $T_A(k,\theta,z)$ is their transfer function with respect to an initial value of $b_x$ or $b_y$. We can then write all auto- and cross-correlations for fluid variables in terms of the PMF power spectrum as
\eq{\label{eq:cross_corr}
\langle A(\bd{x},z)B(\bd{x},z) \rangle &= \frac{1}{2} \int d(\ln k) \, \Delta_B^2(k) \overline{T}_{AB}(k,z).\\
}
Since all fluid perturbations are erased once a mode enters free-streaming, the transfer functions given by Eq.~\eqref{eq:Tfs_b} can be computed in the free-streaming regime by replacing the redshift of horizon crossing, $z_{HC}$, with the redshift of free-streaming crossing, $\zfs$. Having established these conventions, we now present the evolution of our linear modes, starting with the TCR.

\begin{figure*}[!t]
    \includegraphics[width=1\textwidth]{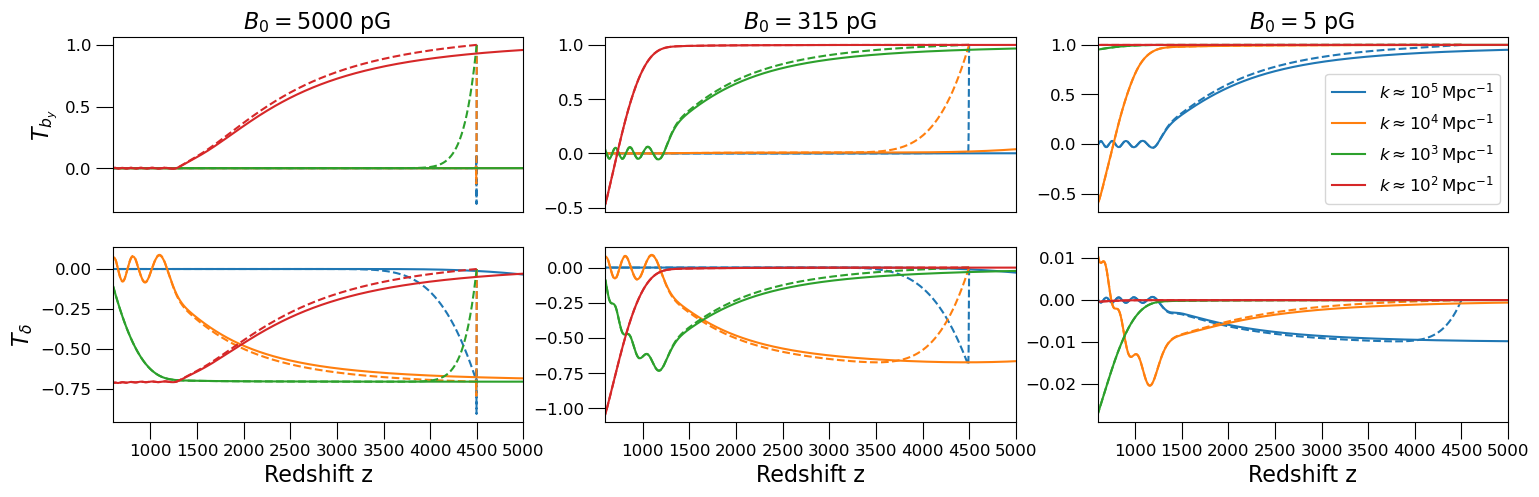} 
    \captionsetup{width=1\linewidth}
    \caption{Transfer functions for magnetosonic plane. Solid lines are initialized at $\zfs$ and dashed lines are initialized at $z = 4500$. All plots are shown for $\theta = \pi/4$. The two results quickly converge due to the terminal velocity attractor state discussed in Section~\ref{sec:TCR_FSR_transition}.}
    \label{fig:z4500_comp}
\end{figure*}

\subsubsection{The tight-coupling regime}
\label{sec:TCR}

In this section, we discuss the evolution of our modes through the TCR. During this period, the fluid is mostly incompressible due to the large sound speed of the photon component. 
PMFs therefore only excite and sustain SM and Alfvén waves in the TCR; any compressible FM waves excited by primordial density fluctuations are rapidly damped away on all the scales relevant to our calculation.

Following~\cite{JKO98}, we define the characteristic damping scale of a mode $\lambda_D(z)$, as the largest wavelength damped by at least one e-fold at any given redshift. 
We can read off the damping scale for each mode from its respective transfer function. 
To obtain the linear transfer functions for FM modes, we can initialize with $b_x = b_y = 0, \delta =1$ (or equivalently $\xi_x = \xi_y = 0, \xi_z = 1$). 
We only consider PMF strengths for which $v_A^2 \ll c_s^2$ during the TCR. 
Therefore, in $\xi_z$'s evolution equation, Eq.~\eqref{eq: LMHD_xi_TC}, the Lorentz force is negligible and FM waves damp like acoustic waves in standard LCDM. 
At the time of recombination, $\lambda_D^{\rm FM}$ (shown in Fig.~\ref{fig:FM_SM_damping}) is always larger than the photon mean free path, and any primordial density perturbations are damped away in the TCR in agreement with the WKB results from Ref.~\cite{JKO98}.

Since we have incompressible MHD during the TCR, SM and Alfvén modes' evolution are nearly identical prior to the FSR. 
From the integrated induction Eq.~\eqref{eq:integ_induc_cont}, we see that the transfer functions for $b_y$, as shown in the top right of Fig.~\ref{fig:Tfs_master}, are proportional to $\xi_y$ during the TCR, i.e. the SM modes.
Depending on the strength of the magnetic field, some SM modes oscillate and decay, while others can become overdamped and survive the TCR. 
Unlike FM modes, the damping of SM and Alfvén modes are sensitive to $B_0$ and $\theta$. 
Their damping scales $\lambda_D^{\rm SM}(z)$ and $\lambda_D^{\rm A}(z)$ are shown in Fig.~\ref{fig:FM_SM_damping}.
The flattening of the damping scales reflects the fact that these rotational modes become overdamped. 
A weaker PMF amplitude leads to a weaker driving term and therefore allows smaller modes to become overdamped and  survive the TCR.

\subsubsection{The transition from tight-coupling to free-streaming}
\label{sec:TCR_FSR_transition}

To accurately handle the transition to free-streaming, we need to evolve the baryon fluid equations coupled to the photon Boltzmann hierarchy. 
We defer this work to a future study, and instead employ a simplified scheme in which we evolve the fluid using the TCR equations given by Eq.~\eqref{eq: LMHD_xi_TC} while $\lambda > \lmfp$. 
Once a mode satisfies $\lambda = \lmfp$, we erase all fluid velocities, and instantaneously transition to the FSR, where Eq.~\eqref{eq: LMHD_xi_FS} applies.

Despite our somewhat haphazard treatment of the regime transition, we can still be reasonably assured in our results because of the presence of a strong attractor state in the highly viscous MHD which governs the FSR evolution. In particular, the drag provided by the photons from the $\alpha \bd{v}$ term in Eq.~\eqref{eq:FSR_LMHD} is very large due to the small mean free path of photons prior to recombination. We therefore expect to have balance between the large photon drag and the Lorentz force term, so that the fluid quickly goes to the terminal velocity state
\eq{\label{eq:term_vel}
\bd{v}_T &\sim -\frac{{\bd{{B}}}_0 \times (\nabla \times {\bd{{b}}})}{4\pi a^5\rho_b \alpha}
}
when it enters the FSR  \cite{banerjee2004evolution}. 

We numerically confirmed that this holds whether we initialize with the velocity set to zero or with the velocity set to its value from the TCR at the time of $\zfs$. This suggests that the terminal velocity state's basin of attraction is large enough to ensure that the simplistic instantaneous regime-transition scheme with all fluid perturbations erased is adequate.

\begin{figure*}[!t]
    \includegraphics[width=1\linewidth]{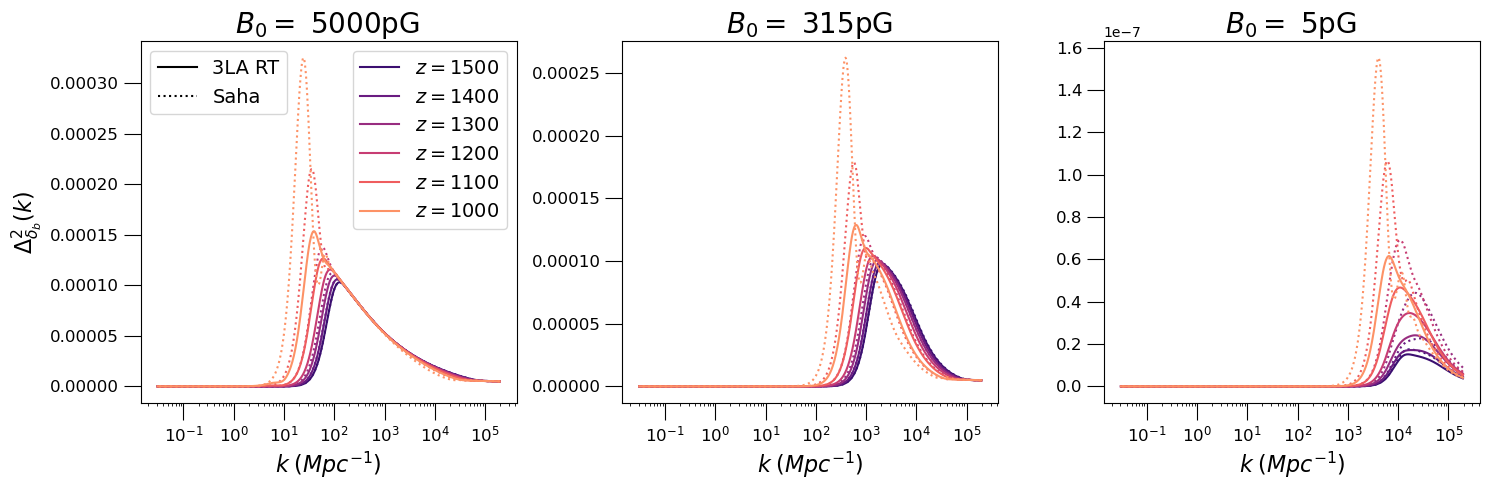} 
    \captionsetup{width=1\linewidth}
    \caption{Dimensionless power per log wavenumber for $\delta_b$. Examples shown for spectral tilt $\epsilon = -0.1$. Dotted lines use the Saha model for recombination (Eqs.~\eqref{eq:saha_rec_homo} and \eqref{eq:saha_rec_first_O}) and solid liens are with $x_e$ set by the three-level atom (3LA) model for recombination (Eqs.~\eqref{eq:rec_TLA_zero_O} and \eqref{eq:rec_TLA_first_O}) with radiative transfer (RT) taken into account. }
    \label{fig:deltam_power}
\end{figure*}

In Fig.~\ref{fig:z4500_comp}, we show the difference in the FSR transfer functions when we initialize at $z = 4500$ instead of $\zfs$, to compare to the initial conditions used by Ref.~\cite{jedamzik2024cosmicrecombinationpresenceprimordial}. 
The results indicate that in general, there is little difference between whether we initialize at $\zfs$ or at $z=4500$. 
This can be ascribed to the presence of the strong attractor state in the FSR, so that even a delayed initialization reaches the correct terminal state well before recombination. 
We find that results tend to converge by $z \approx 2000$, further lending credence and support to the initialization scheme used in Ref.~\cite{jedamzik2024cosmicrecombinationpresenceprimordial}.

\subsubsection{The free-streaming regime}\label{sec:FSR}

During the FSR, the loss of pressure support from the now decoupled photons allows the PMF to source compressional power. The Lorentz force contributes to $\xi_z$'s evolution, found in Eq.~\eqref{eq: LMHD_xi_FS}, in two ways. 
The first is through the magnetic pressure, given by $(k^2/a^2)v_A^2\sin^2\theta \xi_z$, which works with thermal baryonic motion to provide additional pressure support to the fluid and resist compression. 
The second term is the magnetic tension projected onto the compressible $\hat{\bd{z}}$ direction and is given by $(k^2/a^2)v_A^2\sin\theta\cos\theta \, \xi_y$. 
The latter serves as a driving term in the otherwise damped oscillatory evolution of $\xi_z$ -- it tries to compress the fluid when modes enter free-streaming, hence $\xi_z$ (or equivalently $\delta_b$) can grow if the magnetic tension can overcome the various pressure terms. 
Eventually, the driving magnetic tension and the pressure terms balance at which point the compressional mode stops growing. 
As the magnetic field continues to decay, the now dominant pressure causes the compressed fluid to expand leading to the eventual decay in $\xi_z$. Throughout, $\xi_y$ decays as well due to the viscous damping it experiences from the free-streaming photons. Once the fluid returns to the equilibrium, uncompressed state, it briefly oscillates about this state, before lasting equilibrium is achieved, with the remaining power in $\xi_y$ and $\xi_z$ damped away. 
This qualitative evolution history presented above can be seen from the FSR transfer functions shown in Figs.~\ref{fig:Tfs_master} and \ref{fig:z4500_comp}.

\subsubsection{Baryon clumping and compression}
\label{sec:clumping}

In Fig.~\ref{fig:deltam_power}, we show the baryonic density's dimensionless power $\Delta_{\delta_b}^2(k)$ as a function of $k$ for a range of redshifts. 
The first feature of note is the substantial difference between the three-level atom and Saha models for recombination. 
We discuss each of these models in detail in the next section. 
For now, we can simply note that Saha model of recombination overestimates the rate of recombination. 
A faster recombination leads to increased baryon clumping since the loss of free electrons causes $c_s$ to fall by a factor of $1/\sqrt{2}$ very rapidly. 
There is therefore less pressure support in the Saha model of recombination and the fluid can be more easily compressed, as evidenced by the results in Fig.~\ref{fig:deltam_power}.

We can also see that the amplitude of the perturbations and characteristic scale on which we expect to source the most compressible power depends on the strength of the mean field $B_0$. For the range of field strengths we consider, we find that compressible power in our fluid near the time of recombination peaks on $k \sim 0.1 \text{ kpc}^{-1} - 10 \text{ kpc}^{-1}$ scales. This is in good agreement with the findings of Ref.~\cite{jedamzik2023primordialmagneticfieldshubble}, which found similar scales from their full nonlinear compressible MHD simulations. The findings of Ref.~\cite{jedamzik2024cosmicrecombinationpresenceprimordial} suggest uncertainty over the required resolution and dynamical range for nonlinear compressible MHD simulations. LMHD can serve to guide and inform the simulation parameters in future compressible MHD simulations. Our results suggest that a larger than expected dynamical range is required. This is already alluded to in Appendix C in Ref.~\cite{jedamzik2024cosmicrecombinationpresenceprimordial}, which found that their clumping factor did not converge as they increased the resolution of their simulation. From Fig.~\ref{fig:deltam_power}, it can be seen that for particularly weak fields, baryon clumping can take place even at scales as small as $k \approx 10^5 \text{ Mpc}^{-1}$ suggesting that a very fine resolution may be needed in compressible MHD simulations with sub-pG field strengths.

\begin{figure}[!t]
    \centering
    \includegraphics[width=1\linewidth]{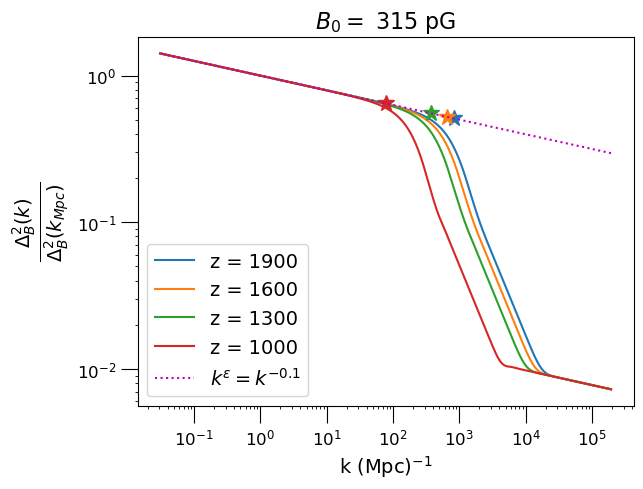} 
    \captionsetup{width=1\linewidth}
    \caption{Dimensionless power per log wavenumber for PMF normalized to its value at $k = 1 \text{ Mpc}^{-1}$. Dotted magenta line is the primordial spectrum. The spectra2 are shown for a few different redshifts, with the corresponding value of $\kcons$ for each redshift shown as a star along the primordial spectrum $k^\epsilon$. }
    \label{fig:Delta_B_evo}
\end{figure}

In Fig.~\ref{fig:Delta_B_evo}, we show $\Delta_B^2(k)$'s evolution as a function of redshift. 
As an example, we plot results for $B_0 = 315 \text{ pG}$, $\epsilon = -0.1$, but the qualitative features are similar for other amplitudes and spectral indices. 
It can be seen that there is a particular scale at which the spectrum begins to deviate from the primordial spectrum, which increases with redshift. 
If we restrict our attention to the FSR, then we can simply estimate the scale at which we expect this to occur. 
From the induction equation, we have $a\dot{\bd{b}} = \nabla \times (\bd{v} \times \mathbf{B_0})$. We therefore expect scales that satisfy $aHb \gg kvB_0$ to be conserved. 
From the terminal velocity approximation, we have that the FSR velocity is approximately given by $v \sim kbv_A^2/(a\alpha B_0)$, so that scales which satisfy $  k^2 \ll a^2\alpha H/v_A^2 $ are expected to have a conserved comoving PMF. 

We denote the smallest scale at which we expect the field to be conserved by 
\eq{ 
\label{eq:k_cons}
\kcons^2 = a^2\alpha H/v_A^2.
}
The colored stars in Fig.~\ref{fig:Delta_B_evo} show $\kcons$ as a function of redshift. 
In both Figs.~\ref{fig:Tfs_master} and \ref{fig:Delta_B_evo}, it can be seen that the actual conservation scale (the wavenumber at which $\Delta_B^2(k)$ begins to deviate form its primordial value) is slightly smaller than $\kcons$, though always within an order of magnitude. 
This is likely because the correct conservation criterion is $k \ll a^2\alpha H/v_A^2 = \kcons$.
Nevertheless, $\kcons$ provides a useful order of magnitude estimate for the conservation scale and the onset of baryon clumping.  
A similar analysis performed in the full nonlinear case for post-recombination evolution found that the $aH \gg kv$, with $v$ approximated by the terminal velocity found by balancing the Lorentz force with the drag term, is a good criterion for conservation of the comoving PMF~\cite{Ralegankar_2024,ralegankar2024matterpowerspectruminduced}. 

The conservation scale also provides a good estimate of the largest scale that can clump as a function of redshift. So long as the magnetic field is conserved, no compressible power can be sourced. However, once the field begins to evolve, power can be transferred from the PMF to compressional modes. This can be seen from the right column in Fig.~\ref{fig:Tfs_master}, in which $\kcons$ decently approximates the onset of compressional power at a particular scale.

\section{Modified recombination in an inhomogeneous Universe}
\label{sec:BE_all}

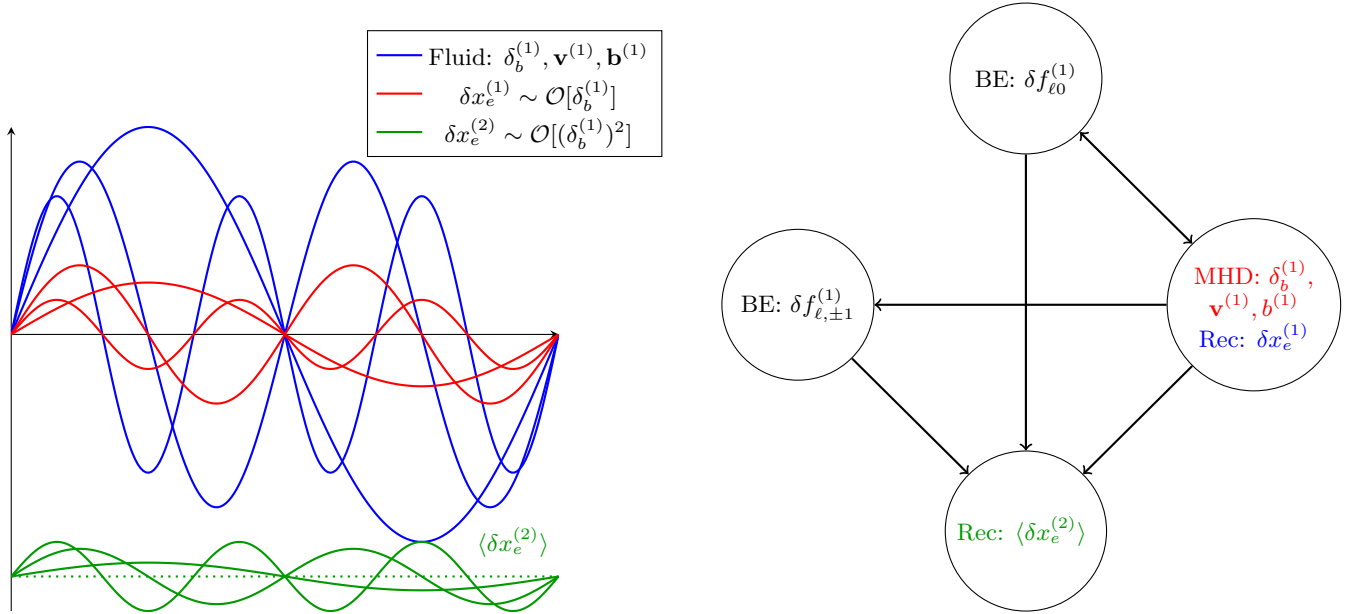
\begin{figure*}[t]  % Use figure* to span across both columns in two-column layout
    \centering
    % Left figure (Sine plot)
    \begin{minipage}[t]{0.49\textwidth}
        \centering
        \begin{tikzpicture}
            \begin{axis}[
                axis lines=middle,
                domain=0:4*pi,
                samples=200,
                width=\textwidth,
                height=8cm,
                legend style={at={(0.65,1.2)}, anchor=north west},
                xtick=\empty, ytick=\empty,
            ]

            \addplot[blue, thick] coordinates {(0,0)}; 
            \addlegendentry{Fluid: $\delta_b^{(1)}, \bd{v}^{(1)}, \bd{b}^{(1)}$}

            \addplot[red, thick] coordinates {(0,0)};
            \addlegendentry{$\delta x_e^{(1)} \sim \mathcal{O}[\delta_b^{(1)}]$}

            \addplot[green!60!black, thick] coordinates {(0,0)};
            \addlegendentry{$\delta x_e^{(2)} \sim \mathcal{O}[(\delta_b^{(1)})^2]$}

            \addplot[blue, thick] {0.5*sin(deg(x))};
            \addplot[blue, thick] {0.4*sin(1.5*deg(x))};
            \addplot[blue, thick] {0.6*sin(0.5*deg(x))};

            \addplot[red, thick] {0.2*sin(deg(x))};
            \addplot[red, thick] {0.1*sin(1.5*deg(x))};
            \addplot[red, thick] {0.15*sin(0.5*deg(x))};

            \addplot[green!60!black, thick] {0.08*sin(deg(x)) - 0.7};
            \addplot[green!60!black, thick] {0.1*sin(1.5*deg(x)) - 0.7};
            \addplot[green!60!black, thick] {0.04*sin(0.5*deg(x)) - 0.7};

            \addplot[green!60!black, dotted, thick, domain=0:4*pi] {-0.7};
            \node[anchor=west, text=green!60!black] at (axis cs: 10.5, -0.6) {$\langle \delta x_e^{(2)} \rangle$};
            \end{axis}
        \end{tikzpicture}
    \end{minipage}
    \hfill
    % Left figure (Bubble diagram)
    \begin{minipage}[t]{0.49\textwidth}
        \centering
        \begin{tikzpicture}[
          bubble/.style={draw, circle, minimum size=2cm, align=center},
          arrow/.style={->, thick},
          doublearrow/.style={<->, thick}
        ]
        
        \foreach \i/\name/\label in {
          0/A/{\textcolor{red}{MHD: $\delta_b^{(1)},$}\\ \textcolor{red}{$\bd{v}^{(1)},b^{(1)}$}\\ \textcolor{blue}{Rec: $\delta x_e^{(1)}$}},
          90/C/{BE: $\delta f_{\ell 0}^{(1)}$},
          180/D/{BE: $\delta f_{\ell,\pm 1}^{(1)}$},
          270/E/{\textcolor{green!60!black}{Rec: $\langle\delta x_{e}^{(2)}\rangle$}}
        } {
          \pgfmathsetmacro\x{3*cos(\i)}
          \pgfmathsetmacro\y{3*sin(\i)}
          \node[bubble] (\name) at (\x,\y) {\label};
        }

        \draw[doublearrow] (A) -- (C);
        \draw[arrow] (A) -- (D);
        \draw[arrow] (A) -- (E);
        \draw[arrow] (C) -- (E);
        \draw[arrow] (D) -- (E);
        \end{tikzpicture}
    \end{minipage}

    \caption{Schematic representations of (left) LMHD modes and the perturbed ionization fraction (right) interdependence of LMHD variables, moments of the Lyman-$\alpha$ photon phase-space density, and the perturbed ionization fraction.}
    \label{fig:combined_mhd_schematic}
\end{figure*}

Now that we have a full account of the linear transfer functions for our LMHD modes, we turn to a detailed discussion of recombination. 
Recombination has two effects in our setup. 
First, during the FSR, the ionization fraction $x_e(z)$ impacts the baryon thermal pressure and hence modifies the evolution of baryonic density modes.
Therefore, once $x_e(z)$ starts to deviate from $1$ ($z \lesssim 2000$), shown as the gray band in Fig.~\ref{fig:FM_SM_damping}, we need to couple our fluid equations to a recombination framework to accurately compute transfer functions. 
Second, a modified recombination history affects observables of interest such as the location of the surface of last scattering, which is one of the primary motivations of this work. 

In the context of cosmological recombination, it was first noted by Ref.~\cite{jedamzik2011weak} that an excess in baryon clumping, such as the one sourced by PMFs, speeds up the overall background recombination rate, since the latter depends in a non-linear fashion on the local matter density field ($\langle n_e^2 \rangle \geq \langle n_e \rangle^2$). 

In the perturbative framework we are employing, all spatial averages of linearly perturbed quantities necessarily vanish, including $\langle \delta x_e^{(1)} \rangle$ (the number in parentheses in the superscript indicates the order in the perturbative expansion). 
Since we are interested in the lowest-order shift to the background ionization fraction, defined as the average of the number of ionized electrons divided by the number of hydrogen nuclei in a sufficiently large patch, we need to go beyond the strictly linearly perturbed quantities we have considered thus far. 
This is illustrated in the left panel of Fig.~\ref{fig:combined_mhd_schematic}. 
The lowest order shift to the background ionization fraction is given by
\eq{\label{eq:rec_second_O}
\Delta x_e = \bigg\langle \frac{n_e^{(2)}}{n_H^{(0)}} \bigg\rangle &= \langle \delta x_e^{(2)} \rangle + \langle \delta x_e^{(1)}\delta_b^{(1)} \rangle.
}
The first term is the second-order perturbed ionization-fraction $\delta x_e^{(2)}$ for long-wavelength modes ($k \rightarrow 0$), and the second term is the cross-correlation of the linear perturbations to the ionization fraction and the baryon density. 
In Fourier space, the latter requires us to track all short-wavelength linear modes, but the former needs only the $k \rightarrow 0$ part of the second-order piece $\delta x_e^{(2)}$.  
In the context of Fig.~\ref{fig:combined_mhd_schematic}, this means that we do not need to compute the spatially varying component of $\delta x_e^{(2)}$ (the green sinusoidal curves) which greatly simplifies the computation. 
In order to compute transfer functions for LMHD variables and $\delta x_e^{(1)}$ near the time of recombination and compute the $k \rightarrow 0$ component of $\delta x_e^{(2)}$, we need a model for recombination. 

Beginning with the seminal work of Refs.~\cite{peebles1968recombination,zeldovich1969recombination} the theory of cosmological recombination has been developed to percent-level precision as needed to interpret the primary anisotropies in the CMB \cite{Seager_2000, Seager_1999, Chluba_2010, PhysRevD.102.083517,PhysRevD.83.043513}. 
The simplest treatment of recombination in the presence of fluctuations, such as the LMHD modes we discussed in the previous section, is to perturb the homogeneous equations for the evolution of $x_e$. 
This corresponds to a `local' treatment of recombination in which inhomogeneities locally change cosmological parameters~\cite{Senatore_2009, Lee_2021}. 

A significant channel for recombination to the ground state of neutral Hydrogen ($n=1$) is via the $2p$ state and the emission of a Lyman-$\alpha$ photon. 
For this process to result in a net recombination, the Lyman-$\alpha$ photon has to redshift out of resonance. 
When the Universe is inhomogeneous on scales that are sufficiently large relative to the Lyman-$\alpha$ photons' mixing scale (as is the case for adiabatic fluctuations that survive until recombination in LCDM), we can disregard their spatial transport so that treating recombination locally is well-justified.
However, the compressible MHD modes of interest have relatively short wavelengths and Lyman-$\alpha$ photons can stream over these length scales.
We therefore need to explicitly account for the non-local radiative transport of Lyman-$\alpha$ photons to accurately compute recombination histories in the presence of these fluctuations \cite{jedamzik2024cosmicrecombinationpresenceprimordial}. 
In the rest of this section, we describe how inhomogeneities change recombination -- first by assuming it is a local process, and next by using the full non-local computation. 

\subsection{Local inhomogeneous recombination}
\label{sec:local_rec_all}

In this section, we discuss how fluid perturbations alter the recombination history using a purely local treatment. 
We will first work with the Saha model, which is a very simplified model that assumes local thermodynamic equlibrium (LTE) sets the ionization fraction. 
We then turn to the three-level atom (3LA) model for recombination, first proposed in Refs.~\cite{peebles1968recombination,zeldovich1969recombination}, which is an effective, yet highly accurate framework for recombination. 
In this section, we only provide a qualitative discussion of inhomogeneous local recombination in the 3LA.
We will make these claims rigorous in Section \ref{sec:lin_pert_BE} by taking the large scale limit of the non-local recombination calculation. 

\subsubsection{Saha model}
\label{sec:local_rec_saha}

The Saha model for recombination assumes LTE between the ground and ionized states of the Hydrogen atom/ion system, with ionizations mediated by blackbody CMB photons ($e^- + p \leftrightarrow H + \gamma$). 
The equilibrium condition for this system sets the ionization fraction to
\eq{\label{eq:saha_rec_homo}
\frac{x_e^2}{1-x_e} &= \frac{(2\pi m_e k_B T)^{(3/2)}}{h^3 n_H}e^{-E_I/k_B T},
}
where $E_I = 13.6 \text{ eV}$ is the ionization energy for Hydrogen. 
LTE is a good assumption at early redshifts, but it breaks down once recombination begins in earnest ($z\lesssim 2000$). 
In Appendix~\ref{app:BE_cont}, we solve for the transport of continuum photons, and recover the well-known fact that in a homogeneous Universe, direct recombination to the ground state has a negligibly small contribution to the overall recombination rate (the ionizing photons have extremely short mean free paths which leads to detailed balance between direct recombination and photoionization).

Before moving to the more realistic 3LA model for recombination, let us compute the shift to the background ionization fraction within the Saha model. 
Perturbing Eq.~\eqref{eq:saha_rec_homo} to linear order, we find
\eq{\label{eq:saha_rec_first_O}
\delta x^{(1)}_e = -x_e \delta^{(1)}_b\frac{1-x_e}{2-x_e}.
}
We can directly substitute this into the FSR MHD equations (Eq.~\eqref{eq: LMHD_xi_FS}) which enables us to compute the relevant transfer functions in this approximation. 

As noted above, to get the correction to the background ionization history, we additionally need the spatially-averaged second-order perturbation. 
Proceeding similarly to the first-order case, we find
\eq{\label{eq:saha_rec_second_O}
\langle \delta x^{(2)}_e \rangle &= x_e\bigg(\frac{1-x_e}{2-x_e} - \frac{1-x_e}{(2-x_e)^3} \bigg) \langle \delta_b^{(1)}\delta_b^{(1)} \rangle,
}
which is a strictly positive quantity. 
Substituting Eqs.~\eqref{eq:saha_rec_first_O} and \eqref{eq:saha_rec_second_O} into Eq.~\eqref{eq:rec_second_O}, we find the mean shift to the ionization fraction to be
\eq{\label{eq:dxe2_saha}
\Delta x_e &= - x_e\frac{1-x_e}{(2-x_e)^3} \langle \delta^{(1)}_b \delta^{(1)}_b \rangle,
}
which is a strictly negative quantity. Within the Saha model, we see that excess baryon clumping does in fact speed up the overall rate of recombination, as expected.

% , which accounts for the Universe's expansion and mediation by non-blackbody photons, two crucial features that the Saha model neglects~\cite{caltechthesis6404}.

\subsubsection{Three-level atom model}
\label{sec:3LA}

Since direct recombination to the ground state is ineffective, recombination primarily proceeds as follows. 
An ionized hydrogen atom first recombines to an excited state and relaxes down to the $n=2$ state, with all excited states in thermal equilibrium. 
A successful recombination to the ground state can then occur either from the $2s$ state via a two-photon emission or from the $2p$ state by a Lyman-$\alpha$ photon redshifting out of resonance. 
Recombination is therefore accurately captured by the 3LA, which approximates the full multi-level hydrogen atom as an ionized state, the $n=2$ state, and the $n=1s$ ground state.

As mentioned earlier, to compute the recombination history, we need to account for Lyman-$\alpha$ photons, which we describe using their phase-space density (PSD) $f(\nu)$. 
In the limit of an infinitesimally thin Lyman-$\alpha$ line, the rate of removal of photons is given by the difference between the line-averaged PSD $\overline{f}$ (we will define line averages later when we work with the Boltzmann equation), and the equilibrium PSD $\feq$. 
Assuming all angular momentum states in the $n=2$ level are equally occupied, the equilibrium PSD is given by
$\feq = x_2/4 x_{1s}$, where $x_2 = (n_{2p} + n_{2s})/n_H$. 
The rate of recombination through the $2p$ channel is then given by
\eq{\label{eq:rec_x2p1s_dot}
\dot{x}_{1s} \vert_{2p, 1s} &= 3A_{\text{Ly}\alpha}x_{1s}[\feq - \overline{f}].
}
For the transition from the $2s$ state to the ground state, the recombination rate is given by
\eq{\label{eq:rec_x2s1s_dot}
\dot{x}_{1s}\vert_{2s,1s} &= \Lambda_{2s,1s} x_{1s}\bigg[\feq - e^{-(h\nu_{\text{Ly}\alpha})/(k_B T)} \bigg],\\
}
where the rate $\Lambda_{2s,1s} \approx 8.22 \text{ s}^{-1}$ \cite{Labzowsky_2005}. 
Since the total recombination rate in the 3LA is the sum of Eqs.~\eqref{eq:rec_x2p1s_dot} and \eqref{eq:rec_x2s1s_dot}, we require an expression for $\feq$ (or equivalently $x_2$) and $\overline{f}$. 

For case-B recombination (recombination to $n=2$), the occupancy of the $n=2$ level evolves according to
\eq{\label{eq:rec_x2_dot}
\dot{x}_2 &= n_Hx_e^2 \alpha_B - x_2\beta_B - \dot{x}_{1s} \vert_{2p,1s} - \dot{x}_{1s}\vert_{2s,1s}\\
&= n_Hx_e^2 \alpha_B - 4 x_{1s}\feq\beta_B - \dot{x}_{1s} \vert_{2p,1s} - \dot{x}_{1s}\vert_{2s,1s},\\
}
where the first two terms account for recombination to and photoionization from $n=2$, and the last two terms account for the two avenues for relaxation to the ground state: from the $n=2p$ and $n=2s$ states. 
Finally, we note that the left hand side of Eq.~\eqref{eq:rec_x2_dot} is of order Hubble ($\dot{x}_2 \sim Hx_2$), while all the atomic transition rates on the right hand side are large compared to $H$. 
We can therefore work in the steady-state approximation in which $\dot{x}_2 \approx 0$, which allows us to relate $\feq$ to $\overline{f}$. 
To derive the 3LA recombination equation, we require an additional equation for the line-averaged PSD $\overline{f}$, which comes from the BE for Lyman-$\alpha$ photons.
We solve the homogeneous BE in Appendix~\ref{app:sob_sol}. 
Substituting this solution back into equation \eqref{eq:rec_x2_dot} and simplifying yields the Peebles 3LA recombination equation
\eq{\label{eq:rec_TLA_zero_O}
\dot{x}_e &= -C(n_Hx_e^2\alpha_B - 4x_{1s}\beta_B e^{-(h\nu_{\text{Ly}\alpha})/(k_B T)}),
}
where $\alpha_B$ and $\beta_B$ are effective case-B recombination and photoionization coefficients, $x_{1s} = n_{1s}/n_H = 1 - x_e$, and $C$ is the Peebles $C$-factor which accounts for the fact that not all atoms which relax to the $n=2$ state successfully reach the ground state, since some will be ionized or excited \cite{peebles1968recombination, zeldovich1969recombination, Seager_2000}. 
The explicit expression for $C$ is given in Eq.~\eqref{eq:peeb_C_hom} and is recovered from the homogeneous BE in Appendix~\ref{app:sob_sol}.

We end this section with a qualitative discussion on how inhomogeneities modify the above picture if we neglect the spatial transport of Lyman-$\alpha$ photons. 
In the presence of local density perturbations, a Lyman-$\alpha$ photon will encounter a different density of neutral hydrogen atoms, i.e. $n_H \to n_H(1+ \delta_b)$.  
Additionally, in the presence of a local velocity divergence, the redshift rate must be modified to account for either the convergence or expansion of the fluid, namely $H \to H+\Theta/3a$ \cite{Senatore_2009}. 
These two effects (a change in the collision rate on account of a modified density field and a change in the Hubble expansion on account of a local velocity divergence) modify the local escape rate of Lyman-$\alpha$ photons. 
Under the local recombination assumption, any patch in the Universe can be treated with the standard 3LA recombination equations so long as the Lyman-$\alpha$ escape rate is modified accordingly. 
We show how this emerges from taking the large scale limit of the BE in Section \ref{sec:lin_pert_BE}.
The prescriptions of Ref.~\cite{Lee_2021} can then be used to compute the lowest order shift to the background ionization fraction due to the presence of fluid perturbations.

\subsection{Non-local recombination}
\label{sec:non-local_Rec}

We now turn to non-local recombination, which is necessary to accurately handle the small-scale perturbations that PMFs source. 

We will first introduce the full Boltzmann equation for Lyman-$\alpha$ photons and highlight some of its key features. 
We then solve the homogeneous BE using the Fokker-Planck approximation for Lyman-$\alpha$ scattering and photon redistribution. Finally, we examine the behavior of the Boltzmann equation in a perturbative setup. 

In the inhomogeneous case, the Lyman-$\alpha$ PSD depends on the direction of photon propagation, so we will expand the perturbed PSD $\delta f^{(1)}$ and $\delta f^{(2)}$ in spherical multipoles. 
Once we account for the PSD's dependence on photon propagation, we need to replace the line average of the PSD in equation \eqref{eq:rec_x2p1s_dot} with the line average of the monopole $\overline{f}_{00}$.

Recall from Eq.~\eqref{eq:rec_second_O} that the average shift to the background ionization fraction depends on the first- and second-order perturbations to the ionization fraction ($\delta x_e^{(1)}(k)$ and $\langle \delta x_e^{(2)} \rangle$). 
%Recall from Eq.~\eqref{eq:rec_second_O} that the average shift to the background ionization fraction has two contributions -- one term which depends on the short wavelength modes of the first-order ionization fraction and density perturbations $\delta x_e^{(1)}$ and $\delta_b^{(1)}$, and one term which requires only the long wavelength ($k\to 0$) mode of the second-order ionization fraction $\langle \delta x_e^{(2)} \rangle$. 
We can see from Eqs.~\eqref{eq:rec_x2p1s_dot} and \eqref{eq:rec_x2s1s_dot} that to compute $\delta x_e^{(1)}$, we require the full spatially-varying first-order perturbed monopole $\delta f^{(1)}_{00}$ and equilibrium PSD $\delta \feq^{(1)}$.
To compute $\langle \delta x^{(2)}_e \rangle$ we require only the second-order, spatially-averaged monopole $\langle \delta f_{00}^{(2)} \rangle$ and equilibrium PSD $ \langle \delta \feq^{(2)} \rangle$. 
We will show below that the first-order multipoles $\delta f^{(1)}_{\ell m}$ source the equation for the second-order monopole $\langle \delta f_{00}^{(2)} \rangle$, so we need to solve for all first-order multipoles using the linearly perturbed BE. 
The scalar and vector perturbations excited by MHD only source $m = 0,\pm 1$ multipoles, so we do not need to solve for the $m\geq 2$ multipoles. 
Figure \ref{fig:combined_mhd_schematic} shows the interdependence between the PSD moments to various orders and the fluid variables. 

Some of the required first-order pieces were first derived in Ref.~\cite{Venumadhav_Hirata_15}; we extend them to account for rotational velocity fields that are naturally sourced in MHD. 
Finally, we will take the linearly perturbed solution for the BE and use it to solve for the second-order jump across the Lyman-$\alpha$ line, which is the major new result in this section. 
Our shift to the background ionization fraction is self-consistent in a perturbative sense, and is valid as long as we remain in the linear regime for radiative transfer. 
We discuss the domain of validity of our linearized scheme in Appendix~\ref{app:LBE}. 

Small-scale inhomogeneities can also perturb detailed balance in the continuum and make direct recombination viable. 
We provide the solution to second order in Appendix~\ref{app:BE_cont} -- the resulting correction to the recombination rate is subdominant to that from the $n=2$ state on the scales of interest. 
We therefore only provide a detailed discussion of the BE across the Lyman-$\alpha$ line in this section. 

\subsubsection{The Boltzmann equation in Lyman-$\alpha$}
\label{sec:boltzmannlya}

The Boltzmann equation is generically given by 
\eq{
Lf = C[f]
}
where $L$ is the diffeomorphism invariant Liouville operator acting on the seven dimensional phase space, $f$ is the PSD, and $C[f]$ is the sum of all of the collisional processes \cite{Senatore_2009}. 
In this equation and moving forward, we will often suppress the explicit dependence of variables on position, frequency, etc. to improve readability. 
We restore the dependencies when needed to emphasize key features of equations.

Parametrizing the PSD of photons $f(t,\bd{x},\nu,\bd{\hat{\bd{n}}})$ in terms of coordinate time, position $\bd{x}$, frequency $\nu$, and direction of photon propagation $\bd{\hat{\bd{n}}}$, the Boltzmann equation is given by
\eq{\label{eq:BE_full}
\frac{\d f}{\d t} + \frac{d\nu}{dt}\frac{\d f}{\d \nu} + \frac{d x^i}{dt}\frac{\d f}{\d x^i} + \frac{d\hat{{n}}^i}{dt}\frac{\d f}{\d \hat{{n}}^i} = C[f].
}
Throughout the rest of this paper, we work in the quasi-steady state approximation, in which the PSD is assumed to have no explicit time dependence and drop the first term above. We will also work in the matter's local rest frame, since writing down the collisional term $C[f]$ is simplest in this frame. Since we are working in the matter's rest frame, we need to perform a Lorentz boost from the Newtonian or background frame. The frequency of a photon in the background frame $\nu_N$ is related to the frequency in the matter rest frame by
\eq{\label{eq:lor_boost}
\nu = \nu_n \bigg[\gamma \bigg(1 + \frac{\bd{v}\cdot\hat{\bd{n}}}{c}\bigg) \bigg]^{-1}.
}
The second term on the left hand side of Eq.~\eqref{eq:BE_full} accounts for the Doppler shift of photons due to cosmic expansion and local peculiar velocities, which enter $d\nu/dt$ from the Lorentz boost of Eq.~\eqref{eq:lor_boost}. The third term accounts for the advection of photons in real space, and the final term accounts for the lensing of photons.

The collisional term includes emission, absorption, and scattering of Lyman-$\alpha$ photons. Under the reasonable set of assumptions detailed in Section VIIA in Ref.~\cite{Venumadhav_Hirata_15}, the collisional term in the matter's rest frame is given by
\eq{\label{eq: coll_Term}
-\frac{1}{H\nu}C[f] = \tau_{s}p_{sc}&\bigg[\phi(\nu) f(\nu,\bd{x}, \hat{\bd{n}})\\
- \int &d\nu' \frac{d\hat{\bd{n}}'}{4\pi}\phi(\nu')p(\nu',\nu)f(\nu',\bd{x},\hat{\bd{n}}') \bigg]\\
+ \tau_s\phi(\nu) &\bigg[ p_{ab}f(\nu,\bd{x},\hat{\bd{n}}) - \feq + p_{sc}\overline{f}_{00}(\bd{x}) \bigg].
}
Here, $\tau_s$ is the Sobolev optical depth, defined in the context of the Sobolev approximation in Appendix \ref{app:sob_sol}; we reproduce it here for clarity
\eq{\label{eq:sob_opt_depth2}
\tau_s = \frac{3 A_{\text{Ly}\alpha} \lambda_{\text{Ly}\alpha}^3}{8\pi H}n_{1s}.
}
The function $\phi(\nu)$ is the Lyman-$\alpha$ line profile, which we take to be a Voigt profile, and the symbol $\overline{f}_{00}$ is the line- and angle-average of the PSD
\eq{
\overline{f}_{00}(\bd{x}) = \frac{1}{4\pi} \int d^2{\hat{\bd{n}}} \int d\nu \, \phi(\nu) f(\nu, \bd{x}, \hat{\bd{n}}).
}
A Lyman-$\alpha$ collision with a neutral hydrogen atom is really a two-photon event in which a ground state atom is first excited to the $2p$ state by absorbing a Lyman-$\alpha$ photon -- $p_{ab}$ is the probability of the metastable $2p$ atom absorbing a second photon which further excites or ionizes it, and $p_{sc}$ is the probability that the $2p$ atom relaxes back down to the ground state and re-emits a Lyman-$\alpha$ photon. %$\tau_s$ is the Sobolev optical depth defined in Eq.~\eqref{eq:sob_opt_depth} in Appendix \ref{app:sob_sol}. $\phi(\nu)$ is the Lyman-$\alpha$ line profile, which we take to be a Voigt profile. The line average of the PSD is given by $\overline{f} = \int d\nu \phi(\nu) f$.

\subsubsection{Homogeneous solution}
\label{sec:BE_homo}

In the homogeneous case, the PSD is only a function of frequency so that we can drop the advection and lensing terms in the Liouville operator.  For the scattering term in Eq.~\eqref{eq: coll_Term}, we can first assume complete redistribution of scattered photons, i.e. $p(\nu,\nu') = \phi(\nu)$, also known as the Sobolev approximation. 
In Appendix~\ref{app:sob_sol}, we solve the BE under this assumption, and substitute it back into Eqs.~\eqref{eq:rec_x2p1s_dot}--\eqref{eq:rec_x2_dot} to recover Eq.~\eqref{eq:rec_TLA_zero_O} and find an explicit expression for the Peebles $C$-factor
\eq{\label{eq:peeb_C_hom}
C &= \frac{3A_{\text{Ly}\alpha}/\tau_s + \Lambda_{2s1s}}{4\beta_B + 3A_{\text{Ly}\alpha}/\tau_s + \Lambda_{2s1s}}.
}
A more realistic treatment of scattering employs the Fokker-Planck approximation \cite{Rybicki_1994,Hirata_2008,unno1955theoretical} instead of complete redistribution in the scattering term in Eq.~\eqref{eq: coll_Term}. We can then write the collisional term using a second-order differential operator to find
\eq{\label{eq: Hom_BE}
\frac{\d f}{\d \nu} = &-\tau_{s}p_{sc}\frac{\nu_{\text{Ly}\alpha}\Delta_H^2}{2} \frac{\d }{\d\nu}\bigg[\phi(\nu) \frac{\d f}{\d\nu} \bigg]\\
&+ \tau_s\phi(\nu) [p_{ab}f(\nu) - \feq + p_{sc}\overline{f}].
}
This second-order differential equation can be numerically solved across the line.
We set the boundary conditions $f(\nu_B) = f_{bb}$ on the blue side of the line and $(\d f/\d\nu) \eval_{\nu_R} = 0$ on the red side of the line. 

\subsubsection{The linearly perturbed Boltzmann equation}
\label{sec:lin_pert_BE}

Our goal in this section is to perturb Eq.~\eqref{eq:BE_full} to first order to find a linear differential equation in frequency for the Fourier amplitudes of the perturbed PSD $\delta f^{(1)}(\nu, \bd{k}, \hat{\bd{n}})$. The result takes the schematic form
\eq{
D_\nu \delta f^{(1)}(\nu, \bd{k}, \hat{\bd{n}}) = \sum_{i}\mathcal{F}_i(\nu)A^{(1)}_i(\bd{k})g_i(\hat{\bd{n}}),
}
where $D_\nu$ is a linear differential operator in frequency, $\mathcal{F}_i(\nu)$ is some known/computable function of frequency such as $\phi(\nu)$, $A^{(1)}_i(\bd{k})$ is a first order perturbed variable, and $g_i(\hat{\bd{n}})$ is a function of the photon's propagation direction. Throughout this and the next section we use the convention that $\hat{\bd{n}} = (\sin\theta\cos\phi,\sin\theta\sin\phi,\cos\theta)$ so that $\hat{\bd{k}}\cdot\hat{\bd{n}}= \cos\theta$.

The lensing term in Eq.~\eqref{eq:BE_full} enters only at second order in the perturbations, so the first-order perturbed BE in the quasi steady-state is given by
\eq{\label{eq:BE_(1)}
&\bigg(\frac{d\nu}{dt}\bigg)^{(1)} \frac{\d f}{\d \nu} + \bigg(\frac{d\nu}{dt}\bigg)^{(0)} \frac{\d \delta f^{(1)}}{\d \nu}\\
&+ \bigg(\frac{d x^i}{dt}\bigg)^{(1)} \frac{\d f}{\d x^i} + \bigg(\frac{d x^i}{dt}\bigg)^{(0)} \frac{\d \delta f^{(1)}}{\d x^i} = C^{(1)}[f].\\
}
We Fourier transform the PSD $f$ and use the results from Appendix~\ref{app:photon_eom} for the terms on the LHS to write this as
\eq{\label{eq:BE_lin_fourier}
\frac{\d \delta f^{(1)}}{\d \nu} &+ \bigg[ \frac{1}{aH} (\Phi_x \sin\theta\cos\theta\cos\phi\\
&+ \Phi_y \sin\theta\cos\theta\sin\phi + \Theta\cos^2\theta) \bigg] \frac{\d f}{\d \nu}\\
&- \frac{ick}{H\nu a} \cos\theta \delta f^{(1)} = -\frac{1}{H\nu}C^{(1)}[f],
}
where we recall that $\Phi_i = ikv_i$ and $\Theta = ikv_z$ as defined in Section \ref{sec:LMHD_system}. 
We can find the first-order collisional term on the RHS by perturbing Eq.~\eqref{eq: coll_Term}:
\eq{\label{eq:coll_lin_fourier}
-\frac{C^{(1)}[f]}{H\nu} &= \bigg(\delta_b + \frac{\delta x_{1s}}{x_{1s}} \bigg) \frac{\d f}{\d \nu}\\
&+ \tau_{s}p_{sc} \bigg[\phi(\nu) \delta f^{(1)}(\nu,k, \hat{\bd{n}})\\
&- \int d\nu' \frac{d\hat{\bd{n}}'}{4\pi}\phi(\nu')p(\nu',\nu) \delta f^{(1)}(\nu', k, \hat{\bd{n}}') \bigg]\\
&+ \tau_s\phi(\nu) \bigg[ p_{ab} \delta f(\nu,k,\hat{\bd{n}}) - \delta \feq^{(1)} + p_{sc}\delta \overline{f}^{(1)}_{00}  \bigg].
}
To handle the angular dependence due to the photon's propagation direction, we further expand the PSD's Fourier amplitudes in spherical harmonics 
\eq{
\delta f^{(1)}(\nu, \bd{k}, \hat{\bd{n}}) &= \sum_{\ell, m} (-i)^{\ell} Y^{\ell m}(\theta,\phi) \delta \tilde{f}^{(1)}_{\ell m}(\nu, \bd{k}).
}
It is convenient to also define 
\eq{
\delta \tilde{f}^{(1)}_{\ell m} = \sqrt{\frac{4\pi}{2\ell+1}\frac{(\ell+m)!}{(\ell - m)!}} \delta f^{(1)}_{\ell m},
} 
since the equations take a simpler form under this redefinition. 
From this expansion, and the form of the source terms in Eqs.~\eqref{eq:BE_lin_fourier} and \eqref{eq:coll_lin_fourier}, we can see that we need to solve three hierarchies: $m=0,\pm 1$. 
We will employ the Fokker-Planck approximation for the collisional term as in the homogeneous case. 

We can project out the $m = 0$ scalar hierarchy from Eq.~\eqref{eq:BE_lin_fourier}, by multiplying it by $Y^{\ell 0}(\hat{\bd{n}})$ and integrating over $\int d^2\hat{\bd{n}}$, which allows us to recover the result of Ref.~\cite{Venumadhav_Hirata_15}
\eq{\label{eq:scalar_BE_(1)}
&\frac{\d \delta f^{(1)}_{\ell 0}}{\d \nu} + \frac{\Theta}{aH}\bigg( -\frac{2}{3}\delta_{\ell,2} + \frac{1}{3}\delta_{\ell,0} \bigg)  \frac{\d f}{\d \nu}\\
&- \frac{ck}{H\nu a} \bigg[ - \frac{\ell}{2\ell-1}\delta f^{(1)}_{\ell-1,0} + \frac{\ell+1}{2\ell+3} \delta f^{(1)}_{\ell+1,0} \bigg]\\
&= \delta_{\ell,0}\bigg\{ \bigg(\delta_b + \frac{\delta x_{1s}}{x_{1s}} \bigg)\frac{\d f}{\d \nu} - \tau_{s}\phi(\nu) [\delta \feq^{(1)} - p_{sc}\overline{\delta f}^{(1)}_{00}]\\
&- \tau_{s}p_{sc}\frac{\nu_{\text{Ly}\alpha}^2\Delta_H^2}{2}\frac{\d}{\d\nu}\bigg[\phi(\nu)\frac{\d \delta f^{(1)}_{00}}{\d \nu} \bigg]+ p_{ab}\tau_s\phi(\nu) \delta f^{(1)}_{00} \bigg\}\\
&+ \tau_s\phi(\nu) \delta f^{(1)}_{\ell 0}\delta_{\ell \neq 0}.\\
}
We can write the solution to this equation in terms of three basis solutions as in Ref.~\cite{Venumadhav_Hirata_15}:
\eq{\label{eq:scalar_basis_solns}
\delta f^{(1)}_{\ell 0}(k,\nu) &= \bigg( \delta_b + \frac{\delta x_{1s}}{x_{1s}} \bigg) \mathcal{A}_{\ell}(k,\nu) + \frac{\Theta}{a H} \mathcal{B}_{\ell}(k,\nu)\\
&+ (\delta \feq^{(1)} - p_{sc}\delta \overline{f}^{(1)}_{00}) \mathcal{C}_{\ell}(k,\nu).
}
Since we require only the perturbed line-averaged monopole and equilibrium PSD to compute the ionization fraction to linear order, we do not need to solve the vector $m = \pm 1$ hierarchies to compute $\delta x^{(1)}_e$. We integrate Eq.~\eqref{eq:scalar_basis_solns} across the line profile for $\ell = 0$, which allows us to relate $\delta \overline{f}^{(1)}_{00}$ to $\delta \feq^{(1)}$, the line-averaged basis solutions $\overline{\mathcal{A}}_0,\overline{\mathcal{B}}_0,\overline{\mathcal{C}}_0$, and the scalar fluid variables. We can then linearly perturb Eqs.~\eqref{eq:rec_x2p1s_dot}--\eqref{eq:rec_x2_dot} and use the fact that the steady state balance $ \dot{x}_2 = 0$ holds order by order in perturbation theory to set $\delta\dot{x}^{(1)}_{2} = 0$, which allows us to find the perturbed recombination equation. The details of this computation can be found in Section VIIB in Ref.~\cite{Venumadhav_Hirata_15}. The final result is
\eq{
\label{eq:rec_TLA_first_O}
\delta &\dot{x}^{(1)}_e \vert_{2p,2s} = \mathcal{P}n_H x_e^2\alpha_B \delta_b\\
&+ \delta x_{1s} \bigg[ (1-\mathcal{P})n_H \frac{x_e^2}{x_{1s}}\alpha_B -2 \mathcal{P}n_Hx_e \alpha_B - 4 \feq\beta_B \bigg]\\
&- 3(1-\mathcal{P})x_{1s}A_{\text{Ly}\alpha}\bigg[ \bigg(\delta_b + \frac{\delta x_{1s}}{x_{1s}} \bigg)\frac{\overline{\mathcal{A}}_0}{1+p_{sc}\overline{\mathcal{C}}_0}\\
& \hspace{4.8cm} + \frac{\Theta}{aH}\frac{\overline{\mathcal{B}}_0}{1+p_{sc}\overline{\mathcal{C}}_0}  \bigg],
}
where $\mathcal{P}(k)$ is the linearly perturbed analog to the Peebles C factor and is given by
\eq{
\mathcal{P}(k) = \frac{3A_{\text{Ly}\alpha}\frac{1-p_{ab}\overline{\mathcal{C}}_0(k)}{1+p_{sc}\overline{\mathcal{C}}_0(k)} + \Lambda_{2s1s}}{3A_{\text{Ly}\alpha}\frac{1-p_{ab}\overline{\mathcal{C}}_0(k)}{1+p_{sc}\overline{\mathcal{C}}_0(k)} + \Lambda_{2s1s} + 4\beta_B}.
}
The sum of Eqs.~\eqref{eq:rec_TLA_first_O} and the continuum contribution to $\delta \dot{x}_e^{(1)}$, which is provided in Appendix~\ref{app:BE_cont}, can be added to the FSR MHD equations (Eqs.~\eqref{eq:FSR_LMHD}) to get a closed system of equations. 
This allows us to evolve LMHD modes with radiative transfer taken into account for the linearly perturbed ionization fraction.

With these results in hand, we can also demonstrate how the local recombination framework presented in Section \ref{sec:3LA} emerges for sufficiently large scale perturbations. 
The comoving mean free path of a Lyman-$\alpha$ photon is
\eq{
\lyamfp(\nu) = \frac{1}{an_{1s}\sigma_{sc}} = \frac{c}{\tau_s \phi(\nu) H\nu_{\text{Ly}\alpha} a}.
}
If the wavelength of a mode is much larger than the mean free path, $\lambda \gg \lyamfp \Leftrightarrow ck/(\tau_s \phi(\nu) H\nu_{\text{Ly}\alpha}a ) \ll 1$, the perturbed BE is well-approximated by taking the $k \to 0$ limit of Eq.~\eqref{eq:BE_(1)} \cite{Venumadhav_Hirata_15}.
In this limit, the advection term vanishes so that the monopole decouples from all other multipoles. The sum of the homogeneous and perturbed monopole $f + \delta f_{00}$ now satisfies the same equation as the homogeneous PSD with the substitution $\tau_s \to \tau_{s} + \delta \tau_s - \Theta/{3aH}$, where the perturbed Sobolev optical depth is given by $\delta \tau_s/\tau_s = \delta_b + \delta x_{1s}/x_{1s}$. 
To linear order, this is equivalent to making the following substitution in the homogeneous BE
\eq{
\tau_s &= \frac{3 A_{\text{Ly}\alpha} \lambda_{\text{Ly}\alpha}^3}{8\pi H}n_{1s}\\
&\to \frac{3 A_{\text{Ly}\alpha} \lambda_{\text{Ly}\alpha}^3}{8\pi H(1 + \Theta/3aH)}n_{1s}\bigg(1 + \delta_b + \frac{\delta x_{1s}}{x_{1s}}\bigg).
}
We have therefore recovered the result from Ref.~\cite{Senatore_2009} that large scale perturbations only modify recombination locally by changing the escape rate of Lyman alpha photons due to perturbations in the local density and ionization fraction fields, and by modifying the Hubble rate to account for the local expansion or convergence of the fluid, i.e. $H \to H + \Theta/3a$. As expected, for large scale perturbations, the non-local transport of Lyman-$\alpha$ photons becomes unimportant so that recombination can be treated as a local process as in the homogeneous case.

As discussed in this section's beginning, we need to solve the vector $m = \pm 1$ hierarchies for the second-order piece $\langle \delta x_e^{(2)} \rangle$. 
We can project out the equations for these hierarchies from  Eq.~\eqref{eq:BE_lin_fourier} by multiplying by $Y^{\ell\pm 1}(\hat{\bd{n}})$ and integrating over $\int d^2\hat{\bd{n}}$. Defining $\Phi_{\pm} = \Phi_x \mp i\Phi_y$, we can write these hierarchies as
\eq{\label{eq:vector_BE_(1)}
\frac{\d \delta f_{\ell 1}}{\d \nu} &= - \frac{1}{6aH} \delta_{\ell,2} \Phi_+ \frac{\d f}{\d \nu}\\
&+ \frac{ck}{H\nu a} \bigg[ -\frac{\ell - 1}{2\ell - 1} \delta f_{\ell-1,1} + \frac{\ell + 2}{2\ell + 3} \delta f_{\ell+1,1} \bigg]\\
&+ \tau_s\phi(\nu) \delta f_{\ell 1}\\
\frac{\d \delta f_{\ell, -1}}{\d \nu} &= \frac{1}{aH} \delta_{\ell,2} \Phi_- \frac{\d f}{\d \nu}\\
&+ \frac{ck}{H\nu a} \bigg[ -\frac{\ell + 1}{2\ell - 1} \delta f_{\ell-1, -1} + \frac{\ell}{2\ell + 3} \delta \tilde{f}_{\ell+1, -1} \bigg]\\
&+ \tau_s\phi(\nu) \delta f_{\ell, -1}.\\
}
We then supplement our basis solutions in Eq.~\eqref{eq:scalar_basis_solns} with the basis solutions
\begin{subequations}
\label{eq:vector_basis_solns}
\begin{align}
  \delta {f}_{\ell,1}(k,\nu) & = \frac{\Phi_{+}}{ a H}\mathcal{D}_{\ell}^{+}(k, \nu), \, {\rm and} \\
  \delta {f}_{\ell,-1}(k,\nu) & = -6\frac{\Phi_{-}}{ a H}\mathcal{D}_{\ell}^{-}(k,\nu).
\end{align}
\end{subequations}
We numerically solve the $m=0,\pm 1$ hierarchies across the line. To solve these equations, we follow a similar strategy as in \cite{Venumadhav_Hirata_15} and discretize in frequency out to $\pm 1000$ Doppler widths, $x = (\nu - \nu_{\text{Ly}\alpha})/(\nu_{\text{Ly}\alpha} \Delta_H)$, with 50 bins per Doppler width. 
We have to close the system at some $\ell_{max}$; we use a simple truncation in which we set all multipoles with $\ell > \ell_{max}$ to zero. 
This can induce unwanted reflections from our boundary back to the lower moments \cite{Ma_1995}. 
To mitigate this effect, we try a range of $\ell_{max}$ until the $\ell \leq 2$ results converge. 
We find that $\ell_{max} = 30$ is sufficient to avoid unwanted reflection back down to the $\ell \leq 2$ moments.

% \js{check whether there is interesting structure in k for the D solutions because then might be worth showing how these look different across scales.}

\begin{figure*}[t]
    \includegraphics[width=1\textwidth]{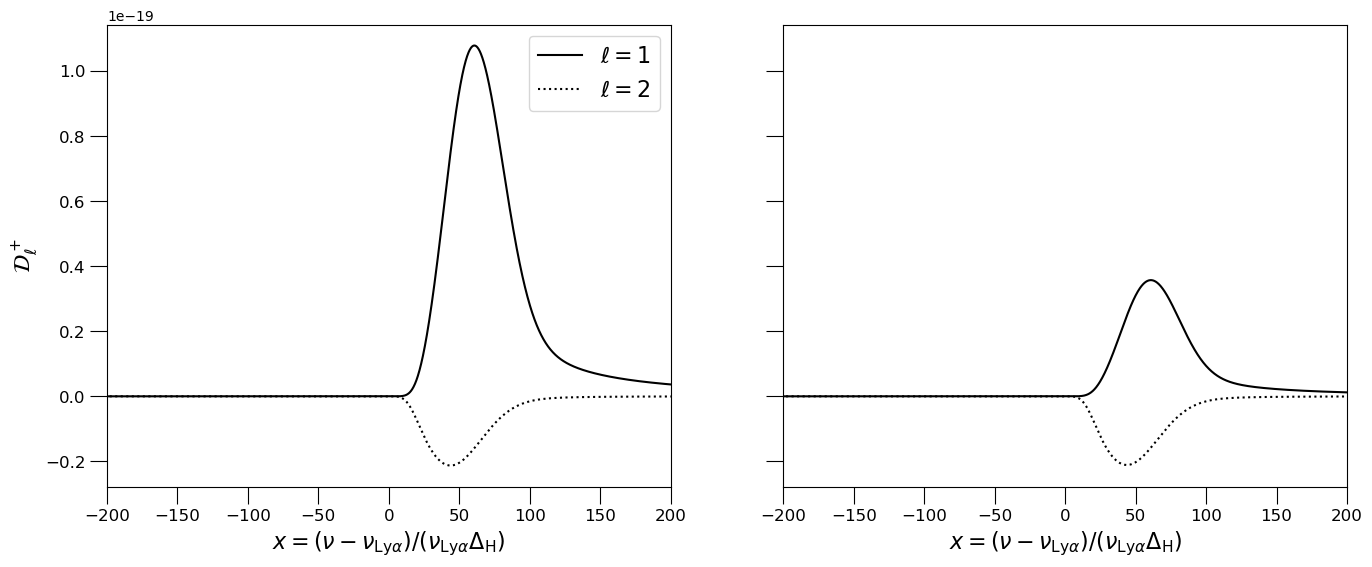}  
    \captionsetup{width=1\linewidth}
    \caption{$D_\ell^{\pm}$ basis solutions for the vector hierarchy $m = \pm 1$ defined in Eq.~\eqref{eq:vector_basis_solns} at $z = 1100$ and $k \approx 10^5 \text{ Mpc}^{-1}$. These bases solutions are sourced by rotational velocity fields which induce a Doppler shift in the matter's rest frame. They are not needed to compute the perturbed ionization fraction to linear order, but do enter as source functions for the second-order, spatially-averaged monopole $\langle \delta f^{(2)}_{00} \rangle$, see Section~\ref{sec:BE_2} and Fig~\ref{fig:combined_mhd_schematic}.
    }
    \label{fig:D_pm}
\end{figure*}

Fig.~\ref{fig:D_pm} shows the $D_{\ell}^\pm$ basis solutions to complement the scalar $m=0$ results provided in Ref.~\cite{Venumadhav_Hirata_15}. The vector basis solutions are sourced by rotational velocity fields which modify the local Doppler shift of photons in the matter rest frame.  %Although to linear order the vector solutions do not impact the recombination history, they are needed to compute $\langle \delta x_e^{(2)} \rangle$ as we discuss in the next section. 

\subsubsection{The second-order perturbed Boltzmann equation}
\label{sec:BE_2}

As we mentioned in the beginning of this section, we need to solve for the spatially-averaged monopole to second order $ \langle \delta f_{00}^{(2)} \rangle$ to compute the lowest-order shift to the ionization fraction. 
To this end, we perturb each term in the BE, Eq.~\eqref{eq:BE_full}, to second order, take a spatial average and perform an angular integral $\int d^2\hat{\bd{n}}$ to project out the monopole. 
As was the case in the first order BE, when we take the $k \to 0$ limit, we decouple the monopole from all other multipoles. We can therefore solve for the spatially homogeneous component of $\delta f^{(2)}_{00}$ without a hierarchy which relates it to other second-order multipoles. Our goal is to ultimately derive and numerically solve a linear differential equation in frequency for $ \langle \delta f_{00}^{(2)} \rangle$, which schematically takes the form
\eq{\label{eq:BE_2_schematic}
D_\nu \langle \delta f_{00}^{(2)}\rangle &= \sum_{i} \langle A_i^{(1)}(\bd{x})B_i^{(1)}(\bd{x}) \rangle \mathcal{F}_i(\nu)\\
&+ \sum_{i} \langle A_i^{(2)}(\bd{x}) \rangle \mathcal{F}_i(\nu) + \mathcal{S}(\nu).
}
Here, $A^{(1)}_i(\bd{x})$ and $B^{(1)}_i(\bd{x})$ are first-order perturbed fluid variables like density and velocity, $A^{(2)}_i(\bd{x})$ is a second-order perturbed variable like the ionization fraction, $\mathcal{F}_i(\nu)$ is some known/computable function of frequency such as $\phi(\nu)$, and $\mathcal{S}(\nu)$ are source terms that are quadratic in linear fluid variables, but whose form does not factor into a function of frequency times a cross-correlation of first-order perturbed variables. 
In general, the variable $A$ in $\langle A^{(2)}_i(\bd{x}) \rangle$ can include all second-order fluid variables, but spatial averages of the perturbed density and velocity fields vanish to all orders of perturbation theory, which significantly simplifies the computation. 
%such as $\langle \delta^{(2)}_b \rangle$. However, 

For illustrative purposes, we will show how to explicitly compute one of the second-order terms in the BE from the Doppler shift contribution. All other terms are provided in Appendix~\ref{app:BE_(2)}.

The second-order Doppler shift term is given by
\eq{
\bigg(\frac{d\nu}{dt} \frac{\d f}{\d\nu}\bigg)^{(2)} &= \bigg(\frac{d\nu}{dt}\bigg)^{(2)} \frac{\d f}{\d \nu}\\
&+ \bigg(\frac{d\nu}{dt}\bigg)^{(1)} \frac{\d \delta f^{(1)}}{\d \nu} + \bigg(\frac{d\nu}{dt}\bigg)^{(0)} \frac{\d \delta f^{(2)}}{\d \nu}.
}
Let us focus on the second term. Once we drop the subdominant contribution from the dipolar velocity derivative term and the derivative of the metric potential in Eq.~\eqref{eq:nu_EOM} in Appendix~\ref{app:photon_eom}, this term is given by
\eq{
-\frac{1}{H\nu}&\bigg(\frac{d\nu}{dt}\bigg)^{(1)} \frac{\d \delta f^{(1)}}{\d \nu} \\
&= \frac{\hat{{n}}^i\hat{{n}}^j}{aH} \frac{\d v_i}{\d x^j}  \sum_{\ell, m}(-i)^{\ell}(\d_\nu \delta \tilde{f}^{(1)}_{\ell m}) Y^{\ell m}(\theta,\phi).
}
Taking an angular integral, we have
\eq{
-\frac{1}{H\nu}&\int \frac{d^2\hat{\bd{n}}}{4\pi} \bigg(\frac{d\nu}{dt}\bigg)^{(1)} \frac{\d \delta f^{(1)}}{\d \nu}\\
&= \bigg[ \frac{1}{30} \frac{\Phi_+}{aH}(\d_\nu \delta f^{(1)}_{2,-1}) - \frac{1}{5} \frac{\Phi_-}{aH} (\d_\nu \delta f^{(1)}_{2,1}) \bigg]\\
&+ \frac{\Theta}{3aH} \bigg[ \frac{2}{5}(\d_\nu \delta f^{(1)}_{20}) - (\d_\nu \delta f^{(1)}_{00})\bigg]  \\
}

We can now use the first-order basis solutions and Eq.~\eqref{eq:cross_corr} to take a spatial average and find
\begin{widetext}
    \eq{\label{eq:red_1_1}
-\frac{1}{H\nu} &\int \frac{d^2\hat{\bd{n}}}{4\pi} \bigg\langle  \bigg(\frac{d\nu}{dt}\bigg)^{(1)} \frac{\d \delta f^{(1)}}{\d \nu} \bigg\rangle =  \int \,d\ln k\ \frac{\Delta^2_B(k)}{2B_0^2} \bigg\{\frac{1}{3aH}\bigg[\bigg(\overline{T}_{\Theta\delta_b} -\frac{\overline{T}_{\Theta\delta \overline{x}_e} }{x_{1s}}\bigg)\bigg(\frac{2}{5}\d_\nu \mathcal{A}_2(k,\nu) - \d_\nu \mathcal{A}_0(k,\nu) \bigg)\\
&+ \frac{\overline{T}_{\Theta\Theta}}{aH}\bigg(\frac{2}{5}\d_\nu \mathcal{B}_2(k,\nu) - \d_\nu \mathcal{B}_0(k,\nu) \bigg) + \bigg( \overline{\alpha} \overline{T}_{\Theta \delta_b} - \frac{\overline{\beta}}{x_{1s}} \overline{T}_{\Theta \delta \overline{x}_e} + \frac{\overline{\gamma}}{aH} \overline{T}_{\Theta \Theta} \bigg) \bigg(\frac{2}{5}\d_\nu \mathcal{C}_2(k,\nu) - \d_\nu \mathcal{C}_0(k,\nu) \bigg)  \bigg]\\
&\hspace{2cm} - \frac{\overline{T}_{\Phi_+\Phi_-}}{5a^2H^2}(\d_\nu\mathcal{D}_2^+(k,\nu) + \d_\nu \mathcal{D}_2^-(k,\nu)) \bigg\}.
}
\end{widetext}
$\overline{\alpha}$, $\overline{\beta}$, and $\overline{\gamma}$ are defined in Eq.\eqref{eq:alpha_bar_def} in Appendix~\ref{app:BE_(2)}. Recall that terms such as $\overline{T}_{\Theta\delta_b}$ are angular integrals of products of transfer functions for first-order fluid variables given by Eq.~\eqref{eq:cross_corr_ang_avg} and are themselves functions of $k$. We restore the explicit dependence on wavenumber $k$ of the first-order basis solutions, e.g $\mathcal{A}(k,\nu)$, to emphasize that the integral over $k$ differs from a simple cross-correlation as in Eq.\eqref{eq:cross_corr}.

Once we include similar computations for the advection, lensing, and collisional terms, the final result is a second-order linear differential equation in frequency for $\langle \delta f^{(2)}_{00} \rangle$. Introducing the linear differential operator
\eq{
D_\nu \langle \delta f_{00}^{(2)}\rangle 
 = &\tau_{s}p_{sc}\frac{\nu_{\text{Ly}\alpha}^2\Delta_H^2}{2}\frac{\d}{\d \nu}\bigg[\phi(\nu) \frac{\d \langle \delta f_{00}^{(2)}\rangle}{\d \nu} \bigg]\\
 &+ \frac{\d \langle \delta f_{00}^{(2)}\rangle}{\d \nu} - \tau_s p_{ab}\phi(\nu) \langle \delta f_{00}^{(2)}\rangle,
}
the second-order BE can be written as
\eq{\label{eq:BE_(2)}
D_\nu \langle \delta f_{00}^{(2)}\rangle = &\mathcal{S}(\nu) - \bigg[ \frac{\Delta x_e }{x_{1s}} \bigg]  \frac{\d f}{\d \nu}\\
&- \tau_s\phi(\nu) [ \langle \delta \feq^{(2)} \rangle - p_{sc}\langle \delta \overline{f}^{(2)}_{00}\rangle ],\\
}
where the source terms $\mathcal{S}(\nu) = \sum_i\mathcal{S}_i(\nu)$ are of the form
\eq{\label{eq:source-term-general}
% \mathcal{S}_i(\nu) &= \frac{1}{(2\pi)^6} \int d^3\bd{k}_1 d^3\bd{k}_2 \langle A(\bd{k}_1)B(\bd{k}_2) \rangle \mathcal{F}(\bd{k}_2,\nu)  e^{i(\bd{k}_1+\bd{k}_2)\bd{x}}\\
\mathcal{S}_i(\nu) &= \int d \ln k \frac{\Delta^2_B(k)}{2B_0^2} \mathcal{F}(k,\nu) \overline{T}_{AB}(k).\\
}
$\mathcal{F}(k,\nu)$ can be any of the first-order basis solutions to the BE from the previous section. The source terms can be read off from Eqs.~\eqref{eq:red_1_1},\eqref{eq:coll_(2)},\eqref{eq:red_0_2},\eqref{eq:red_20},\eqref{eq:advec_(2)}, and \eqref{eq:lensing_(2)}. 

The source terms in Eq.~\eqref{eq:source-term-general} involve $\bar{T}_{AB}(k)$ and hence depend on the first-order fluid transfer functions. 
They must be solved independently for every PMF spectrum, and at every redshift. 
They involve integrals across wavenumbers for every frequency. 
To reduce the number of required integrations, we first compute source functions across the line using a sparse binning in frequency (only 2800 in total), with bins concentrated near the line center where the integrand has the most support. 
We then use splines to interpolate the source functions across the line. 
Once we have these results, we numerically solve for $\langle \delta f_{00}^{(2)} \rangle $ using the same binning from the first-order numerical solution, i.e., $\pm 1000$ Doppler widths, $50$ bins per width. 
We introduce the following second-order basis solutions
\eq{\label{eq:2nd_O_basis_sol}
\langle \delta f^{(2)}_{00} \rangle(\nu) = &s(\nu) + \mathcal{A}^{(2)}(\nu) \bigg[ \frac{\langle \delta x_{1s}^{(2)} \rangle - \langle \delta_b^{(1)} \delta x_e^{(1)} \rangle }{x_{1s}} \bigg]\\
&+ \mathcal{C}^{(2)}(\nu)[ \langle \delta \feq^{(2)} \rangle - p_{sc} \langle \delta \overline{f}^{(2)}_{00}\rangle ],
}
where our basis terms satisfy
\eq{
D_\nu s(\nu) &= \mathcal{S}(\nu)\\
D_\nu \mathcal{A}^{(2)}(\nu) &= \frac{\d f}{\d \nu}\\
D_\nu \mathcal{C}^{(2)}(\nu) &= -\tau_s\phi(\nu).
}
Note that the basis solutions $\mathcal{A}^{(2)}(\nu)$ and $\mathcal{C}^{(2)}(\nu)$ can be computed independent of the fluid variables, but $s(\nu)$ is unique in that it can only be computed given the particular set of fluid perturbations.
This arises because the associated source term $\mathcal{S}(\nu)$ in Eq.~\eqref{eq:BE_2_schematic} does not factorize into the product of a known function of frequency times a spatial average of fluid variables.
Integrating Eq.~\eqref{eq:2nd_O_basis_sol} across the line, we find
\eq{\label{eq:monopole_second_O_consistency}
&\langle \delta \overline{f}^{(2)}_{00}\rangle = \frac{1}{1 + p_{sc}\mathcal{\overline{C}}^{(2)}} \\
&\times \bigg\{ \overline{s} + \mathcal{\overline{A}}^{(2)} \bigg[ \frac{\langle \delta x_{1s}^{(2)} \rangle - \langle \delta_b^{(1)}\delta x_{e}^{(1)}\rangle}{x_{1s}} \bigg] + \mathcal{\overline{C}}^{(2)} \langle \delta \feq^{(2)} \rangle \bigg\}.
}

\subsubsection{Perturbed recombination with radiative transfer}
\label{sec:rec_results}

Equation \ref{eq:monopole_second_O_consistency} is a key result relating the second-order line-averaged monopole $\langle \delta \overline{f}^{(2)}_{00}\rangle$ to $\langle \delta \feq^{(2)}\rangle$. 
At this point, we have all the necessary ingredients to derive the recombination rate for the second-order, spatially-averaged ionization fraction. 
Perturbing Eqs.~\eqref{eq:rec_x2p1s_dot}--\eqref{eq:rec_x2_dot} to second order and taking a spatial average, we find
\begin{widetext}
    \eq{
    \langle \delta\dot{x}^{(2)}_{1s}   \rangle &=  \bigg[ n_H \frac{x_e^2}{x_{1s}}\alpha_B - 4 \feq\beta_B \bigg] \langle \delta x^{(2)}_{1s}   \rangle + \bigg[ 3A_{\text{Ly}\alpha} \frac{1 - p_{ab}\mathcal{\overline{C}}^{(2)}}{1 + p_{sc}\mathcal{\overline{C}}^{(2)}} + \Lambda_{2s1s}  \bigg] x_{1s} \langle \delta \feq^{(2)}  \rangle +(3A_{\text{Ly}\alpha} + \Lambda_{2s1s} ) \langle \delta \feq^{(1)} \delta x_{1s}^{(1)} \rangle\\
    &\quad - 3 x_{1s} A_{\text{Ly}\alpha} \bigg[ \frac{ \overline{s} }{1 + p_{sc}\mathcal{\overline{C}}^{(2)} } + \frac{\mathcal{\overline{A}}^{(2)}}{1 + p_{sc}\mathcal{\overline{C}}^{(2)} } \frac{\langle \delta x_{1s}^{(2)} \rangle - \langle \delta_b^{(1)}\delta x_{e}^{(1)}\rangle}{x_{1s}} \bigg]   - 3A_{\text{Ly}\alpha} \langle \delta \overline{f}_{00}^{(1)} \delta x_{1s}^{(1)} \rangle.\\
    \langle\delta \dot{x}^{(2)}_2\rangle
    &= -\langle \delta\dot{x}^{(2)}_{1s}   \rangle +   n_H\alpha_B \langle \delta x_e^{(1)} \delta x_e^{(1)}\rangle - [2n_H x_e \alpha_B + 4  \feq\beta_B]\langle\delta x^{(2)}_{1s}\rangle - 4\beta_B x_{1s}\langle \delta \feq^{(2)} \rangle\\
    &- 2 n_H x_e \alpha_B \langle \delta^{(1)}_b \delta x^{(1)}_{1s} \rangle - 4\beta_B \langle \delta \feq^{(1)} \delta x_{1s}^{(1)} \rangle.\\
    }
\end{widetext}
We again use the steady state approximation and set $\langle\delta \dot{x}^{(2)}_2\rangle = 0$, which allows us to finally write the evolution equation for $\langle\delta x^{(2)}_{e}\rangle$ as

\begin{widetext}
    \eq{\label{eq:dxe2_RT}
    \langle &\delta \dot{x}^{(2)}_{e}\rangle 
    = \bigg[ (1-\mathcal{P}^{(2)})n_H \frac{x_e^2}{x_{e}} \alpha_B - 2 \mathcal{P}^{(2)} n_Hx_e\alpha_B- 4 \feq\beta_B \bigg]  \langle \delta x_{e}^{(2)} \rangle\\
    &+ 3(1-\mathcal{P}^{(2)}) x_{1s} A_{\text{Ly}\alpha} \bigg[ \frac{ \overline{s} }{1 + p_{sc}\mathcal{\overline{C}}^{(2)} } - \frac{\langle \delta \overline{f}_{00}^{(1)}\delta x_{e}^{(1)}\rangle}{x_{1s}} - \frac{\mathcal{\overline{A}}^{(2)}}{1 + p_{sc}\mathcal{\overline{C}}^{(2)} } \frac{\langle \delta x_{e}^{(2)} \rangle + \langle \delta_b^{(1)}\delta x_{e}^{(1)}\rangle}{x_{1s}}  \bigg]\\
    &- 2 \mathcal{P}^{(2)} n_H x_e \alpha_B \langle \delta^{(1)}_m \delta x^{(1)}_{e}\rangle + \mathcal{P}^{(2)} n_H\alpha_B \langle \delta x_e^{(1)} \delta x_e^{(1)}\rangle + \bigg[(1-\mathcal{P}^{(2)})(3 A_{\text{Ly}\alpha} + \Lambda_{2s1s}) - 4\mathcal{P}^{(2)} \beta_B \bigg] \langle \delta \feq^{(1)} \delta x_{e}^{(1)} \rangle,\\
    }
\end{widetext}
where $\mathcal{P}^{(2)}$ is the second-order analog to the modified Peebles C-factor
\eq{
\mathcal{P}^{(2)} &= \frac{3A_{\text{Ly}\alpha} \frac{1 - p_{ab}\mathcal{\overline{C}}^{(2)}}{1 + p_{sc}\mathcal{\overline{C}}^{(2)}} + \Lambda_{2s1s}}{3A_{\text{Ly}\alpha} \frac{1 - p_{ab}\mathcal{\overline{C}}^{(2)}}{1 + p_{sc}\mathcal{\overline{C}}^{(2)}} + \Lambda_{2s1s} + 4\beta_B }.
}

\begin{figure*}[t!]
    \centering
    \includegraphics[width=1\textwidth]{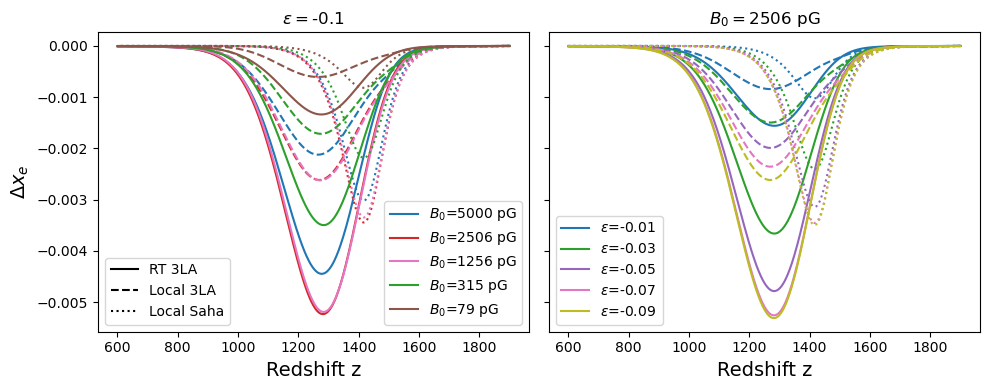}  
    \captionsetup{width=1\linewidth}
    \caption{Shift to the background ionization fraction in the presence of PMFs. Solid lines show the result when radiative transfer is included using Eq.~\eqref{eq:dxe2_RT}. Dotted lines show the result when the Saha model for recombination is perturbed as in Eq.~\eqref{eq:dxe2_saha}. Dashed lines show the result when the local treatment for modified recombination is used as described in Ref.~\cite{Lee_2021}. }
    \label{fig:dxe_TLA_saha}
\end{figure*}

To find the overall enhancement to the background ionization fraction due to the presence of small-scale inhomogeneities as given in Eq.~\eqref{eq:rec_second_O}, we need only integrate this ODE and add $\langle \delta x_e^{(1)} \delta_b^{(1)} \rangle$.

\section{Modified recombination results}
\label{sec:mod_rec}

We can now use our transfer functions from Section \ref{sec:MHD_evo} to compute the relevant spatial averages and cross-correlations of fluid variables by integrating over $k$.
We then substitute these results into Eq.~\eqref{eq:dxe2_RT} and use Eq.\eqref{eq:rec_second_O} to find the shift to the background ionization fraction. 

In Fig.~\ref{fig:dxe_TLA_saha}, we show results for the change to the ionization fraction for a few choices of PMF spectra.
We compare the full calculation with non-local radiative transport taken into account as in Section~\ref{sec:rec_results} to the Saha model as in Section~\ref{sec:local_rec_saha} and to the local perturbed 3LA as in Section \ref{sec:3LA} (the last of which follows the scheme in Ref.~\cite{Lee_2021}, with their equation (12) modified by integrating over the angular dependence in the transfer functions as in Eq.~\eqref{eq:cross_corr_ang_avg} above). 
We choose to show the absolute shift to the background ionization fraction, in contrast to Refs.~\cite{jedamzik2024cosmicrecombinationpresenceprimordial,jedamzik2025hints,mirpoorian2024modifiedrecombinationhubbletension,mirpoorian2025dynamicaldarkenergynecessary} which present their results using $\Delta x_e/x_e$. 
% \js{move to discussion - just say that it casues a spectral distortion (cite jens and Yacine). We note that a modified recombination history would also leave an imprint in the cosmological recombination lines in CMB spectral distortions, which could in principle be detected by future CMB instruments. (Need references and to discuss wording)}
% %In order to compute $\Delta x_e$ using a 3LA local recombination scheme following 

We previously found that baryon clumping was reduced when radiative transfer (RT) is taken into account as compared to Saha in Section~\ref{sec:clumping} (see Fig.~\ref{fig:deltam_power}). 
Nevertheless, the recombination rate is enhanced beyond Saha, as well as beyond the local scheme of Ref.~\cite{Lee_2021}. 
These results agree with the findings of Ref.~\cite{jedamzik2024cosmicrecombinationpresenceprimordial}, which also found that radiative transfer speeds up the recombination rate beyond a no-mixing model for Lyman-$\alpha$ photons. 
We ascribe the enhanced recombination rate to the ability of Lyman-$\alpha$ photons to escape over-dense regions. 
Conversely, recombination is slowed down in under-dense regions, but the net effect is that the mean rate is sped up beyond what would be expected from a local treatment.

We wish to determine how a modified and inhomogeneous recombination history changes the visibility function.
One might think that this simply amounts to replacing the homogeneous free-electron number density $\overline{n}_e$ in the Thomson optical depth by $\langle n_e \rangle = \overline{n}_e + \langle n_e^{(2)}\rangle = \overline{n}_e + \overline{n}_H\Delta x_e $. 
However, as noted in Ref.~\cite{chluba2025itocalculusmeetshubble}, time-time correlations in the optical depth $\tau = \int d\eta \dot{\tau}$, where $\eta$ is conformal time, can further distort the visibility function and shift its peak. 
To determine the visibility function in the presence of an inhomogeneous recombination history, we need to compute
\eq{
\langle g(\bd{x},z) \rangle = \langle \dot{\tau}(\bd{x},z)e^{-\tau(\bd{x},z)} \rangle. 
}
For linear perturbations, both $\dot{\tau}$ and $\tau$ are Gaussian random fields near recombination. Therefore, $e^{-\tau}$ is log-normally distributed and its expectation is given in terms of the optical depth's mean $\mu_\tau = \langle \tau(\bd{x},z) \rangle$ and variance $\sigma_\tau^2 = \langle \tau(\bd{x},z) \tau(\bd{x},z) \rangle -  \langle \tau(\bd{x},z) \rangle^2$:
\eq{\label{eq:e_tau}
\langle e^{-\tau(\bd{x},z)} \rangle = e^{-\mu_\tau + \sigma_\tau^2/2}.
}
Using $\langle g(\bd{x},z) \rangle = -d(\langle e^{-\tau(\bd{x},z)} \rangle)/d\eta$, the spatially averaged visibility function is given by
\eq{\label{eq:visib_avg}
\langle g(\bd{x},z) \rangle = \bigg(\frac{d}{d\eta}\mu_\tau - \frac{1}{2}\frac{d}{d\eta}\sigma_{\tau}^2\bigg)e^{-\mu_\tau + \sigma_\tau^2/2}.
}
Since $\tau$ is a temporally integrated quantity, both space and time correlations in the ionization fraction are needed to accurately compute the spatially averaged visibility function. 
Studies such as Refs. \cite{lynch2024desihubbletensionlight,lynch2024reconstructing,mirpoorian2024modifiedrecombinationhubbletension,mirpoorian2025dynamicaldarkenergynecessary,jedamzik2025hints} which only utilize the average ionization fraction in order to compute the visibility function and infer cosmological parameters from the CMB might therefore need to be revisited.
We note that Eq. \eqref{eq:visib_avg} is only valid so long as $\tau$ is a Gaussian random field and therefore breaks down in the presence of non-linear perturbations. 
If Gaussianity is broken, all cumulants of the optical depth are required to compute $\langle g \rangle$.

\begin{figure*}[t]
    \centering
    \includegraphics[width=1\textwidth]{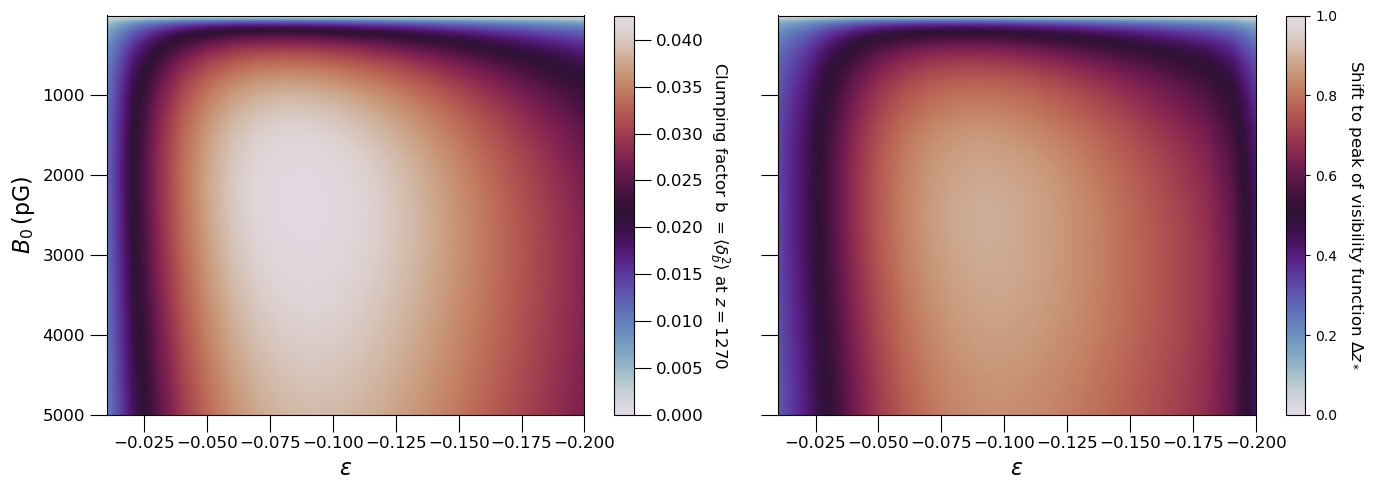}  
    \captionsetup{width=1\linewidth}
    \caption{(Left) Clumping factor at $z = 1270$, which is approximately the redshift at which $\Delta x_e$ is largest. (Right) Shift in the peak of the visibility function $\Delta z_*$ for different PMF spectra. An earlier peak in the visibility function shrinks the sound horizon and alleviates the Hubble tension.}
    \label{fig:b_and_dz_star}
\end{figure*}

In Fig.~\ref{fig:b_and_dz_star}, we show the change in the redshift of the peak of the visibility function $\Delta z_*$ as a function of the amplitude $B_0$ and tilt $\epsilon$ of our spectra. 
For the parameters we consider, the largest shift we get is $\Delta z_* \approx 1$ which occurs for $B_0 \approx 2.96 \text{ nG}, \epsilon \approx -0.09$. 
This modification to the recombination history is about an order of magnitude smaller than the shift to the peak of the visibility function from the best fit modified recombination histories to Planck + DESI or Planck + DESI + PP$M_b$ obtained by Ref.~\cite{mirpoorian2024modifiedrecombinationhubbletension}, where PP$M_b$ is the Pantheon+ supernovae dataset \cite{Brout_2022} using the $\text{SH}_0\text{ES}$ calibration of the supernovae magnitude \cite{Riess_2022}.
However, this is not entirely unexpected. In Fig.~\ref{fig:b_and_dz_star} we show the clumping factor $b = (\langle\rho_b^2\rangle - \langle\rho_b\rangle^2)/\langle\rho_b\rangle^2 = \langle \delta_b^2\rangle$ at $z = 1270$ (the redshift at which we get the largest change to the recombination history). 
It can be seen that it is on the order of $1\%$. 
This can be contrasted with the results from Ref.~\cite{jedamzik2024cosmicrecombinationpresenceprimordial} which considered PMF spectra that source nonlinear baryon clumping and can significantly alleviate the Hubble tension. 
Moreover, our modified ionization fractions are about an order of magnitude smaller than the modified recombination histories considered by Refs.~\cite{jedamzik2025hints,lynch2024desihubbletensionlight,lynch2024reconstructing}, which suggested that current cosmological data shows a slight preference for PMFs over LCDM. 
This is all consistent with the very modest clumping ratios that are sourced by the PMF spectra we consider here.

\section{Discussion}
\label{sec:Discussion}

Baryon clumping sourced by PMFs remains a viable and intriguing avenue to reduce the Hubble tension \cite{mirpoorian2024modifiedrecombinationhubbletension} and provide an improved fit to the DESI BAO data \cite{mirpoorian2025dynamicaldarkenergynecessary}. A complete account of the modified recombination history in the presence of PMFs is a formidable task that requires detailed understanding of the evolution of small-scale fluctuations up until the time of recombination and careful treatment of Lyman-$\alpha$ radiative transport. In this paper, we self-consistently address both of these matters using a linearized treatment of the MHD equations and radiative transport BE. 

We have extended the LMHD framework in an expanding spacetime, first introduced by Refs.~\cite{Brandenburg_1996,SB98,JKO98}, so that modes can be evolved up until the time of recombination, while taking care to accurately account for the scale-dependent speed of sound and its sensitivity to the recombination history. We have shown that when linear modes are evolved with all the relevant dissipative effects accounted for, that density fluctuations in the early Universe remain firmly in the linear regime for the case of $n_B < -3$, with a clumping factor $b \lesssim 0.1$ prior to recombination for magnetic field strengths $B_0 = [5 \text{ pG}, 5 \text{ nG}]$. Beyond the applicability of our extended LMHD framework to red-tilted spectra for which a linearized treatment is well-justified, the linear results suggest a characteristic scale and amplitude for density fluctuations as a function of magnetic field strength and spectral index that can help inform the dynamical range and resolution of future nonlinear compressible MHD simulations. 

% Throughout our analysis, we have made the simplifying assumption that the baryon and photon temperature are equal $T_b = T_\gamma$. In reality, the LMHD equations need to be supplemented with an equation for $\dot{T}_b$ which accounts for Thomson, adiabatic and volumetric (from the dissipation of kinetic and magnetic energy) heating. In Appendix \ref{app:bar_heating} we treat these effects using our linearized formalism.  and adiabatic heating/cooling and volumetric heating from the dissipation of kinetic and magnetic energy \cite{jedamzik2024cosmicrecombinationpresenceprimordial}. 
% This 
% We find that when we compute $T_b(z)$ using the output from our LMHD results, that Thomson scattering overwhelms the other two terms and maintains $T_b \approx T_\gamma$ over the redshifts we consider. We therefore do not expect the inclusion of a baryon temperature evolution equation to significantly alter our results, though we plan to include this in future work.

The conservation scale discussed in Section \ref{sec:FSR} indicates that the LMHD approach may be applicable for a broader range of spectra which also admit a separation of scales between a large scale conserved field and a small-scale perturbation. Generically, we expect phase transition fields to take the form of a Batchelor spectrum $(n_B = 2)$ on scales larger than the integral scale and a turbulent, Kolmogorov like spectrum on scales below the integral scale $(n_B = -11/3)$ \cite{durrer2013cosmological}
\eq{
P_B(k) = A\begin{cases}
    (k/k_I)^{2}, & k < k_I\\
    (k/k_I)^{-11/3}, & k>k_I
\end{cases}.
}

If the integral scale is much larger than the clumping scale, then we can similarly decompose our magnetic field on small scales into a conserved mean field $\mathbf{B_0}$, which is comprised of the large scale modes satisfying $k< \kcons$, and a small-scale perturbed field $\bd{b}$. The FSR is highly viscous and expected to not be turbulent for most of its evolution so that discarding the advection term in the Euler equation is also reasonable in this regime \cite{banerjee2004evolution}. This opens up a much larger region of parameter space for which the LMHD formalism can directly be applied. We plan to revisit this in a future study and to use compressible MHD simulations to test our assumptions about the domain of applicability of LMHD. 

The LMHD framework is also particularly well-suited for implementation in CMB codes. In the LMHD scheme we have presented here, the entire linear evolution can be incorporated by introducing one additional dynamical variable, the perturbed small-scale magnetic field, which couples linearly to all other fluid variables. This will hopefully allow for better understanding of CMB constraints on PMFs, as well as better inform MHD evolution in the early Universe by coevolving the MHD equations with the photon Boltzmann hierarchy to self-consistently handle the evolution of modes across different regimes in the early Universe. 

In the second part of our work, we developed a fully self-consistent treatment of modified recombination in the presence of linear fluctuations. 
A modified recombination history not only changes the visibility function's shape and location of its peak, making it a promising avenue for alleviating the Hubble tension, but also introduces spectral distortions from the recombination lines that could in principle be detected by future CMB instruments \cite{sunyaev2009signals,Chluba_2016,chluba2019spectraldistortionscmbprobe}.
Although presented within the context of PMF induced perturbations, this work is applicable to any mechanism which sources small-scale inhomogeneities, such as small-scale isocurvature fluctuations \cite{Lee_2021}. 
To self-consistently handle modified recombination in the presence of small-scale inhomogeneities, we developed a flexible framework which solves the linearized radiative transfer BE across the Lyman-$\alpha$ line. 
Although the PMF spectra we have considered here are incapable of resolving the Hubble tension, the general approach may also find applicability to other PMF spectra so long as the density fluctuations remain linear and $\delta \tau_s \ll \tau_s$. 

In a future study, we  plan to implement a hybrid approach in which MHD simulations are used to solve for fluid perturbations for generic spectra. For the class of spectra which only source linear density perturbations, our modified recombination scheme can be implemented alongside these simulations to extract a self-consistent treatment of the modified recombination history. 
This should provide better constraints on the PMF spectra which can resolve the Hubble tension as well as allow for improved compressible MHD simulations, which accurately track the ionization fraction and the changing speed of sound as recombination proceeds. 

\section*{Acknowledgments}
The authors have benefited from discussions with Yacine Ali-Haïmoud, Jens Chluba, Axel Brandenburg, Kandaswamy Subramanian, Rishi Khatri, and Matias Zaldarriaga. 
TV acknowledges support from NSF grants 2012086 and 2309360, the Alfred P. Sloan Foundation through grant number FG-2023-20470, the BSF through award number 2022136, and the Hellman Family Faculty Fellowship.

\appendix
\onecolumngrid % Switch to single column for appendices

\section{Further details on PMFs and linear MHD}

\subsection{MHD and linear MHD derivation}\label{app:MHD_deriv}
In this appendix, we rederive the LMHD equations, which we use to evolve PMFs to and through recombination. We draw heavily on the work of Refs.~\cite{Brandenburg_1996,SB98,JKO98,subramanian2016origin}. To obtain the LMHD equations in the FLRW spacetime, we use the conservation of the stress-energy tensor. We decompose the stress-energy tensor into three pieces: $T^{\mu\nu} = T^{\mu\nu}_{I} + T^{\mu\nu}_{NI} + T^{\mu\nu}_{EM}$, where $T^{\mu\nu}_I$ is the stress-energy tensor of an ideal fluid, $T^{\mu\nu}_{NI}$ is the non-ideal contribution to the stress-energy tensor and gives rise to diffusive damping in the Euler equation, and $T^{\mu\nu}_{EM}$ is the electromagnetic contribution to the stress-energy tensor \cite{SB98}. 

\subsubsection{Metric and tetrads}

We denote components of tensors in comoving coordinates by Greek letters in the middle of the alphabet (e.g. $\mu,\nu\,\ldots)$ and components of tensors projected via the tetrads into the orthonormal frame with hatted Greek letters from the beginning of the alphabet (e.g. $\hat{\alpha},\hat{\beta},\ldots$). For spatial coordinates we use the same conventions but with Latin letters. In the comoving frame, the unperturbed FLRW metric takes the familiar form
\eq{\label{eq:FLRW_metric}
ds^2 = -dt^2 + a^2(t)\delta_{ij}dx^idx^j.
}
The tetrad vector components with respect to the this metric are
\eq{\label{eq:tetrads}
(e_{\hat{0}})^\mu &= (1,0,0,0)^\mu\\
(e_{\hat{a}})^\mu &= \frac{1}{a}\delta^\mu_{\hat{a}}.\\
}

For the perturbed metric, we work in the Newtonian gauge and consider only scalar fluctuations so that the perturbed metric is given by
\eq{\label{eq:pert_FLRW}
ds^2 = -e^{2\varphi}dt^2 + a^2[e^{-2\psi}\delta_{ij}] dx^idx^j.
}
We ignore anisotropic stress which allows us to relate the two scalar degrees of freedom in the perturbed metric via $\psi = -\varphi$. The linearized Einstein field equations give the Poisson equation
\eq{\label{eq:poisson}
\nabla^2 \varphi = 4\pi G a^2 T^{00}.
}

\subsubsection{Stress-energy tensor of a perfect fluid}

From homogeneity and isotropy of the metric, the unperturbed stress-energy tensor must take the form of a perfect fluid
\eq{
T_I^{\mu\nu} = (\rho + p)U^\mu U^\nu + p g^{\mu\nu} = \text{diag}[\rho,a^{-2}p,a^{-2}p,a^{-2}p],
}
where we have introduced the four velocity $U^\mu = dx^\mu/dt $. Linearly perturbing our fluid variables, we have
\eq{
\rho &= \rho + \delta\rho, \; \; p = p + \delta p\\
U^\mu &= U^\mu + \delta U^\mu = (e_{\hat{0}})^\mu + (0,\frac{dx^i}{dt}),
}
where $\rho = \rho_\gamma + \rho_b$ and $p = p_\gamma + p_b \approx p_\gamma$ in the TCR and $\rho = \rho_b$ and $p = p_b$ in the FSR.
The proper peculiar velocity is given by $v^i = adx^i/dt$ \cite{bertschinger1995cosmologicaldynamics}, so that to linear order we can write the components of the stress-energy tensor for an ideal fluid as
\eq{
T_I^{00} &= \rho + \delta \rho\\
T_I^{0i} &= (\rho + p)\frac{v^i}{a}\\
T_I^{ii} &= a^{-2}(p + \delta p).\\
}

We now apply the conservation of the stress-energy tensor $\nabla_\mu T_I^{\mu\nu} = 0$ to find the ideal fluid equations. Taking $\nu = 0$, we find the continuity equation, whose zeroth-order contribution gives the usual $\rho \propto a^{-3(1+w)}$ solution if we have a single fluid with an equation of state $p = w\rho$. Using conservation of baryon number and defining $\Theta = \nabla \cdot \bd{v}$, $\delta_b = \delta\rho_b/\overline{\rho}_b$, and $\delta_\gamma = 4\delta\rho_\gamma/\overline{\rho}_\gamma$, the first-order continuity equation gives
\eq{\label{eq:linearized_cont}
\dot{\delta}_\gamma &= -\frac{\Theta}{3a}, \; \; \dot{\delta}_b = -\frac{\Theta}{a}.\\
}
When including metric perturbations, we must add $-3\dot{\varphi}$ to the right hand side of this equation. 
We drop this term because, as argued in Appendix~\ref{app:photon_eom}, it is subdominant compared to $\Theta/a$ during matter domination when we are most concerned with sourcing compressional power that alters recombination. 
The spatial components of the conservation equation ($\nu = i$) have no zeroth-order term. 
The first-order term gives
\eq{
a\nabla_\mu T_I^{\mu i} &=  \frac{1}{a^4}\frac{\d}{\d t}[a^4(\rho + p)\bd{v} ] + \frac{\nabla \delta p}{a} + \rho \frac{\nabla\varphi}{a}.\\
}
During the TCR, when we have incompressible MHD, we drop the gravitational potential from the Euler equation. Having found the first-order continuity and Euler equation for an ideal fluid, we now turn to the electromagnetic contribution to the the stress-energy tensor and fluid equations.

\subsubsection{The electromagnetic contribution to the fluid equations}
In an inertial frame, the components of the field strength tensor are
\eq{\label{eq:physical_fields_strength}
F^{\hat{\alpha}\hat{\beta}} = \begin{pmatrix}
    0 & E_{x,p} & E_{y,p} & E_{z,p}\\
    -E_{x,p} & 0 & B_{z,p} & -B_{y,p}\\
    -E_{y,p} & -B_{z,p} & 0 & B_{x,p}\\
    -E_{z,p} & B_{y,p} & -B_{x,p} & 0\\
\end{pmatrix}.
}
We denote the electromagnetic fields projected into an orthonormal frame by $\bd{E}_p,\bd{B}_p$, with the subscript $p$ meant to indicate that these are the physical fields that an inertial observer would measure. These can be contrasted with the comoving fields $\bd{E} = a\bd{E}_p$ and $\bd{B} = a^2\bd{B}_p$. We note that our comoving electric field differs from that of Ref.~\cite{SB98}, since they define their tetrad with respect to the conformally flat FLRW metric ($dt = a d\eta$). 

Using our orthonormal tetrad from Eq.~\eqref{eq:tetrads}, the components in comoving coordinates are given by:
\eq{\label{eq:comov_fields_strength}
F^{\mu\nu} = \begin{pmatrix}
    0 & E_{x,p}/a & E_{y,p}/a & E_{z,p}/a\\
    -E_{x,p}/a & 0 & B_{z,p}/a^2 & -B_{y,p}/a^2\\
    -E_{y,p}/a & -B_{z,p}/a^2 & 0 & B_{x,p}/a^2\\
    -E_{z,p}/a & B_{y,p}/a^2 & -B_{x,p}/a^2 & 0\\
\end{pmatrix}.
}
We see then that going from the orthonormally projected field-strength tensor in Eq.~\eqref{eq:physical_fields_strength} to the comoving field-strength tensor in Eq.~\eqref{eq:comov_fields_strength} simply amounts to replacing physical fields with comoving ones. 

To find Maxwell's equations in a FLRW universe, we start with their covariant form
\eq{
\nabla_{\nu}F^{\mu\nu} &= 4\pi J^\mu\\
\d_{[\lambda} F_{\mu\nu]} &= 0.
}
The components of the four-current density in an orthonormal frame are given by  $ J^{\hat{\alpha}} = (\rho_e,\bd{J}) $, where $\rho_e$ is the charge density and $\bd{J}$ is the current density. 
In the comoving frame we have $J^\mu = (\rho_e, \bd{J}/a)$. 
We can now write down Maxwell's equations in terms of the physical electric and magnetic fields
\eq{
\nabla \cdot \bd{E}_p &= 4\pi a \rho_e\\
\nabla \times (a^2 \bd{B}_p) &= 4\pi a^3 \bd{J} + a \frac{\d}{\d t} (a^2 \bd{E}_p)\\
\frac{\d}{\d t} (a^{2} \bd{B}_p) &= -a\nabla \times \bd{E}_p\\
\nabla \cdot \bd{B}_p &= 0.
}
In the non-relativistic case we consider here, we drop the displacement current $\d_t(a^2 E)$ in the Ampere-Maxwell equation \cite{thorne2021elasticity}. To close this system we need to add the non-relativistic limit of the generalized Ohm's law. In the rest frame of the conducting plasma, Ohm's law is given by $\bd{J} = \sigma \bd{E}$. Boosting to the comoving frame, we find                                                                  
\eq{\label{eq:ohm_law}
\bd{J} &= \sigma(\bd{E}_p + \bd{v} \times \bd{B}_p),
}
where $\sigma$ is the conductivity of the fluid \cite{SB98}\cite{JKO98}. Defining the magnetic diffusivity $\eta_M = 1/(4\pi\sigma)$, we can rewrite our induction equation in terms of the comoving magnetic field
\eq{\label{eq:induction}
\dot{\bd{B}} &= \frac{1}{a}\nabla \times (\bd{v} \times \bd{B}) + \frac{\eta_M}{a^2}\nabla^2\bd{B}.
}

Throughout this work, we use the ideal MHD limit, in which we assume the primordial plasma to have infinite conductivity, or equivalently $\eta_M \to 0$. This is justified so long as we consider length scales above the comoving magnetic diffusion scale $k_{MD}(t) = (\int_0^t [\eta_D(t')/a^2(t')]dt')^{-1/2} $. Using the Spitzer resistivity $\eta_M \simeq m_e^{1/2}e^2/T^{3/2}$ for temperatures below the mass of an electron \cite{Brandenburg_2005,Jedamzik_2011}, we find that all wavenumbers that we consider satisfy $k \ll k_{MD}$ for the redshifts over which we evolve the LMHD equations. We are therefore safe to work in the ideal MHD limit. 

The electromagnetic contribution to the stress-energy tensor is given by
\eq{
T^{\mu\nu}_{EM} &= \frac{1}{4\pi}[F^{\mu\sigma}F^{\nu}_\sigma - \frac{1}{4}g^{\mu\nu}F^{\rho\sigma}F_{\rho\sigma}].
}
Decomposing the comoving magnetic field into our mean and perturbed fields ($\bd{B} = \bd{B_0} + \bd{b}$), we find the electromagnetic contribution to the Euler equation
\eq{
a\nabla_\mu T^{\mu i}_{EM} = \frac{\bd{B}_{0} \times (\nabla \times \bd{b})}{4\pi a^5} . \\
}

\subsubsection{Fluid regimes and non-ideal damping effects}\label{app:regimes_damping}

Having derived the ideal MHD equations for a perfect fluid, we can now consider the various dissipative effects that the primordial plasma experiences after neutrino decoupling. Thomson scattering between photons and electrons causes diffusive damping which produces a non-ideal damping term that must be accounted for in the Euler equation. We initially ignore PMFs and work within the standard LCDM framework to illustrate the key aspects of diffusive damping. We only consider scalar perturbations to simplify the discussion and denote the velocity projected along the direction of the wavector by $\bd{v} = \hat{\bd{k}}v$ for both baryons and photons.

To account for diffusive damping effects, we use the Boltzmann hierarchy for photons coupled to the baryon fluid equations. The equations can be written in terms of the brightness function $\Delta = \delta T/T$ and the Stokes polarization parameter $Q$. The normal modes for $\Delta $ and $Q$ can then be expanded in terms of their angular dependence, e.g. $\Delta(t,\bd{k},\hat{\bd{n}}) = \sum_{\ell}(-i)^{\ell}\Delta_{\ell}(t,k) P_{\ell}(\mu) $, with $\mu = \hat{\bd{k}}\cdot \hat{\bd{n}}$. This gives rise to an infinite hierarchy of coupled ordinary differential equations for $\Delta_{\ell}, Q_{\ell}$ \cite{bond1984cosmic, Ma_1995}. 

We now focus on the hierarchy for $\Delta_{\ell}$ and drop terms proportional to the metric potential and its derivatives, as well as terms which couple $\Delta_\ell$ to $Q_\ell$ since this simplifies the discussion and does not effect the overall qualitative behavior. 
The hierarchy in $\Delta_{\ell}$ as well as the continuity and Euler equations for the baryon fluid can then be written as \cite{Hu_1997, dodelson2020modern}
\eq{\label{eq:fluid_Boltz_coupled}
\dot{\Delta}_0 &= -\frac{k}{3a}\Delta_1\\
\dot{\Delta}_1 &= \frac{k}{a}\bigg[\Delta_0 - \frac{2}{5}\Delta_2\bigg] - \frac{\dot{\tau}}{a}(\Delta_1 - v_b)\\
\dot{\Delta}_2 &= \frac{k}{a}\bigg[\frac{2}{3}\Delta_1 - \frac{3}{7}\Delta_3\bigg] - \frac{9\dot{\tau}}{10a}\Delta_2\\
\dot{\Delta}_\ell &= \frac{k}{a}\bigg[\frac{\ell}{2\ell - 1}\Delta_{\ell -1} - \frac{\ell + 1}{2\ell + 3}\Delta_{\ell+1}\bigg] - \frac{\dot{\tau}}{a}\Delta_{\ell} \text{, for ($\ell \geq 3$)}\\
\dot{\delta}_b &= -\frac{1}{a}\nabla \cdot \bd{v}_b\\
\dot{v}_b &= -\frac{\dot{a}}{a}v_b + \frac{\dot{\tau}}{aR}(\Delta_1 - v_b).
}

Let us first consider $\dot{\tau} \gg k$ limit of these equations. In this limit, we can drop the time derivative term in the $\ell \geq 2$ equations to see that these moments all satisfy $\Delta_\ell \sim (k/\dot{\tau})\Delta_{\ell-1} \ll \Delta_{\ell - 1}$ \cite{baumann2022cosmology}. Since higher moments are suppressed by a factor of $k/\dot{\tau} \ll 1$, we can set $\Delta_\ell = 0$ for $\ell \geq 3$ and close our system using the $\Delta_2$ equation. When polarization is taken into account, we find $\Delta_2 = (8k/9\dot{\tau})\Delta_1$ \cite{Hu_1997}. We therefore have a fluid description for photons for modes satisfying $\dot{\tau} \gg k$ with $\Delta_0 = \delta_\gamma, \Delta_1 = v_\gamma$.  

At early times, the mean free path of a photon is very short so that to lowest order in $1/\dot{\tau}$, we can set $v_\gamma \approx v_b$, which is known as the no-slip condition or tight-coupling approximation between baryons and photons. One can perform higher order expansions in $1/\dot{\tau}$ in the baryon Euler equation leading to the various tight-coupling approximations that are commonly employed in CMB Boltzmann solvers \cite{Diego_Blas_2011}. Performing a first-order expansion in $1/\dot{\tau}$, and working in the WKB approximation in which the time dependence of the Boltzmann moments is assumed to be $\Delta_0,\Delta_1 \sim \exp[i\int \omega(t')dt]$, we can find the imaginary component of $\omega$ and define a characteristic damping scale $k_D$, which is given by equation \eqref{eq:diff_damp_WKB}. 

In LCDM, modes which satisfy $k > k_D$ are exponentially damped by the diffusing photons. As mentioned in Section \ref{sec:diff_damp}, a mode will also be damped if it satisfies $k \sim \dot{\tau}$. From Eq.~\eqref{eq:fluid_Boltz_coupled}, we can see why this is. Once a mode's wavelength becomes comparable to the photon mean free path, i.e. $k \sim \dot{\tau}$, the truncation at $\ell = 3$ and closure of the hierarchy relating $\Delta_2$ to $\Delta_1$ we employed above breaks down. Instead, power is transferred from the $\ell = 0,1$ moments to higher moments, which are then exponentially suppressed \cite{Hu_1997}.

From Eq. \eqref{eq:fluid_Boltz_coupled}, we can also see how the damping terms of Eq. \eqref{eq:euler_F_d} emerge. In the TCR regime, the truncation and closure scheme at $\Delta_2$ gives rise to a damping term of the form $(k^2/\dot{\tau})\bd{v}$ in the Euler equation from substituting in $\Delta_2 = (8k/9\dot{\tau})\Delta_1$ into $\Delta_1$'s evolution equation. In the FSR, $\Delta_1$ is erased so that in the baryon Euler equation we are left with a term of the form $-\dot{\tau}/aR v_b$, which is precisely the $\alpha \bd{v}$ damping term in Eq.\eqref{eq:euler_F_d}.

In the TCR, we define the non-ideal part of the stress-energy tensor $T^{\mu\nu}_{NI}$ to incorporate the $(k^2/\dot{\tau})\bd{v}$ damping term caused by diffusing photons. The non-ideal stress-energy contribution is given by \cite{WeinbergGC}
\eq{
T_{NI}^{\mu\nu} &= -\eta H^{\mu\alpha}H^{\nu\beta}W_{\alpha\beta}\\
H^{\alpha\beta} &= g^{\alpha\beta} + U^{\alpha}U^{\beta}\\
W_{\alpha\beta} &= \nabla_{\alpha}U_{\beta} + \nabla_{\beta}U_{\alpha} - \frac{2}{3}g_{\alpha\beta}\nabla_\gamma U^{\gamma}.
}
We can now find the non-ideal contribution to the Euler equation
\eq{
a\nabla_\mu T^{\mu i}_{NI} &= -\frac{\eta}{a^2}[\nabla^2 v^i + \frac{1}{3}\nabla^i (\nabla \cdot \bd{v})].
}
For a compressional mode, this damping term is given by $(16/45)(k^2/a)\rho_\gamma \lmfp\bd{v}$ in agreement with the damping term obtained from a Boltzmann like treatment presented above \cite{Hu_1997}.

\subsection{Linear MHD - shortcomings and applicability}
\label{appendix:LMHD_validity}

In this appendix, we seek to determine which, if any, PMF spectra admit a reasonable separation of scales and linearization of the MHD equations. 
% The PMF two point correlator and power spectrum are given by Eqs.~\eqref{eq:B_spec} and \eqref{eq:P_B(k)}. 
The LMHD scheme we have employed assumes that the comoving mean-field, ${\bd{B}}_0$, is constant until recombination. 
If ${\bd{B}}_0$ has non-trivial dynamics, then not only could the amplitude of the comoving mean-field vary with time, but the field itself could rotate direction so that the angle it makes with the wavevector would also become time dependent, i.e. $\hat{\bd{B}}_0 \cdot \hat{\bd{k}} = \cos\theta(t)$. 
We would then need to solve for the evolution of ${B}_0(t)$ and $\theta(t)$, which we wish to avoid. 
A conserved comoving mean-field is most easily achieved if it is dominated by super-horizon modes $(k\ll aH)$, since the super-horizon limit of the induction Eq.~\eqref{eq:induction} gives the conservation of the comoving field. 
To this end, we set our PMF spectrum so that it has a red tilt $(n_B<-3)$ with a super-horizon integral scale. 
We can then regard the small-scale field $\bd{{b}}(\bd{x},t)$ to be a dynamically evolving, spatially varying perturbation on top of the constant homogeneous mean-field $\bd{{B}}_0$. 
As mentioned in Section \ref{sec:Discussion}, this formalism may also be applicable to the FSR for some causally generated fields with subhorizon integral scales, a matter we plan to revisit in future work.

Since the LMHD scheme requires spectra which allow the Lorentz force in the Euler equation $\sim \bd{B} \times (\nabla \times \bd{B})$ to be reasonably approximated by $\mathbf{B_0} \times (\nabla \times \bd{b})$, we have to check that red-tilted spectra are consistent with a linearized approximation of the Lorentz force. In general, if we decompose our field into $\bd{B} = \mathbf{B_0} + \bd{b}$, four terms contribute to the Lorentz force 
\eq{
\bd{B} \times (\nabla \times \bd{B}) &= \mathbf{B_0} \times (\nabla \times \mathbf{B_0}) + \mathbf{B_0} \times (\nabla \times \bd{b}) + \bd{b} \times (\nabla \times \mathbf{B_0}) + \bd{b} \times (\nabla \times \bd{b}).
}
We can already anticipate that the Lorentz force will be dominated by $\mathbf{B_0} \times (\nabla \times \bd{b})$ for spectra with power concentrated on larger scales. In Fourier space, the gradient weights the Lorentz force by a factor of the wavenumber $k$ of the mode of the field on which it operates, i.e. $\nabla \times \bd{b} \gg \nabla \times \mathbf{B_0}$. Additionally, by construction we have $\abs{\mathbf{B_0}} \gg \abs{\bd{b}}$, so that we expect $\mathbf{B_0} \times (\nabla \times \bd{b})$ to dominate over the other terms in the Lorentz force.

We now analyze the statistical features of the Lorentz force to see whether our suspicions are borne out and to provide a more rigorous justification for its linearization. In Fourier space, the Lorentz force is given by
\eq{
 \bd{L_B}(\bd{k}) &= \int d^3x [\bd{B}(\bd{x}) \times (\nabla \times \bd{B}(\bd{x}) ) ] e^{-i\bd{k}\cdot \bd{x}} = \int d^3q \bd{B}(\bd{k} - \bd{q}) \times  (i\bd{q} \times \bd{B}(\bd{q})).
}
Since $ \bd{L_B}(\bd{k})$ is itself a stochastic variable, we need to characterize its statistical properties to determine whether it receives a dominant contribution from a certain configuration of modes $\bd{q}$ and $\bd{k}$. 

$ \bd{L_B}$'s mean vanishes and its variance is given by
\eq{
&\langle  \bd{L_B}(\bd{k})\cdot  \bd{L^*_B}(\bd{k}')\rangle \propto \delta(\bd{k}-\bd{k}')\int \,d^3q\ k q \bigg(\frac{kq\sin^2\theta}{k^2 + q^2 - 2kq\cos\theta} + 2\cos\theta\bigg)P_B(q)P_B(k-q).
}
Substituting in Eq. \eqref{eq:P_B(k)}, we find
\eq{\label{eq:varMHD}
&\langle  \bd{L_B}(\bd{k})\cdot \bd{L^*_B}(\bd{k}')\rangle \propto \int_0^\infty \int_0^\pi  \,d \ln q \,d\theta \bigg( \frac{q}{k} \bigg)^{n_B + 3}\bigg(1 + \frac{q^2}{k^2} - 2\frac{q}{k}\cos\theta \bigg)^{n_B/2} \sin\theta\bigg(\frac{\frac{q}{k}\sin^2\theta}{1 + \frac{q^2}{k^2} - 2\frac{q}{k}\cos\theta} + 2\cos\theta\bigg) . \\
}

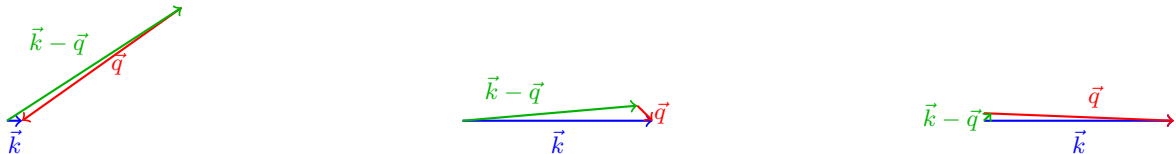
\begin{figure}[t!]
\label{fig:triangles}
    \centering
    \begin{minipage}[b]{0.3\textwidth}
        \centering
        \begin{tikzpicture}
            \draw[->, thick, blue] (0,0) -- (0.2,0) node[midway, below] {$\vec{k}$};
            \draw[->, thick, red] (2.3,1.5) -- (0.2,0) node[midway, right] {$\vec{q}$};
            \draw[->, thick, green!70!black] (0,0) -- (2.3,1.5) node[midway, above left] {$\vec{k}-\vec{q}$};
        \end{tikzpicture}
    \end{minipage}%
    \hfill
    \begin{minipage}[b]{0.3\textwidth}
        \centering
        \begin{tikzpicture}
            \draw[->, thick, blue] (0,0) -- (2.5,0) node[midway, below] {$\vec{k}$};
            \draw[->, thick, red] (2.3,0.2) -- (2.5,0) node[midway, right] {$\vec{q}$};
            \draw[->, thick, green!70!black] (0,0) -- (2.3,0.2) node[midway, above left] {$\vec{k}-\vec{q}$};
        \end{tikzpicture}
    \end{minipage}%
    \hfill
    \begin{minipage}[b]{0.3\textwidth}
        \centering
        \begin{tikzpicture}
            \draw[->, thick, blue] (0,0) -- (2.5,0) node[midway, below] {$\vec{k}$};
            \draw[->, thick, red] (0,0.1) -- (2.5,0.0) node[midway, above right] {$\vec{q}$};
            \draw[->, thick, green!70!black] (0,0) -- (0.1,0.1) node[midway, left] {$\vec{k}-\vec{q}$};
        \end{tikzpicture}
    \end{minipage}%

    \vspace{0.5cm}  % Adds space between the figures and the caption
    
    % Full-width caption using \caption* to avoid extra numbering
    \captionsetup{width=1\linewidth}
    \caption{Three limits for the integrand in Eq. \eqref{eq:varMHD}: (left) $q/k \to \infty$, (middle) $k/q \to 0$, and (right) $\bd{q} \to \bd{k}$. The only configuration that allows us to linearize $\bd{L_B}$ is $\bd{q} \to \bd{k}$.}
    \label{fig:squeezed_triangles}
\end{figure}

If the integrand above is dominated by a certain configuration of modes, then we expect this configuration to dominate the Lorentz force as well. In Fig.~\ref{fig:squeezed_triangles}, we pictorially depict the three limiting cases we consider. If the integrand is dominated by $q/k \to \infty$, then $ \bd{L_B}$ can be approximated by
\eq{
 \bd{L_B}(\bd{k}) &= \int d^3q \bd{B}(\bd{k} - \bd{q}) \times  [i\bd{q} \times \bd{B}(\bd{q})] \approx \int d^3q \bd{B}(-\bd{q}) \times  [i\bd{q} \times \bd{B}(\bd{q})].
}
In this case, all antiparallel modes contribute, which is not consistent with our linearization scheme. For this configuration, the integrand of the variance goes like $(q/k)^{2 n_B+3}$ and would diverge for for $n_B > -3/2$.

If the integrand is dominated by $q/k \to 0$, then $ \bd{L_B}$ can be approximated by 
\eq{
 \bd{L_B}(\bd{k}) &= \int d^3q \bd{B}(\bd{k} - \bd{q}) \times  [i\bd{q} \times \bd{B}(\bd{q})] \approx \int d^3q \bd{B}(\bd{k}) \times  [i\bd{q} \times \bd{B}(\bd{q})].
}
In this case, a particular Fourier mode $\bd{k}$ couples to all other modes, so that we again would not have a reasonable linearization of the Lorentz force. For this configuration, the integrand goes like $(q/k)^{n_B+3}$ and would diverge for $n_B<-3$.

Finally, if we find that the integrand is sharply peaked for $\bd{k} \to \bd{q} \Leftrightarrow q/k \to 1, \theta = 0$ then we can approximate the integral in $ \bd{L_B}(\bd{k})$ by inserting a delta function $\delta(\bd{k} - \bd{q})$ to capture the sharp peak of the integrand. $ \bd{L_B}$ would then be given by
\eq{
 \bd{L_B}(\bd{k}) &= \int d^3q \bd{B}(\bd{k} - \bd{q}) \times  [i\bd{q} \times \bd{B}(\bd{q})] \approx \bd{B}(\bd{0}) \times [i \bd{k} \times \bd{B}(\bd{k})].
}

\begin{figure*}[!t]
    {\includegraphics[width=1\textwidth,height=0.6\textheight]{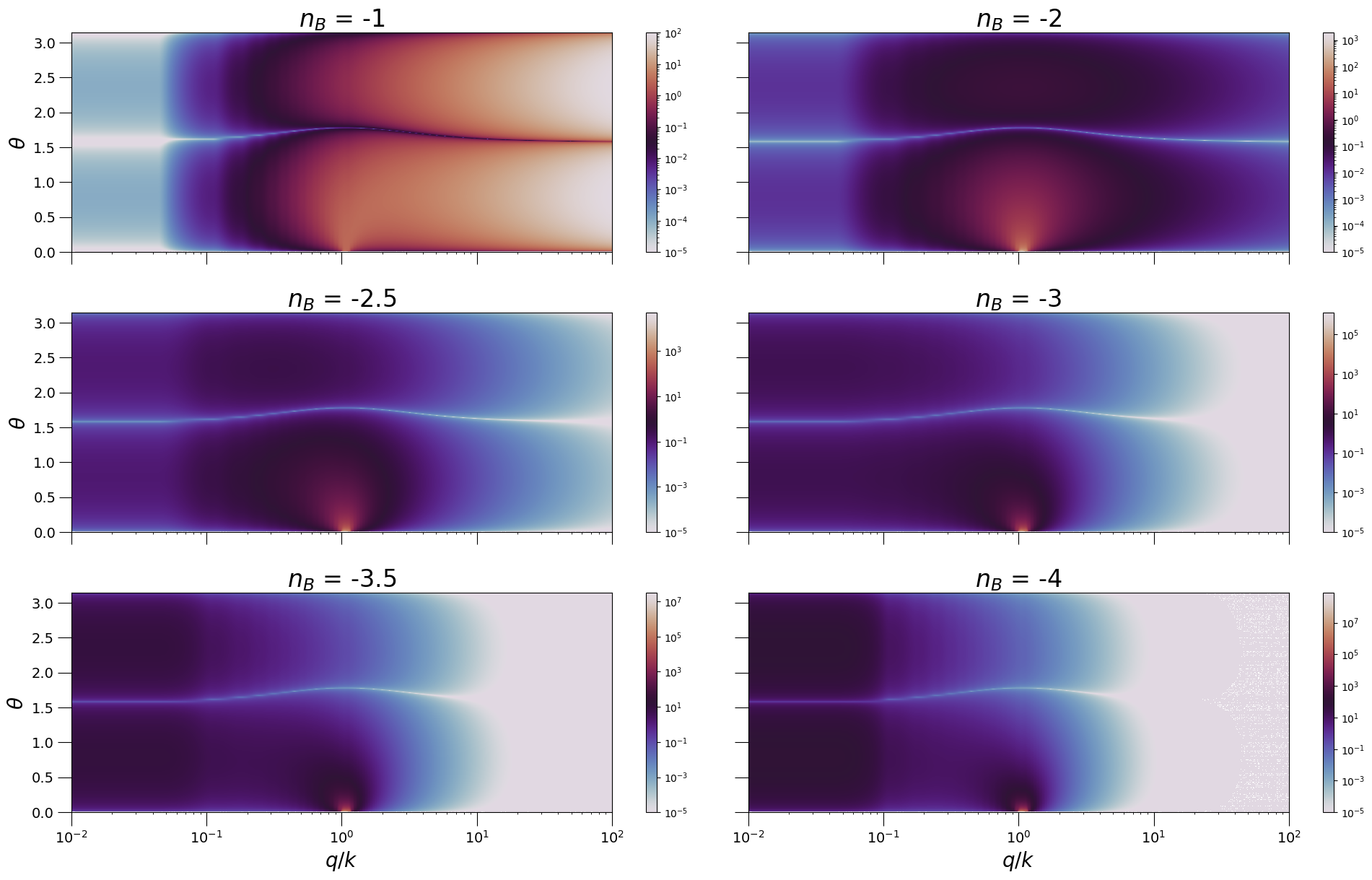}}
    \captionsetup{width=1\linewidth}
    \caption{Integrands from Eq. \eqref{eq:varMHD} for various values of spectral indices $n_B$. For $n_B = -1$, we see that that the integrand is dominated by $q/k \gg 1$ as expected. For all other $n_B$ values, we see the emergence and sharpening of the divergence at $\bd{q} \to \bd{k}$ as expected.}
    \vspace{-10pt}
    \label{fig:MHD_integrand}
\end{figure*}

Here, $\bd{B}(\bd{0})$ is a slight abuse of notation and is simply meant to capture that this is the Fourier amplitude of very long wavelengths. Since $\bd{B}(\bd{0})$ is precisely the large scale magnetic field and $\bd{B}(\bd{k})$ is the field at a particular, small-scale, the Lorentz force reduces to $\mathbf{B_0} \times (\nabla \times \bd{b})$ as desired. In this case, the integrand goes like $\theta^{n_B+1}$ and diverges for $n_B<-1$.

In Fig.~\ref{fig:MHD_integrand} we show the integrand from Eq.~\eqref{eq:varMHD} as a function of $k$ and $\theta$ for a few choices of $n_B$. We see that for $n_B < -1$, the integrand has a peak at $\bd{k} \to \bd{q}$, which sharpens as $n_B$ is decreased. The red-tilted spectrum we have proposed for the LMHD scheme has two divergences in the integrand of the variance: one as $q/k \to 0$ and one as $\bd{q} \to \bd{k}$. However, the second divergence is much faster than the former so that we expect it to dominate in the case of red-tilted spectra.

\subsection{PMF spectral conventions}\label{app:PMF_conventions}
For the red-tilted spectra we consider, we have an infrared (IR) divergence in magnetic energy density $\rho_B$. We introduce a super-horizon IR cutoff scale $\Lambda = 2\pi/k_{\Lambda} = 1 \text{ Gpc}$ to regulate $\rho_B$. We can use the energy density of our magnetic field $\rho_B$ to relate the normalization $A$ in Eq.~\eqref{eq:P_B(k)} to the mean field $B_0$. Since we expect the energy density to be dominated by our large-scale mean field, we have $8\pi\rho_B = \langle \bd{B}(\bd{x}) \bd{B}(\bd{x}) \rangle \approx B_0^2 $. 
Using Eqs.~\eqref{eq:B_spec} and \eqref{eq:P_B(k)}, we find
\eq{
8\pi \rho_B &= \frac{1}{(2\pi)^6} \int d^3k d^3k' \langle \bd{B}(\bd{k}) \bd{B}^*(\bd{k}') \rangle e^{i(\bd{k}-\bd{k}')\bd{x}}\\
&= \frac{A}{\pi^2 k_0^{\epsilon -3}}\int_{k_{\Lambda}}^k dk k^{\epsilon - 1} \approx -\frac{A}{\pi^2 \epsilon k_0^{\epsilon -3}} k_{\Lambda}^{\epsilon}.  \\
}
Equating this with $B_0^2$, we can rewrite our power spectrum as
\eq{\label{eq:P_B(k)_normalized}
P_B(k) &= \pi^2 B_0^2 \abs{\epsilon} k_{\Lambda}^{-\epsilon} k^{\epsilon-3}.
}
The dimensionless power per log wavenumber, can then be written as
\eq{\label{eq:dppl_B}
\Delta_B^2(k) = \frac{\abs{\epsilon}B_0^2}{2}\bigg( \frac{k}{k_\Lambda} \bigg)^{\epsilon}.
}

\subsection{Linear MHD in the presence of primordial density perturbations}
\label{appendix:IC_PCR}
In this appendix, we demonstrate how to evolve primordial adiabatic fluctuations in the presence of PMFs. 
As mentioned in Section \ref{sec:TCR}, the compressible FM modes damp almost identically to acoustic sound waves in LCDM and are therefore exponentially suppressed below the diffusion damping scale. 
However, in the presence of a mean-field ${B}_0$, density perturbations can source fluctuations in the PMF as they damp. 
We can see that this is the case by considering initial conditions with no PMF fluctuations and only initial power in the radiation density field, i.e. ${b}_y = 0, \delta_r = \delta_r(\bd{k},z_{HC})$. The integrated induction equation is given by
\eq{
{b}_y(\bd{k},z_{HC}) &= {B}_0(ik \xi_y(\bd{k},z_{HC})\cos\theta + 3\delta(\bd{k},z_{HC})\sin\theta),
}
so that these initial conditions are equivalent to $ik\xi_y(\bd{k},z_{HC})\cos\theta = -3\delta_r (\bd{k},z_{HC})\sin\theta $. 
Since $\xi_y(\bd{k},z_{HC}) \neq 0$, initial density perturbations in the presence of a mean field excite SM modes in the TCR even without initial power in ${b}_y$. 
The amplitude of PMF fluctuations sourced by adiabatic scalar perturbations when a mode transitions to free-streaming is given by
\eq{
{b}_y(\bd{k},\zfs) &= ik {B}_0\xi_y(\bd{k},\zfs)\cos\theta\\
&= -(3{B}_0T_{\xi_y}(\bd{k},\zfs)\sin\theta)\delta_r(\bd{k},z_{HC})\\
&\equiv T_{{b}_y,\delta_r}(\bd{k},\zfs)\delta_r(\bd{k},z_{HC}).
}

\begin{figure}[!t]
    \centering
    {\includegraphics[width=1.0\textwidth]{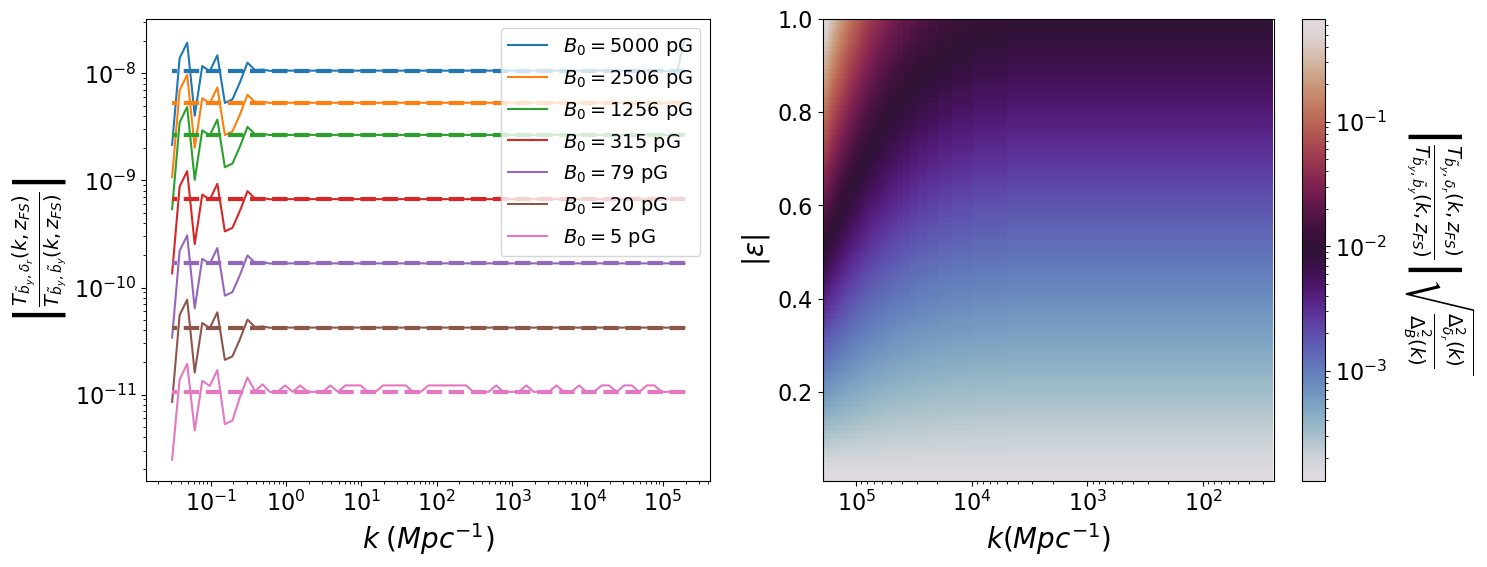}}
    \captionsetup{width=1\linewidth}
    \caption{(Left) Ratio of transfer functions for fluctuations in the PMF sourced by initial perturbations in the density field and in the PMF itself. The dashed lines are $3{B}_0\sin\theta$, showing that the ratio is given by the fraction of the initial power in each perturbation that goes into SM modes. We set $\theta = \pi/4$, but the same behavior is observed for all values of $\theta$. (Right) Ratio of power in $b_y$ sourced from PMFs and from primordial curvature fluctuations as defined in Eq.~ \eqref{eq:b_y_rel_contrib_2}. We set $\theta = \pi/2$ so as to allow for the largest possible contribution from initial density perturbations. With the exception of the smallest scales and reddest tilts, the contribution from initial density fluctuations is subdominant to the contribution from initial PMF fluctuations.}
    \label{fig:T_b_T_d_r}
\end{figure}

We can compare this to the power sourced in $b_y$ from initial perturbations in the PMF. Setting the initial conditions to $b_y = b_y(\bd{k},z_{HC}), \delta_r = 0$ and using $\xi_z(\bd{k},\zfs) = 0$, Eq.~\eqref{eq:integ_induc_cont} gives
\eq{
{b}_y(\bd{k},\zfs) &= T_{\xi_y}(\bd{k},\zfs){b}_y(\bd{k},z_{HC})\\
&\equiv T_{{b}_y,{b}_y}(\bd{k},\zfs) {b}_y(\bd{k},z_{HC}).
}

The characteristic size of the initial fluctuations in either primordial field is $I(k,z_{HC}) \sim \sqrt{\Delta_I^2(k)}$ where $I$ can be $b_y$ or $\delta_r$. The relative contribution to $b_y(\bd{k},\zfs)$ from each initial fluctuation is therefore given by
\eq{\label{eq:b_y_rel_contrib_1}
\abs{\frac{T_{{b}_y,\delta_r}(\bd{k},\zfs)}{T_{{b}_y,{b}_y}(\bd{k},\zfs)}}\sqrt{\frac{\Delta_{\delta_r}^2(k)}{\Delta_B^2(k)}} &= 3{B}_0\sin\theta\sqrt{\frac{\Delta_{\delta_r}^2(k)}{\Delta_B^2(k)}}.
}

To confirm our expectation that $\abs{\frac{T_{{b}_y,\delta_r}(\bd{k},\zfs)}{T_{{b}_y,{b}_y}(\bd{k},\zfs)}} \approx 3{B}_0\sin\theta$, which emerged form our assumption that all density perturbations rapidly damp away during the TCR, we initialize our modes in the TCR with $\delta_r = 1, {b}_y = 0$ to find $T_{{b}_y,\delta_r}$ and with $\delta_r = 0, {b}_y = 1$ to obtain $T_{{b}_y,{b}_y}$. On the left of Fig.~\ref{fig:T_b_T_d_r}, we show the numerical results for the ratio of our transfer functions and see that it is in fact well-approximated by $3{B}_0\sin\theta$.

To find $\Delta^2_{\delta_r}$, we first parametrize the primordial comoving curvature fluctuation's dimensionless power per log wavenumber by
\eq{\label{eq:PDF_spec}
\Delta_\zeta^2(k) &= A_s\bigg(\frac{k}{k_0}\bigg)^{n_s-1}.
}
The super-horizon solution for primordial density fluctuations is $\delta_r = -8/5\zeta$ \cite{baumann2022cosmology}, which allows us to find 
\eq{\label{eq:dppl_delta_r}
\Delta_{\delta_r}^2(k) &= \frac{64}{25} A_s\bigg(\frac{k}{k_0}\bigg)^{n_s-1}.
}
We use the best fit TT,TE,EE+lowE+lensing+BAO value for $A_s$ and $n_s - 1$, and follow the usual convention of setting the pivot scale to $k_0 = 0.05 \text{ Mpc}^{-1}$ \cite{2020}. Substituting Eqs.~\eqref{eq:dppl_B} and \eqref{eq:dppl_delta_r} into Eq.~\eqref{eq:b_y_rel_contrib_1}, we find
\eq{\label{eq:b_y_rel_contrib_2}
\abs{\frac{T_{b_y,\delta_r}(\bd{k},\zfs)}{T_{b_y,b_y}(\bd{k},\zfs)}}\sqrt{\frac{\Delta_{\delta_r}^2(k)}{\Delta_B^2(k)}} &= \frac{24}{5}\sin\theta \sqrt{\frac{A_s}{2}\frac{k^{n_s-1-\epsilon}}{k_0^{n_s-1}k_{\Lambda}^{-\epsilon}}}.
}
On the right of Fig.~\ref{fig:T_b_T_d_r}, we show this ratio for a range of scales and spectral tilts. We see that the contribution to small-scale fluctuations in PMFs from primordial density fluctuations are subdominant to the contribution from initial fluctuations in the PMFs themselves. This justifies our decision to ignore initial density fluctuations in the TCR.

\subsection{Baryon heating}
\label{app:bar_heating}

The baryon temperature's evolution follows from the first law of thermodynamics and is given by
\eq{\label{eq:Tb_evo}
\dot{T}_b + 2HT_b =  \Gamma_c(T_\gamma - T_b) + \frac{2}{3}\frac{\dot{\delta}_b}{1+\delta_b}T_b + \frac{\Gamma_{\text{diss}}}{(3/2)nk_B},
}
where the terms on the right hand side account for Compton heating, adiabatic heating, and any additional heating due to the dissipation of kinetic and magnetic energy \cite{chluba2015effectprimordialmagneticfields,Lee_2021,jedamzik2024cosmicrecombinationpresenceprimordial}. $\Gamma_c$ is the Compton heating rate and is given by
\eq{
\Gamma_c = \frac{8 \rho_\gamma x_e\sigma_T}{3 m_e(1+x_e + f_{He})}.
}
$\Gamma_{\text{diss}}$ is the baryon heating rate per unit volume due to kinetic and magnetic energy dissipation. 
Since some of the dissipated kinetic and magnetic energy can go into CMB spectral distortions, working out the precise form of $\Gamma_{\text{diss}}$ is not straightforward. 

Prior to recombination, the Compton heating term dominates and enforces $T_\gamma \approx T_b$. 
Since $x_e$ falls by several orders of magnitude over the course of recombination, both $\Gamma_c$ and $\alpha$, which are proportional to $x_e$, fall as well. 
The reduced rate of Thomson scattering allows baryons to thermally decouple from the CMB.
Moreover, since $\alpha$ falls significantly, turbulence can be developed in the post-recombination epoch.
The turbulent cascade can transfer kinetic and magnetic power to smaller scales, where it is ultimately dissipated and heats the baryons \cite{Trivedi_2018}.
Additionally, dissipation from ambipolar diffusion has been shown to heat the baryons \cite{sethi2005primordial}.
Post-recombination baryon heating shifts the value of case-B recombination coefficient $\alpha_B$ and can therefore slow down recombination at redshifts of $z\lesssim 1000$ and cause a larger freezeout value for $x_e$ \cite{chluba2015effectprimordialmagneticfields}.

Dissipation due to the linear drag term in the FSR sources y-type spectral distortions in the CMB \cite{Trivedi_2018}. 
Since a y-type distortion shifts $T_\gamma$, the Compton heating term can then heat the baryons as well. 
Understanding the detailed form of the spectral distortions from MHD dissipative processes is a challenging problem and beyond the scope of this paper.
For an overview of CMB spectral distortions see Ref. \cite{Chluba_2011} and for attempts to compute CMB spectral distortions from MHD dissipation in the FSR see Refs. \cite{Trivedi_2018}.

In lieu of a complete treatment of CMB spectral distortions with Thomson heating coupling $T_\gamma$ to $T_b$, we assume that $T_\gamma$'s only evolution is due to cosmic expansion ($T_\gamma \propto a^{-1}$) and make the simplifying assumption that all dissipation of kinetic and magnetic energy goes directly into heating the baryons. 
To incorporate the $\Gamma_{\text{diss}}$ heating rate in our LMHD framework, we follow Ref. \cite{jedamzik2024cosmicrecombinationpresenceprimordial} and set a uniform heating rate from the dissipation of kinetic and magnetic energy
\eq{
\Gamma_{\text{diss}} &= \frac{d}{dt} \bigg[ \frac{1}{2} \langle \rho_b v_b^2 \rangle   + \frac{ \langle B^2 \rangle}{8\pi} \bigg]. 
}
This homogeneous heating rate can at best serve as an upper limit for the expected baryon heating.

% Here, we will be primarily concerned with perturbations to $T_b$ in the post-recombination Universe.
% Baryon heating is most pronounced in the post-recombination Universe.

% In the context of our linear analysis in which we ignore ambipolar diffusion, we cannot fully account for baryon heating. 
% To incorporate linear linear heating effects in our LMHD framework, we follow Ref. \cite{jedamzik2024cosmicrecombinationpresenceprimordial} and set a uniform heating rate from the dissipation of kinetic and magnetic energy
% \eq{
% \Gamma_{\text{diss}} &= \frac{d}{dt} \bigg[ \frac{1}{2} \langle \rho_b v_b^2 \rangle   + \frac{ \langle B^2 \rangle}{8\pi} \bigg]. 
% }
% This homogeneous heating rate does not account for spatial variations in the kinetic and magnetic energy dissipation and assumes that all dissipation from the photon drag $\alpha$ goes into heating the baryons and not into spectral distortions. 
% Therefore, it can at best serve as an upper limit for the expected heating in the context of LMHD. 

We can now perform a perturbative expansion for Eq. \ref{eq:Tb_evo} to solve for linear transfer functions for the linearly perturbed baryon temperature $\delta T_b^{(1)}$ and find the lowest order homogeneous shift to the baryon temperature $\langle \delta T_b^{(2)} \rangle$. We then have
\eq{\label{eq:Tb_evo}
\dot{\overline{T}}_b &= -2H \overline{T}_b + \overline{\Gamma}_c(T_\gamma - \overline{T}_b) \\
\delta \dot{T}^{(1)}_b &= -2H\delta T_b^{(1)} + \delta \Gamma^{(1)}_c(T_\gamma - \overline{T}_b) - \overline{\Gamma}_c \delta T^{(1)} - \frac{2}{3}\frac{\Theta^{(1)}}{a} \overline{T}_b\\
\langle \delta \dot{T}_b^{(2)} \rangle &= -2H\langle \delta T_b^{(2)} \rangle + \langle \delta \Gamma^{(2)}_c \rangle (T_\gamma - \overline{T}_b) - \overline{\Gamma}_c \langle \delta T^{(2)}_b \rangle - \langle \delta \Gamma_c^{(1)} \delta T^{(1)}_b \rangle + \frac{ \Gamma_{\text{diss}} }{(3/2)nk_B} + \frac{2}{3}\frac{\langle \Theta^{(1)}\delta_b^{(1)} \rangle }{a}\overline{T}_b + \frac{2}{3}\frac{\langle \Theta^{(1)}\delta T_b^{(1)} \rangle }{a}, \\
}
where we have used the fact that the uniform heating rate $\Gamma_{\text{diss}}$ only enters at second order in the perturbations. 

We couple the equation for $\delta T_b^{(1)}$ to the FSR LMHD equations. 
Since the case-B recombination coefficient depends on the local matter temperature, we also require $\delta \alpha_B$ which will modify the perturbed recombination equations for $\delta x_e$. 
We use the expression for $\alpha_B$, as given in Ref. \cite{pequignot1991total}, which we perturb to second order in $T_b$. 

\begin{figure*}[!t]
    \centering

    \includegraphics[width=1\linewidth]{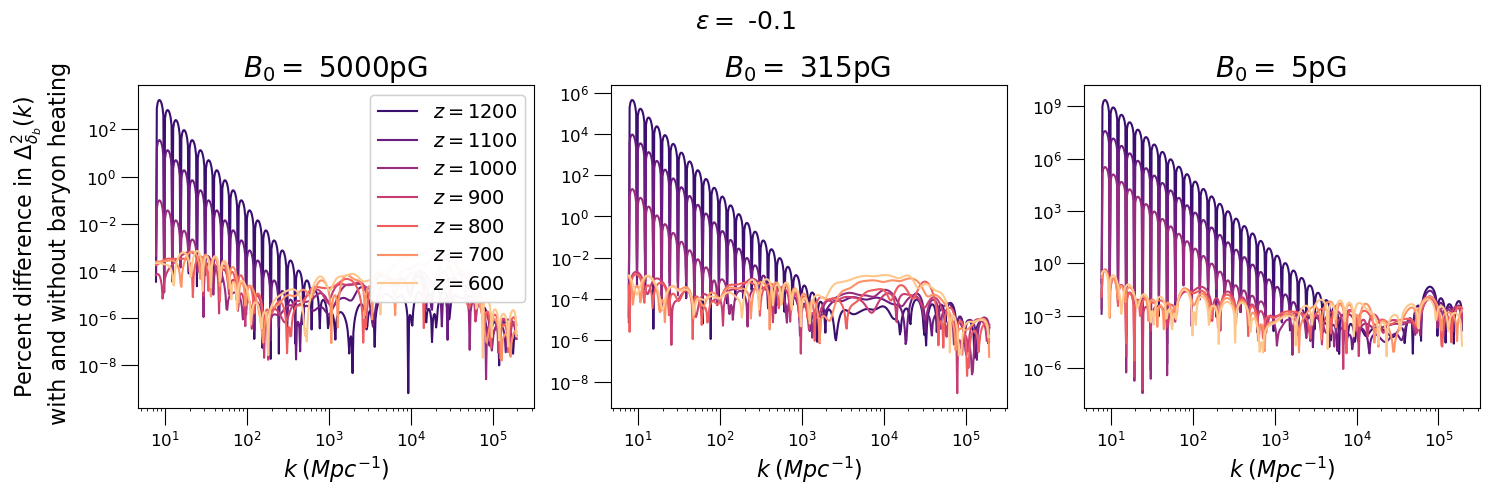}

    \vspace{1em}

    \includegraphics[width=1\linewidth]{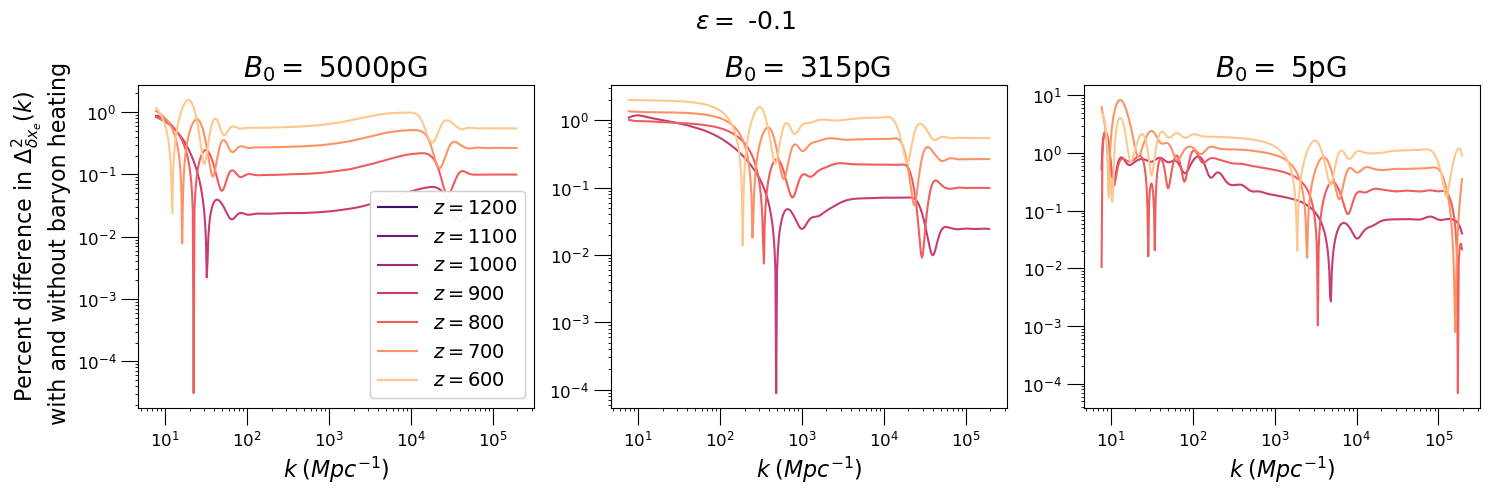}

    \caption{Percent difference in the dimensionless power for (top) $\delta_b^{(1)}$ and (bottom) $\delta x_e^{(1)}$ when baryon heating is and is not taken into account}
    \label{fig:PS_deltab_xe}
\end{figure*}

% \begin{figure*}[!t]
%     \centering

%     \includegraphics[width=1\linewidth]{Figures/PS_xe.png}

%     \caption{Percent difference in the dimensionless power per log wavenumber for $\delta x_e^{(1)}$ when baryon heating is and is not taken into account}
%     \label{fig:PS_xe}
% \end{figure*}

Since we expect $T_b = T_\gamma$ at early redshifts, we only introduce the temperature perturbations into the system at a redshift of $z = 1900$. 
We initialize with $T_b = T_\gamma$ and all LMHD variables' transfer functions at their respective values at $z = 1900$. 
In Fig. \ref{fig:PS_deltab_xe} we show the percent shift in the power spectra for $\delta^{(1)}_b$ and $\delta x_e^{(1)}$ when baryon heating is and is not taken into account. 
For the most part, the change due to baryon heating is below percent level, with the exception of low $k$ values for $\delta_b^{(1)}$. 
The large shift for low $k$ in $\delta_b^{(1)}$'s power spectrum at redshifts $z \sim 1000-1200$ can also be neglected since the absolute value of the power spectrum at these wavenumbers is exceedingly small, as can be seen in Fig. \ref{fig:deltam_power}.

\begin{figure*}[!t]
    \centering

    \includegraphics[width=1\linewidth]{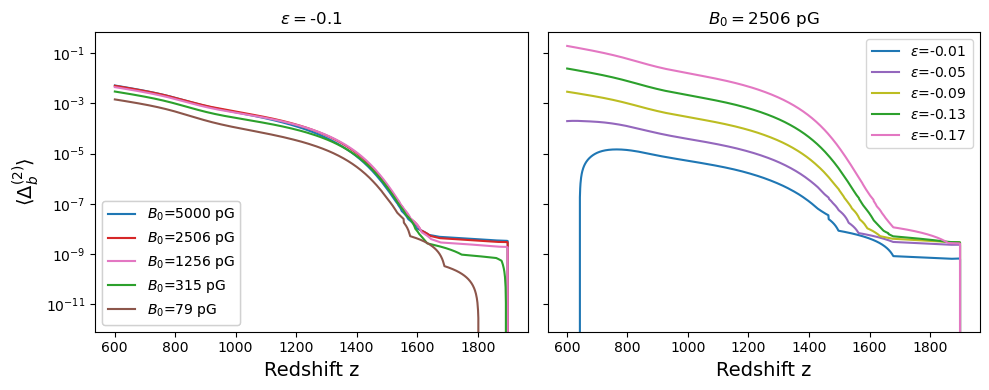}

    \caption{Lowest order shift to the background baryon temperature $\Delta_b^{(2)} = \delta T_b^{(2)}/\overline{T}_b $, whose evolution equation is given in Eq. \eqref{eq:Tb_evo}.}
    \label{fig:dTb}
\end{figure*}

\begin{figure*}[!t]
    \centering

    \includegraphics[width=1\linewidth]{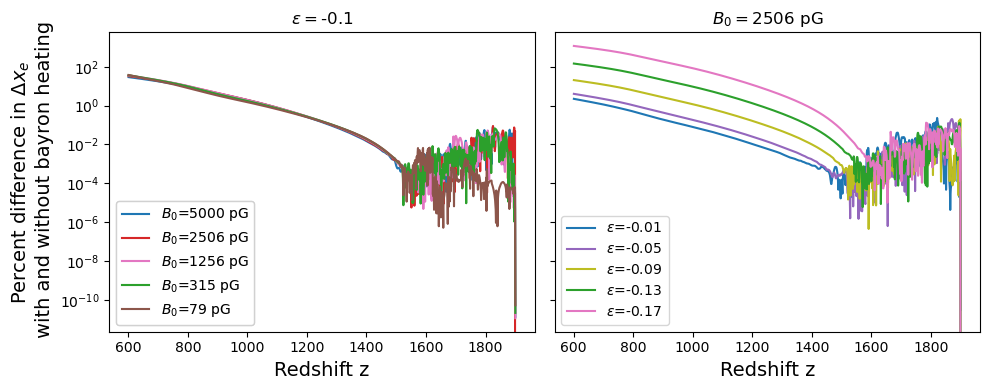}

    \caption{Percent difference in the lowest order shift to the background ionization fraction $\Delta x_e$ as defined in Eq. \eqref{eq:rec_second_O} when baryon heating is and is not taken into account.}
    \label{fig:Dx_e_perc_diff}
\end{figure*}

In Fig. \ref{fig:dTb} we present the lowest order shift to the background baryon temperature, which is given by $\langle \delta T_b^{(2)}/\overline{T}_b \rangle \equiv \langle \Delta_b^{(2)} \rangle$. 
We see that at later redshifts and for redder spectra, this deviation can start to grow to $\sim 10 \%$. 
Finally, in Fig. \ref{fig:Dx_e_perc_diff}, we present the shift to $\Delta x_e$ when baryon heating is and is not taken into account. 
We see that at later redshifts, there can be an $\mathcal{O}(1)$ shift for certain spectra, in agreement with expectations that post-recombination baryon heating delays recombination and increases the freezeout value of $x_e$.
Ultimately, the simplified treatment of linearized baryon heating presented here is inadequate. 
Further work is required to more carefully handle spectral distortions in the pre-recombination Universe and baryon heating from MHD dissipation in the post-recombination Universe. 
% \js{Do I need concluding sentence here?}

% We conclude our baryon heating discussion by once more emphasizing that the simplified linearized treatment of baryon heating presented here needs to be revisited and improved in future studies.
% In particular, nonlinear simulations which can account for the turbulent evolution in the post-recombination epoch and more accurately treat baryon heating are needed. 
% \eq{
% \alpha_B = 10^{-13} \frac{4.309(\frac{T_b}{10^4 K})^{-0.6166}}{1 + 0.6703(\frac{T_b}{10^4})^{0.53}},
% }

\section{Boltzmann equation further details}
\label{app:BE}

\subsection{Photon equations of motion}
\label{app:photon_eom}

In this appendix, we solve for the equations of motion $d\nu/dt$, $dx^i/dt$, $d\hat{n}^i/dt$. From Eqs.~\eqref{eq:BE_(1)} and \eqref{eq:BE_(2)}, we see that while we require $d\nu/dt$ to second order, we only require $dx^i/dt$ and $d\hat{n}^i/dt$ to first order. Defining our scalar fluctuations to be given by $\varphi = \varphi^{(1)} + \varphi^{(2)}$, with $\psi = -\varphi$, we can use Eq.~\eqref{eq:pert_FLRW} to write our perturbed metric to second order
\eq{
ds^2 &= -(1+2\varphi^{(1)} + 2(\varphi^{(1)})^2 + 2\varphi^{(2)}) dt^2 + a^2 (1+2\varphi^{(1)} + 2(\varphi^{(1)})^2 + 2\varphi^{(2)})\delta_{ij}dx^idx^j.
}    

We denote the photon's four-momentum by $P^\mu = dx^\mu/d\lambda$, with $\lambda$ an affine parameter. For the three momentum, we can write $P^i = C\hat{n}^i$. Using the plane wave expansion for photons, $exp[i(-\nu_N t + P_i x^i)/2\pi ] = exp[i P_\mu x^\mu/2\pi]$, we can write the four-momentum as $P^\mu = (-g^{00}\nu_N,C\hat{n}^i)$, wher $\nu_N$ is the frequency of a photon in the Newtonian or comoving frame. Since $P^\mu$ is a null vector, we have
\eq{
P^0 &= (1-2\varphi^{(1)}-2\varphi^{(2)})\nu_N\\
P^i &= \nu_N\frac{\hat{n}^i}{a}.
}
For clarity, we restore factors of $c$ in the equations of motion. Let us start with $dx^i/dt = P^i/P^0$ so that we have
\eq{\label{eq:advec_eom}
\bigg(\frac{dx^i}{dt} \bigg)^{(0)} &= c\frac{\hat{n}^i}{a}\\
\bigg(\frac{dx^i}{dt} \bigg)^{(1)} &= 2c\frac{\hat{n}^i}{a}\varphi^{(1)}.\\
}
To solve for $d\nu/dt$ and $d\hat{n}^i/dt$, we need to use the geodesic equation
\eq{
\frac{dP^\mu}{d\lambda} = -\Gamma^\mu_{\rho\sigma}P^\rho P^\sigma.
}
For $d\hat{n}^i/dt$, we can use the $\mu = i$ geodesic equation to find
\eq{\label{eq:lensing_eom}
\bigg(\frac{d \hat{n}^i}{dt} \bigg)^{(0)} &= 0\\
\bigg(\frac{d \hat{n}^i}{dt} \bigg)^{(1)} &= -\frac{2}{ac}(\delta^{ij} - \hat{n}^i\hat{n}^j)\d_j\varphi^{(1)}.
}
For $d\nu_N/dt$ we can use the $\mu = 0$ geodesic equation to find
\eq{\label{eq:redshift_eom}
\bigg(\frac{1}{\nu_N}\frac{d \nu_N}{dt} \bigg)^{(0)} &= -H\\
\bigg(\frac{1}{\nu_N}\frac{d \nu_N}{dt} \bigg)^{(m)} &= 2\dot{\varphi}^{(m)} \text{ for m = 1,2}.
}

Since we work in the matter rest frame, we have to perform a Lorentz boost using Eq.~\eqref{eq:lor_boost} to relate the frequency in the Newtonian frame to the frequency in the matter's rest frame. Arranging order by order in our perturbations, we find
\eq{\label{eq:nu_EOM}
\bigg( \frac{1}{\nu}\frac{d\nu}{dt} \bigg)^{(0)} &= \bigg( \frac{1}{\nu_N}\frac{d\nu_N}{dt} \bigg)^{(0)}\\
\bigg( \frac{1}{\nu}\frac{d\nu}{dt} \bigg)^{(1)} &= \bigg( \frac{1}{\nu_N}\frac{d\nu_N}{dt} \bigg)^{(1)} - \frac{\hat{n}^i}{c}\dot{v}_i^{(1)} - \frac{\hat{n}^i \hat{n}^j}{a} \frac{\d v_i^{(1)}}{\d x^j}\\
\bigg( \frac{1}{\nu}\frac{d\nu}{dt} \bigg)^{(2)} &= \bigg( \frac{1}{\nu_N}\frac{d\nu_N}{dt} \bigg)^{(2)} - \frac{\hat{n}^i}{c}\dot{v}^{(2)}_i  - \frac{\hat{n}^i \hat{n}^j}{a} \frac{\d v^{(2)}_i}{\d x_j} + \frac{v_i^{(1)}}{ca} (\delta^{ij} - \hat{n}^i\hat{n}^j)\d_j(\Psi^{(1)} -\Phi^{(1)})\\
&+ \frac{\hat{n}^i\hat{n}^j v^{(1)}_i }{c^2}\dot{v}^{(1)}_j + \frac{\hat{n}^i\hat{n}^j\hat{{n}}^k v^{(1)}_i }{ca} \frac{\d v^{(1)}_j}{\d x^k} + \frac{v_i^{(1)}}{c^2} (\dot{v}^i)^{(1)} + \frac{\hat{n}^j v_i^{(1)}}{ca}  \frac{\d (v^i)^{(1)}}{\d x^j}.
}
We ultimately drop all time derivatives of the metric potentials and velocity fields in the final computation since these are subdominant to the other terms during the matter domination era. 

For the time derivative of the metric potential, we can directly show that it is subdominant to $(\hat{n}^i \hat{n}^j/a)(\d v_i/\d x_j)$. Since the time derivative of the metric potential only has a monopole contribution, we can compare it to the monopole contribution of the gradient of the velocity term, which in Fourier space is given by
\eq{
\int \frac{d^2\hat{\bd{n}}}{4\pi}\frac{\hat{n}^i \hat{n}^j}{a} ik_j v_i^{(1)} &= \frac{\Theta}{3a}.
}
During matter domination, the derivative of the metric potential is given by
\eq{
\dot{\varphi} &= 4\pi G\rho_b \frac{a^2}{k^2} \bigg[\frac{\Theta_b}{a} + H\delta_b \bigg] = \frac{3}{2}\frac{H_0^2 \Omega_{b,0}}{ak^2}\bigg[\frac{\Theta_b}{a} + H\delta_b \bigg],
}
where we use the Friedmann equation during matter domination ($H^2 = 8\pi G\rho_m/3 = H^2_0 \Omega_{m,0}a^{-3}$ ) in the last equality and introduce $\Omega_{i,0}$ defined as the ratio of the present day energy density of some element $i$ with respect to the present day critical energy density: $\rho_i/\rho_{cr}$. From the continuity equation, we have $\dot{\delta}_b = -\Theta/a$ and we have $\dot{\delta}_b \sim H\delta_b$. We therefore focus on the $\Theta/a$ term and compare it to the monopole of the velocity gradient term. We see then that the time derivative of the potential can be discarded for all modes which satisfy $(9\Omega_{b,0}/2\Omega_{m,0})H^2(a^2/k^2) \ll 1$. Computing this ratio, we find
\eq{
\frac{9}{2}H^2 \frac{\Omega_{b,0}}{\Omega_{m,0}}\frac{a^2}{k^2} \approx \bigg(\frac{\lambda}{350 \text{ Mpc}}\bigg)^2 \bigg(\frac{1+z}{1100}\bigg).
}
We can therefore safely ignore derivatives of the gravitational potential with respect to $\Theta/a$ for the modes we consider.

For the time derivative of the velocity field, so long as $\dot{\bd{v}} \sim H \bd{v}$, it too can be shown to be subdominant to $(\hat{n}^i \hat{n}^j/a)(\d v_i/\d x_j)$. Once PMFs are introduced, the Lorentz force in the Euler equation opens up the possibility that the velocity field can evolve on faster than cosmological time scales, particularly when a mode transitions to the FSR and rapidly reaches the terminal velocity state. We therefore explicitly check for a few representative examples of the spectra that we consider that the inclusion of the $\dot{\bd{v}}$ terms in either the first- or second-order BE does not significantly alter the results. 

\subsection{The Sobolev solution for the homogeneous Boltzmann equation}
\label{app:sob_sol}

The Sobolev approximation provides a solution to the BE for photons traveling through an expanding medium. 
It is a simplified treatment which neglects resonant scattering and assumes the only collisional processes are absorption and emission as the photons redshift in the expanding background. 
In the context of cosmological recombination in a homogeneous universe, the local expansion is set by the Hubble expansion. 
% \tv{Do we need to say this here?} In an inhomogeneous Universe, the local expansion receives an additional contribution from the local velocity divergence, as discussed in Section \ref{sec:lin_pert_BE}. 

Under these assumptions, the equation of radiative transfer can be put into the form
\eq{\label{eq:sob_BE}
\frac{\d f}{\d \nu} = \tau_s\phi(\nu)[f(\nu) - \feq],
}
where $\tau_s$ is the Sobolev optical depth given by \cite{rybicki1993asp, Seager_2000}
\eq{\label{eq:sob_opt_depth}
\tau_s = \frac{3 A_{\text{Ly}\alpha} \lambda_{\text{Ly}\alpha}^3}{8\pi H}n_{1s}.
}
We can recover Eq.~\eqref{eq:sob_BE} from the more sophisticated treatment including resonant scattering in Sec.~\ref{sec:boltzmannlya} by specializing to a homogenous Universe, and setting $p_{sc} = 0$ or equivalently $p(\nu,\nu') = \phi(\nu)$ (i.e. assuming complete redistribution of the scattered photons) in the collisional term $C[f]$ of Eq.~\eqref{eq: coll_Term}.

%approximation can be for the homogeneous Lyman-$\alpha$ BE can be recovered from Eqs.~\eqref{eq:BE_full} and \eqref{eq: coll_Term} by setting $p_{sc} = 0$ or equivalently $p(\nu,\nu') = \phi(\nu)$ (i.e. assuming complete redistribution of the scattered photons) in the collisional term $C[f]$, yielding

We can integrate Eq.~\eqref{eq:sob_BE} and impose the boundary condition that on the far blue side of the line the PSD goes to a blackbody, $f(\nu_B) = f_{bb} = e^{-h\nu_{\text{Ly}\alpha}/k_B T}$, to find an analytical expression for the PSD under the Sobolev approximation
\eq{\label{eq:Sob_PSD}
f(\nu) &= \feq + [f_{bb} - \feq ]e^{-\chi}.
}
where $\chi = \tau_s\int_\nu^\infty d\nu' \phi(\nu')$.
To find the jump across the line in the Sobolev approximation, we can integrate the PSD over the line profile to find
\eq{\label{eq:Sob_jump}
\feq - \overline{f} &= \frac{1-e^{-\tau_s}}{\tau_s} [\feq - f_{bb}] .
}
Prior to recombination the Sobolev optical depth is very large so that we can approximate $1 - e^{-\tau_s} \approx 1$. Substituting Eq.~\eqref{eq:Sob_jump} into Eq.~\eqref{eq:rec_x2p1s_dot} and using the steady-state approximation to set $\dot{x}_2 \approx 0$ results in Eq.~\eqref{eq:rec_TLA_zero_O}. This procedure allows us to find an explicit expression for the Peebles $C$-factor, which is given by Eq. \eqref{eq:peeb_C_hom}.
% \eq{
% C &= \frac{3A_{\text{Ly}\alpha}/\tau_s + \Lambda_{2s1s}}{4\beta_B + 3A_{\text{Ly}\alpha}/\tau_s + \Lambda_{2s1s}}.
% }

\begin{figure*}[!t]
    \includegraphics[width=1\textwidth]{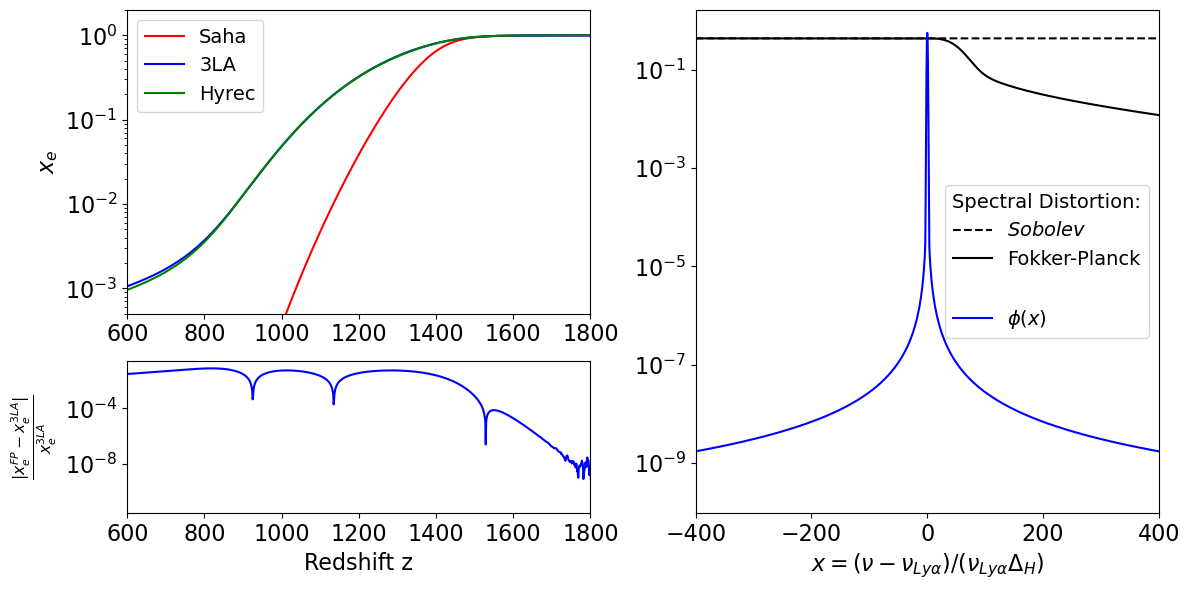} 
    \captionsetup{width=1\linewidth}
    \caption{(Top Left) Ionization fraction computed using the Saha and 3LA models (Eqs.~\eqref{eq:saha_rec_homo} and \eqref{eq:rec_TLA_zero_O}) compared to $x_e$ from HYREC \cite{PhysRevD.102.083517,PhysRevD.83.043513}. (Bottom Left) Difference in ionization fraction when the Sobolev and Fokker-Planck approximations are used for the BE (Eqs.~\eqref{eq:sob_BE}\eqref{eq: Hom_BE}). (Right) Spectral distortion with respect to the CMB blackbody distribution defined as $\frac{8\pi\nu^3}{c^3 n_H}(f(\nu) - f_{bb})$ for each approximation and the Voigt profile at a redshift of $z = 1100$. }
    \label{fig:hom_xe_PSD}
\end{figure*}

The difference in the PSD and the homogeneous recombination rate when scattering is treated with the Sobolev and Fokker-Planck approximations are shown in Fig.~\ref{fig:hom_xe_PSD}. 
Although the recombination rates are nearly identical in each approximation, the PSD across the line, and particularly in the blue wing differs substantially. 
If the spectral distortion in the wings, where the Voigt profile has the least support, is inaccurate, this will not significantly impact the computation of $\overline{f}$ and leads to very small errors in $x_e$. 
We return to this point in Appendix~\ref{app:LBE} in which we justify why a linearization of the BE can still accurately capture the recombination rate, even if it fails to accurately capture the PSD across the entire line.

\subsection{Linearized Boltzmann equation - shortcomings and validity}
\label{app:LBE}

In this appendix, we consider the circumstances that allow for a reasonable linearization of the BE.
We have assumed thus far that if the fluid perturbations are small enough such that $\delta \tau_s \ll \tau_s$, then a linearization and perturbative expansion can be employed to solve the BE. 
However, since $\tau_s \sim 10^8$ at $z = 1100$, even if we only source perturbations on the order of $0.01$ as above, we will still have $\delta \tau_s \gg 1$. 
We may then be concerned that the exponentiation of the optical depth in the Sobolev solution in Eq.~\eqref{eq:Sob_PSD} will cause significant disagreement between the true result and the linearly approximated result. 

To scrutinize this claim more carefully, let us consider the simplest case in which we operate under the Sobolev approximation and simply introduce a uniform shift to the Sobolev optical depth $ \delta \tau_s$. In this case, we can find the PSD and jump across the line by making the substitution of $\tau_s \to \tau_s + \delta \tau_s$ in Eqs.~\eqref{eq:Sob_PSD} and \eqref{eq:Sob_jump}. The BE in the Sobolev approximation in the case of a uniform shift to $\tau_s$ is given by
\eq{
\frac{\d f}{\d \nu} &= \tau_s\bigg(1 + \frac{\delta \tau_s}{\tau_s} \bigg)\phi(\nu)[f(\nu) - \feq].
}
If we define $F = f - \feq$, then we can rewrite this equation as
\eq{
\frac{d F}{d \chi} = -\bigg(1 + \frac{\delta \tau_s}{\tau_s} \bigg) F.
}
This has an exact solution $F_{\text{exact}} = A \exp[-(1+ \delta\tau_s/\tau_s)\chi]$, as expected. By imposing the boundary condition that the PSD go to a blackbody on the far blue side of the line we can solve for the integration constant to obtain
\eq{
F_{\text{exact}} = [f_{bb} - \feq]e^{-\chi}e^{-\delta \tau_s\chi/\tau_s}.
}
If we perform a perturbative treatment and assume the PSD to be given by $F + \delta F$, then the zeroth- and first-order equations are
\eq{
\frac{dF}{d\chi} &= -F\\
\frac{d\delta F}{d\chi} &= -\delta F - \frac{\delta \tau_s}{\tau_s} F
}
The zeroth-order solution is then $F = Aexp[-\chi]$. Substituting this into the first-order equation, we can find the solution for $\delta F$ to be
\eq{
\delta F &= A\chi \frac{\delta \tau_s}{\tau_s} e^{-\chi} + Be^{-\chi}.
}
If we again impose the boundary condition that $f$ go to a blackbody and $\delta f$ vanishes on the blue side of the line, we have
\eq{
F + \delta F = [f_{bb} - \feq]e^{-\chi}\bigg( 1 - \chi \frac{\delta \tau_s}{\tau_s} \bigg),
}
which we could have equivalently found by Taylor expanding the exact solution to first order. The error between the exact solution and the linear result is
\eq{
F_{\text{exact}} - (F + \delta F) = \sum_{n=2}^\infty (-1)^n\bigg(\chi\frac{\delta \tau_s}{\tau_s}\bigg)^n\frac{1}{n!}.
}
At linear order, our result will only be valid on the part of the domain which satisfies $\chi \delta \tau_s/\tau_s \ll 1 \implies \int_\nu^\infty d\nu'\phi(\nu') \ll (\delta \tau_s )^{-1}$, which for large $\delta \tau_s$ would only occur on the blue side of the line. We can therefore have circumstances in which we satisfy $1 \ll \delta \tau_s \ll \tau_s$, but nevertheless linear perturbation theory will perform poorly for frequencies that satisfy $(\delta\tau_s)^{-1} \lesssim \int_\nu^\infty d\nu' \phi(\nu')$. However, to compute the recombination history, we do not need to accurately capture the PSD across the entire line; we only require the jump across the line $\overline{F} = \overline{f} - \feq$. Let us now compute the jump across the line in both the exact result and in the linearly perturbed result to see whether linear perturbations accurately capture the recombination rate.

Integrating across the line, we find
\eq{
\overline{F}_{exact} &= [f_{bb} - \feq]\frac{1 - e^{-\tau_s - \delta\tau_s}}{\tau_s + \delta \tau_s} \approx [f_{bb} - \feq]\frac{1 - e^{-\tau_s}e^{- \delta\tau_s}}{\tau_s }\bigg(1 - \frac{\delta \tau_s}{\tau_s}\bigg)\\
\overline{F} + \delta \overline{F} &= [f_{bb} - \feq]\frac{1 - e^{-\tau_s}}{\tau_s } + [f_{bb} - \feq]\frac{\delta \tau_s}{\tau_s^2}\bigg[e^{-\tau_s}(1+\tau_s) - 1  \bigg]
}
Taking the difference, we find
\eq{
\overline{F}_{exact} - (\overline{F} + \delta \overline{F}) &\propto \frac{e^{-\tau_s}}{\tau_s}\bigg[ 1 - e^{-\delta \tau_s}  - \frac{\delta \tau_s}{\tau_s}(1 + \tau_s^{-1} - e^{-\delta \tau_s}) \bigg]\\
&\approx \frac{e^{-\tau_s}}{\tau_s}\bigg[ 1 - \frac{\delta \tau_s}{\tau_s} \bigg] \approx \frac{e^{-\tau_s}}{\tau_s},\\
}
where we have assumed $1 \ll \delta \tau_s \ll \tau_s$. We see that the error in the jump across the line between the exact and linearly perturbed result is exponentially suppressed by $\tau_s \gg 1$. This suggests that even though linear perturbation theory inaccurately captures the PSD across the line, it does very well in capturing the rate of recombination. 

If we now allow for $\delta\tau_s(t, \bd{x})$ as is the case for perturbations sourced by PMFs, we can no longer provide a rigorous justification for linearizing the BE. 
However, the homogeneous case in which we artificially and uniformly shifted $\tau_s \to \tau_s +\delta \tau_s$ and still accurately computed the rate of recombination using a linearized analysis, suggests that so long as locally we satisfy $\delta \tau_s(t,\bd{x}) \ll \tau_s$, then perturbation theory will perform well in capturing the recombination history, even if it inaccurately predicts the PSD across the Lyman-$\alpha$ line.

\subsection{Second-order Boltzmann equation source functions}
\label{app:BE_(2)}
In this appendix, we derive all the source terms $\mathcal{S}(\nu)$ for the second-order BE, Eq.~\eqref{eq:BE_(2)}. We will utilize the fact that $\langle \delta x_{1s}^{(1)} \rangle = 0$ and that to all orders in perturbation theory, $\langle \delta_b^{(n)} \rangle = \langle v_i^{(n)} \rangle = 0$. For each term we take a spatial average and integrate over $d^2\hat{\bd{n}}/4\pi$ to project out the monopole.
Throughout, we suppress the explicit dependence of transfer functions and first-order basis solutions on wavenumber and frequency to improve readability. 

Let us start with the second-order collisional term $C^{(2)}[f]$. From the definition of the Sobolev optical depth in Eq.~\eqref{eq:sob_opt_depth}, we can define the second-order perturbed Sobolev optical depth
\eq{
\frac{\delta \tau_s^{(2)}}{\tau_s} &= \delta_b^{(2)} + \frac{\delta x_{1s}^{(2)}}{x_{1s}} + \frac{ \delta_b^{(1)}\delta x_{1s}^{(1)}}{x_{1s}} \\
&= \delta_b^{(2)} + \frac{\delta x_{1s}^{(2)}}{x_{1s}} - \frac{ \delta_b^{(1)}\delta x_{e}^{(1)}}{x_{1s}}.
}

Perturbing Eq.~\eqref{eq: coll_Term} to second order and Fourier transforming, we find the second-order collisional term
\eq{
-\frac{1}{H\nu}C^{(2)}[f] &= \frac{\delta \tau_s^{(2)}}{\tau_s}\frac{\d f}{\d \nu} + \frac{\delta \tau_s^{(1)}}{\tau_s}\bigg\{ \tau_{s}p_{sc}\bigg[\phi(\nu) \delta f^{(1)}(\nu, \hat{\bd{n}}) - \int d\nu' \frac{d^2\hat{\bd{n}}'}{4\pi}\phi(\nu')p(\nu',\nu) \delta f^{(1)} (\nu',\hat{\bd{n}}') \bigg]\\
&+ \tau_s\phi(\nu) \bigg[ p_{ab} \delta f^{(1)} (\nu,\hat{\bd{n}}) - \delta \feq^{(1)} + p_{sc}\delta \overline{f}^{(1)}_{00}  \bigg] \bigg\} + \tau_s\phi(\nu) \bigg[ p_{ab} \delta f^{(2)} (\nu,\hat{\bd{n}}) - \delta \feq^{(2)} + p_{sc}\delta \overline{f}^{(2)}_{00}  \bigg] \\
&+ \tau_{s}p_{sc}\bigg[\phi(\nu) \delta f^{(2)}(\nu, \hat{\bd{n}}) - \int d\nu' \frac{d^2\hat{\bd{n}}'}{4\pi}\phi(\nu')p(\nu',\nu) \delta f^{(2)}(\nu',\hat{\bd{n}}') \bigg].
}    

We can now take a spatial average and project out the monopole to find
\eq{
-\frac{1}{H\nu} \int \frac{d^2\hat{\bd{n}}}{4\pi} \langle C^{(2)}[f] \rangle &= \frac{\langle \delta x_{1s}^{(2)} \rangle - \langle \delta_b^{(1)}\delta x_{e}^{(1)}\rangle}{x_{1s}}\frac{\d f}{\d \nu} + \tau_s\phi(\nu)p_{ab}\langle \delta f^{(2)}_{00} \rangle - \tau_s\phi(\nu) [ \langle \delta \feq^{(2)}\rangle - p_{sc}\langle \delta \overline{f}^{(2)}_{00}\rangle ] \\
&+\frac{1}{(2\pi)^6} \int d^3\bd{k}_1 d^3\bd{k}_2 \bigg\langle \frac{\delta \tau_s^{(1)}(\bd{k}_1)}{\tau_s} \bigg\{ \tau_s p_{sc} \bigg[ \phi(\nu) \delta f^{(1)}_{00}(\bd{k}_2,\nu) - \int d\nu' \phi(\nu') p(\nu',\nu) \delta f^{(1)}_{00}(\bd{k}_2,\nu') \bigg]\\
&+ \tau_s\phi(\nu)p_{ab}\delta f^{(1)}_{00}(\bd{k}_2,\nu) - \tau_s\phi(\nu)[\delta \feq^{(1)}(\bd{k}_2) - p_{sc}\delta \overline{f}^{(1)}_{00}(\bd{k}_2)] \bigg\} \bigg\rangle e^{i(\bd{k}_1 + \bd{k}_2)\bd{x}} \\
&+ \tau_s p_{sc} \bigg\langle \bigg[ \phi(\nu) \delta f^{(2)}_{00} - \int d\nu' \phi(\nu') p(\nu',\nu) \delta f^{(2)}_{00}(\nu') \bigg] \bigg\rangle.
}
We again utilize the Fokker-Planck approximation to rewrite this as
\eq{
-\frac{1}{H\nu} \int \frac{d^2\hat{\bd{n}}}{4\pi} \langle C^{(2)}[f] \rangle &= \frac{1}{x_{1s}}[ \langle \delta x_{1s}^{(2)} \rangle - \langle \delta_b^{(1)}\delta x_{e}^{(1)}\rangle ]\frac{\d f}{\d \nu} + \tau_s\phi(\nu)p_{ab}\langle \delta f^{(2)}_{00} \rangle - \tau_s\phi(\nu) [ \langle \delta \feq^{(2)}\rangle - p_{sc}\langle \delta \overline{f}^{(2)}_{00}\rangle ] \\
&+ \frac{1}{(2\pi)^6} \int d^3\bd{k}_1 d^3\bd{k}_2 \bigg\langle  \bigg(\delta_b^{(1)}(\bd{k}_1) + \frac{\delta x_{1s}^{(1)}(\bd{k}_1)}{x_{1s}} \bigg) \bigg\{ -\tau_{s}p_{sc} \frac{\nu_{\text{Ly}\alpha}^2 \Delta_H^2}{2} \frac{\d}{\d \nu} \bigg[ \phi(\nu) \frac{\d \delta f^{(1)}_{00}(\bd{k}_2)}{\d \nu} \bigg]\\
&+ \tau_{s}p_{ab}\phi(\nu)\delta f^{(1)}_{00}(\bd{k}_2,\nu) - \tau_s\phi(\nu)[\delta \feq^{(1)}(\bd{k}_2) - p_{sc}\delta \overline{f}^{(1)}_{00}(\bd{k}_2)] \bigg\} \bigg\rangle e^{i(\bd{k}_1 + \bd{k}_2)\bd{x}} \\
&- \tau_s p_{sc} \frac{\nu_{\text{Ly}\alpha}^2 \Delta_H^2}{2} \frac{\d}{\d \nu} \bigg\langle \bigg[ \phi(\nu) \frac{\d \delta f^{(2)}_{00}}{\d \nu} \bigg] \bigg\rangle.\\
}

To simplify the final expression, let us define
\eq{\label{eq:alpha_bar_def}
\delta \overline{f}_{\text{eq}} - p_{sc}\delta f_{00} &= \overline{\alpha}\delta_b + \overline{\beta}\frac{\delta x_{1s}}{x_{1s}} + \overline{\gamma}\Theta,
}
where $\overline{\alpha},\overline{\beta}, \overline{\gamma}$ can be worked out from Eq.~\eqref{eq:scalar_basis_solns} and from setting $\delta \dot{x}^{(1)}_2 = 0$. The second-order, spatially-averaged monopole of the collisional term can then be written as
\eq{\label{eq:coll_(2)}
-\frac{1}{H\nu} \int \frac{d^2\hat{\bd{n}}}{4\pi} \langle C^{(2)}[f] \rangle  &= \frac{\langle \delta x_{1s}^{(2)} \rangle - \langle \delta_b^{(1)}\delta x_{e}^{(1)}\rangle}{x_{1s}} \frac{\d f}{\d \nu} + \tau_s p_{ab}\phi(\nu)\langle \delta f^{(2)}_{00} \rangle - \tau_s\phi(\nu) [ \langle \delta \feq^{(2)}\rangle - p_{sc}\langle \delta \overline{f}^{(2)}_{00}\rangle ]\\
&- \tau_{s}p_{sc}\frac{\nu_{\text{Ly}\alpha}^2\Delta_H^2}{2} \frac{\d}{\d \nu}\bigg\langle \bigg[ \phi(\nu) \frac{\d \delta f^{(2)}_{00}}{\d \nu} \bigg] \bigg\rangle \\
&+ \int d \ln k \frac{\Delta_B^2(k)}{2B_0^2} \bigg\{ \bigg[ \overline{T}_{\delta_b \delta_b} + \frac{1}{x_{1s}^2}\overline{T}_{\delta \overline{x}_e \delta \overline{x}_e} - \frac{2}{x_{1s}}\overline{T}_{\delta_b \delta \overline{x}_e} \bigg]\\
&\times \bigg[\tau_{s}p_{ab}\phi(\nu) \mathcal{A}_0  -  \tau_{s}p_{sc}\frac{\nu_{\text{Ly}\alpha}^2\Delta_H^2}{2}\frac{\d}{\d\nu} \bigg(\phi(\nu) \frac{\d \mathcal{A}_{0}}{\d\nu} \bigg) \bigg] \\
&+  \frac{1}{aH} \bigg[ \overline{T}_{\delta_b\Theta} - \frac{1}{x_{1s}}\overline{T}_{\delta \overline{x}_e \Theta} \bigg]\bigg[\tau_{s}p_{ab}\phi(\nu) \mathcal{B}_0  -  \tau_{s}p_{sc}\frac{\nu_{\text{Ly}\alpha}^2\Delta_H^2}{2}\frac{\d}{\d\nu} \bigg( \phi(\nu) \frac{\d \mathcal{B}_{0}}{\d\nu} \bigg) \bigg] \\
&+\bigg[ \overline{\alpha} \overline{T}_{\delta_b\delta_b} - \frac{\overline{\alpha} + \overline{\beta}}{x_{1s}}\overline{T}_{\delta_b \delta \overline{x}_e} + \frac{\overline{\beta}}{x_{1s}^2}\overline{T}_{\delta \overline{x}_e \delta \overline{x}_e} + \frac{\overline{\gamma}}{aH} \overline{T}_{\delta_b \Theta} - \frac{\overline{\gamma}}{aH x_{1s}}\overline{T}_{\delta \overline{x}_e \Theta}  \bigg]\\
&\times \bigg[\tau_{s}p_{ab}\phi(\nu) \mathcal{C}_0  -  \tau_{s}p_{sc}\frac{\nu_{\text{Ly}\alpha}^2\Delta_H^2}{2}\frac{\d}{\d\nu}\bigg( \phi(\nu) \frac{\d \mathcal{C}_{0}}{\d\nu}  \bigg) -\tau_s\phi(\nu) \bigg] \bigg\} .
}
The second-order contribution to the redshift term is given by
\eq{
\bigg(\frac{d\nu}{dt}\frac{\d f}{\d \nu}\bigg)^{(2)} &= \bigg(\frac{d\nu}{dt}\bigg)^{(0)}\frac{\d \delta f^{(2)}}{\d \nu} + \bigg(\frac{d\nu}{dt}\bigg)^{(1)}\frac{\d \delta f^{(1)}}{\d \nu} + \bigg(\frac{d\nu}{dt}\bigg)^{(2)}\frac{\d f}{\d \nu}. 
}
For the first term, we have
\eq{\label{eq:red_0_2}
-\frac{1}{H\nu}\int \frac{d \hat{\bd{n}}}{4\pi} \bigg\langle \bigg(\frac{d\nu}{dt}\bigg)^{(0)} \frac{\d \delta f^{(2)}}{\d \nu} \bigg\rangle &= \frac{\d \langle \delta f^{(2) }_{0 0}\rangle}{\d \nu}.
}
The second term is given in Eq.~\eqref{eq:red_1_1}. Once we drop derivatives of the metric potential and of the velocity field, we can write the third term as
\eq{
\bigg( \frac{1}{\nu}\frac{d\nu}{dt} \bigg)^{(2)} &= - \frac{\hat{n}^i}{c} \bigg( \frac{d x^j}{dt} \bigg)^{(0)} \frac{\d v^{(2)}_i}{\d x_j} - \frac{v_i^{(1)}}{c} \bigg(\frac{d \hat{n}^i}{d t} \bigg)^{(1)} + \frac{\hat{n}^i\hat{n}^j v^{(1)}_i }{c^2} \bigg( \frac{dx^k}{dt} \bigg)^{(0)} \frac{\d v^{(1)}_j}{\d x^k} + \frac{v_i^{(1)}}{c^2} \bigg( \frac{dx^j}{dt} \bigg)^{(0)} \frac{\d (v^i)^{(1)}}{\d x^j}.\\
}
Only the second term has a monopole. We therefore find
\eq{\label{eq:red_20}
\frac{1}{H\nu}\bigg\langle \int \frac{d^2\hat{\bd{n}}}{4\pi} \bigg(\frac{d\nu}{dt}\bigg)^{(2)} \bigg\rangle \frac{\d f}{\d\nu} &= \frac{ 16\pi G \rho_b a^2 }{3aHc^2} \int_0^\infty d \ln k \frac{\Delta^2_B(k)}{2k^2B_0^2} \overline{T}_{\delta_b \Theta}(k)\frac{\d f}{\d \nu}. \\
}

The second-order advection term is given by
\eq{
\bigg( \frac{d x^i}{dt} \frac{\d f}{dx}\bigg)^{(2)} &= \bigg(\frac{d x^i}{dt}\bigg)^{(0)} \frac{\d \delta f^{(2)}}{\d x^i} + \bigg(\frac{d x^i}{dt}\bigg)^{(1)} \frac{\d \delta f^{(1)}}{\d x^i}.
}
The first term vanishes upon spatial averaging due to the gradient. We then find
\eq{\label{eq:advec_(2)}
\frac{1}{H\nu}\int \frac{d^2\hat{\bd{n}}}{4\pi}\bigg\langle \bigg( \frac{d x^i}{dt} \frac{\d f}{dx}\bigg)^{(2)} \bigg\rangle
= \frac{8\pi G \rho a }{3cH\nu} \int  d \ln k \frac{\Delta_B^2(k)}{2kB_0^2} \bigg\{ &\overline{T}_{\delta_b\delta_b}[\mathcal{A}_1 + \overline{\alpha}\mathcal{C}_1] - \frac{1}{x_{1s}} \overline{T}_{\delta_b\delta \overline{x}_e} [\mathcal{A}_1 + \overline{\beta}\mathcal{C}_1]\\
&+ \frac{1}{aH}\overline{T}_{\delta_b \Theta} [\mathcal{B}_1 + \overline{\gamma}\mathcal{C}_1]\bigg\}.
}
The second-order lensing term is given by
\eq{
\bigg(\frac{dn^i}{dt}\frac{\d f}{\d n^i}\bigg)^{(2)}&= \bigg(\frac{dn^i}{dt}\bigg)^{(1)}\frac{\d \delta f^{(1)}}{\d n^i}.
}
Substituting in Eq.~\eqref{eq:lensing_eom}, we find
\eq{
 \bigg( \frac{dn^i}{dt}\bigg)^{(1)}\bigg( \frac{\d \delta f^{(1)}}{\d n^i}\bigg) &= i\frac{8\pi G \rho a}{c}\frac{1}{(2\pi)^6}\int d^3\bd{k}_1 d^3\bd{k}_2 \frac{1}{k_1} \delta_b(\bd{k}_1) a^i\frac{\d \delta f^{(1)}(\bd{k}_2)}{\d n^i}e^{i(\bd{k}_1 + \bd{k}_2)\bd{x}},
}
where we have introduced the vector
\eq{
\bd{a} &= \hat{\bd{z}} - \hat{\bd{n}}\cos\theta = -\sin\theta\hat{\theta}.
}
Since $\d \delta f^{(1)}/\d\hat{n}^i$, is tangent to the unit sphere, we can expand it in terms of vector spherical harmonics 
\eq{
\frac{\d \delta f^{(1)}}{\d \hat{\bd{n}}} &= \sum_{\ell, m} (-i)^{\ell} \delta f^{(1)}_{\ell m}  \sqrt{ \frac{4\pi}{2\ell + 1}\frac{(\ell + m)!}{(\ell - m)!} } \bm{\Psi}_{\ell m}.
}
Writing $\bd{a} = \sqrt{4\pi/3}\bm{\Psi}_{10}$ and using the orthogonality of the vector spherical harmonics, we find that the spatially-averaged monopole of the lensing term is given by
\eq{\label{eq:lensing_(2)}
\frac{1}{H\nu}\int \frac{d\hat{\bd{n}}}{4\pi}  \bigg \langle \bigg( \frac{dn^i}{dt}\bigg)^{(1)}\bigg( \frac{\d \delta f^{(1)}}{\d n^i}\bigg) \bigg\rangle = -\frac{16\pi G \rho a}{3 cH\nu} \int d \ln k \frac{\Delta_B^2(k)}{2kB_0^2} \bigg\{&\overline{T}_{\delta_b \delta_b}(k)[\mathcal{A}_1 + \overline{\alpha}\mathcal{C}_1] - \frac{1}{x_{1s}}\overline{T}_{\delta_b \delta \overline{x}_e}[\mathcal{A}_1 + \overline{\beta}\mathcal{C}_1]\\
&+ \frac{1}{aH}\overline{T}_{\delta_b \Theta}(k)[\mathcal{B}_1 + \overline{\gamma}\mathcal{C}_1] \bigg\}.\\
}

We can now find the $\mathcal{S}$ in Eq.~\eqref{eq:BE_(2)} from Eqs.~\eqref{eq:coll_(2)},\eqref{eq:red_0_2},\eqref{eq:red_1_1},\eqref{eq:red_20},\eqref{eq:advec_(2)}, and \eqref{eq:lensing_(2)}.

\subsection{The Boltzmann equation in Lyman-continuum}
\label{app:BE_cont}
For the continuum channel, the BE reduces to \cite{Venumadhav_Hirata_15}
\eq{\label{eq:BE_cont}
\frac{\hat{n}^i}{a}\frac{\d f}{\d x^i} &= -n_{1s}\sigma_a(\nu)f(\nu, \bd{x},\hat{\bd{n}}) + \frac{c^2}{8\pi \nu^2}n_e^2\alpha_{1s}(\nu)\phi(\nu),
}
where $\sigma_a(\nu)$ is the the continuum photon absorption cross-section, $\alpha_{1s}$ is the direct recombination coefficient, and $\phi(\nu)$ is the probability distribution for the emitted photons' frequency. We also define the integrated recombination coefficient $\alpha_{1s} = \int_{\nu_c}^\infty d\nu \alpha_{1s}(\nu)\phi(\nu)$ where $\nu_c$ is the photoionization frequency for hydrogen. Let us now perform a perturbative expansion of Eq.~\eqref{eq:BE_cont} up to second order
\eq{\label{eq:BE_cont_exp}
\frac{1}{a}\hat{\bd{n}}\cdot \nabla &[f + \delta f^{(1)} + \delta f^{(2)}] 
= - n_{H}x_{1s}\sigma_a(\nu)f^{(0)} \bigg[ 1 + \delta_b^{(1)} + \frac{\delta x_{1s}^{(1)}}{x_{1s}} + \delta_b^{(2)} + \frac{\delta x_{1s}^{(2)}}{x_{1s}} + \frac{\delta x_{1s}^{(1)}\delta_b^{(1)}}{x_{1s}} \bigg] \\
&- n_{H}x_{1s}\sigma_a(\nu)\delta f^{(1)} \bigg[ 1 + \delta_b^{(1)} + \frac{\delta x_{1s}^{(1)}}{x_{1s}} \bigg] - n_{H}x_{1s}\sigma_a(\nu)\delta f^{(2)} \\
&+ \frac{c^2}{8\pi\nu^2}n_H^2x_e^2\alpha_{1s}(\nu)\phi(\nu) \bigg( 1 + 2\delta_b^{(1)} + 2\frac{\delta x_{e}^{(1)}}{x_{e}} + 4\delta_b^{(1)}\frac{\delta x_{e}^{(1)}}{x_{e}} + \delta_b^{(1)}\delta_b^{(1)} + \frac{\delta x_{e}^{(1)}}{x_{e}}\frac{\delta x_{e}^{(1)}}{x_{e}}  + 2\delta_b^{(2)} + 2\frac{\delta x_{e}^{(2)}}{x_{e}} \bigg).\\
}
Since the zeroth-order PSD has no gradient, we find the expected detailed balance result for the homogeneous continuum BE
\eq{
0 = \frac{1}{a}\hat{\bd{n}}\cdot \nabla f &= - n_{H}x_{1s}\sigma_a(\nu)f + \frac{c^2}{8\pi\nu^2}n_H^2x_e^2\alpha_{1s}(\nu)\phi(\nu)\\
}
For the first-order perturbation, we have
\eq{
\frac{1}{a}\hat{\bd{n}}\cdot \nabla \delta f^{(1)} &= - n_{H}x_{1s}\sigma_a(\nu)f \bigg[ \delta_b^{(1)} + \frac{\delta x_{1s}^{(1)}}{x_{1s}} \bigg] - n_{H}x_{1s}\sigma_a(\nu)\delta f^{(1)} + 2\frac{c^2}{8\pi\nu^2}n_H^2x_e^2\alpha_{1s}(\nu)\phi(\nu) \bigg[ \delta_b^{(1)} + \frac{\delta x_{e}^{(1)}}{x_{e}} \bigg].
}
Substituting in the homogeneous result, we find
\eq{\label{eq:BE_cont_(1)}
\frac{1}{a}\hat{\bd{n}}\cdot \nabla \delta f^{(1)} + n_{H}x_{1s}\sigma_a(\nu)\delta f^{(1)} &=  \frac{c^2}{8\pi\nu^2}n_H^2x_e\alpha_{1s}(\nu)\phi(\nu) \bigg[ x_e\delta_b^{(1)} + \frac{2-x_e}{1-x_e}\delta x_{e}^{(1)} \bigg].
}
We now introduce the total number flux of continuum photons $N(\bd{x},\hat{\bd{n}}) = = \int_{\nu_c}^{\infty} d\nu \frac{8\pi\nu^2}{c^2}f(\nu,\bd{x},\hat{\bd
n})$
As in Ref.~\cite{Venumadhav_Hirata_15}, we approximate $\sigma_a(\nu)$ in the integrals by its evaluation at the ionization frequency $\sigma(\nu_c)$. We can then rewrite the detailed balance in the homogeneous case as
\eq{
x_{1s}\sigma_a(\nu_c) N^{(0)} &= n_Hx_e^2\alpha_{1s}.
}
We now multiply Eq.~\eqref{eq:BE_cont_(1)} by $8\pi\nu^2/c^2$ and integrate over frequency. Defining
$A = n_Hx_{1s}\sigma_a(\nu_c), B_1 = n_Hx_e^2\alpha_s$  and $B_2 = n_Hx_e\alpha_s\frac{2-x_e}{1-x_e}$,
we can recover the result of Ref.\cite{Venumadhav_Hirata_15} for the Fourier coefficients of the first-order number flux
\eq{\label{eq:dN1_cont}
\delta N^{(1)}(\bd{k}) &= n_H\frac{B_1 \delta_b^{(1)} + B_2 \delta x_e^{(1)}}{A + i(\hat{\bd{n}} \cdot \bd{k}/a)}.
}

For second-order terms in Eq.~\eqref{eq:BE_cont_exp}, we have
\eq{
\frac{1}{a}\hat{\bd{n}}\cdot \nabla \delta f^{(2)} &= - n_{H}x_{1s}\sigma_a(\nu)f^{(0)} \bigg[ \delta_b^{(2)} + \frac{\delta x_{1s}^{(2)}}{x_{1s}} + \frac{\delta x_{1s}^{(1)}\delta_b^{(1)}}{x_{1s}} \bigg] - n_{H}x_{1s}\sigma_a(\nu)\delta f^{(1)} \bigg[\delta_b^{(1)} + \frac{\delta x_{1s}^{(1)}}{x_{1s}} \bigg] - n_{H}x_{1s}\sigma_a(\nu)\delta f^{(2)} \\
&+ \frac{c^2}{8\pi\nu^2}n_H^2x_e^2\alpha_{1s}(\nu)\phi(\nu) \bigg[4\delta_b^{(1)}\frac{\delta x_{e}^{(1)}}{x_{e}} + \delta_b^{(1)}\delta_b^{(1)} + \frac{\delta x_{e}^{(1)}}{x_{e}}\frac{\delta x_{e}^{(1)}}{x_{e}}  + 2\delta_b^{(2)} + 2\frac{\delta x_{e}^{(2)}}{x_{e}} \bigg].\\
}
Substituting in the zeroth-order, homogeneous result, we have
\eq{
\frac{1}{a}\hat{\bd{n}}\cdot \nabla \delta f^{(2)} &= - n_{H}x_{1s}\sigma_a(\nu)\delta f^{(1)} \bigg[\delta_b^{(1)} + \frac{\delta x_{1s}^{(1)}}{x_{1s}} \bigg] - n_{H}x_{1s}\sigma_a(\nu)\delta f^{(2)} \\
&+ \frac{c^2}{8\pi\nu^2}n_H^2x_e\alpha_{1s}(\nu)\phi(\nu) \bigg[ \frac{4 - 3x_e}{1-x_e}\delta_b^{(1)}\delta x_{e}^{(1)} + x_e\delta_b^{(1)}\delta_b^{(1)} + \delta x_{e}^{(1)}\frac{\delta x_{e}^{(1)}}{x_{e}} + x_e\delta_b^{(2)} + \frac{2- x_e}{1-x_e} \delta x_{e}^{(2)} \bigg].\\
}
Once again, we can multiply by $8\pi\nu^2/c^2$ and integrate over frequency to find
\eq{
\frac{1}{a}\hat{\bd{n}}\cdot \nabla \delta N^{(2)} + &n_{H}x_{1s}\sigma_a(\nu_c)\delta N^{(2)} =  - n_{H}x_{1s}\sigma_a(\nu_c)\delta N^{(1)} \bigg[\delta_b^{(1)}  + \frac{\delta x_{1s}^{(1)}}{x_{1s}} \bigg]\\
&+ n_H^2x_e\alpha_{1s} \bigg[ \frac{4 - 3x_e}{1-x_e}\delta_b^{(1)}\delta x_{e}^{(1)} + x_e\delta_b^{(1)}\delta_b^{(1)} + \delta x_{e}^{(1)}\frac{\delta x_{e}^{(1)}}{x_{e}} + x_e\delta_b^{(2)} + \frac{2- x_e}{1-x_e} \delta x_{e}^{(2)} \bigg].\\
}
Defining $B_3 = n_Hx_e\alpha_{1s}\frac{4 - 3x_e}{1-x_e}$ and $B_4 = n_H\alpha_{1s}$, we find
\eq{
\bigg[A + \frac{1}{a}\hat{\bd{n}}\cdot \nabla  \bigg] \delta N^{(2)} &= - A\delta N^{(1)} \bigg[\delta_b^{(1)}  - \frac{\delta x_{e}^{(1)}}{x_{1s}} \bigg] + n_H\bigg[ B_1 ( \delta_b^{(1)}\delta_b^{(1)} + \delta_b^{(2)}) + B_2 \delta x_{e}^{(2)} + B_3\delta_b^{(1)}\delta x_{e}^{(1)} + B_4\delta x_{e}^{(1)}\delta x_{e}^{(1)}  \bigg].\\
}
Let us now turn to the continuum recombination equation, which is given by
\eq{\label{eq:cont_rec}
\dot{x}_e\vert_{\text{cont}} &= x_{1s}\int_{\nu_c}^\infty d\nu \frac{8\pi\nu^2}{c^2}\sigma_a(\nu) f_{0}(\nu,\bd{x}) - n_H x_e^2\alpha_{1s}.
}
The homogeneous continuum contribution to recombination term is negligibly small compared to the $n=2$ channel. Perturbing this equation to linear order, we recover the linear result from Ref.~\cite{Venumadhav_Hirata_15}
\eq{
\delta \dot{x}^{(1)}_e\vert_{\text{cont}} &= -\frac{1}{4\pi a n_H} \int d\hat{\bd{n}} \;  \hat{\bd{n}} \cdot \nabla \delta N^{(1)}.
}
Substituting in Eq.~\eqref{eq:dN1_cont}, we find
\eq{\label{eq:rec_lin_cont}
\delta \dot{x}^{(1)}_e\vert_{\text{cont}} &= -\bigg\{ 1 - \frac{Aa}{k} \arctan\bigg(\frac{k}{Aa} \bigg) \bigg\}\bigg[ B_1\delta_b^{(1)} + B_2\delta x_{e}^{(1)} \bigg].
}

Perturbing \eqref{eq:cont_rec} to second order, we find
\eq{
\delta \dot{x}^{(2)}_e\vert_{\text{cont}} &= \frac{1}{n_H} \bigg\{ A\delta N^{(2)}_0 + A\delta N^{(1)}_0 \bigg[\delta_b^{(1)}  - \frac{\delta x_{e}^{(1)}}{x_{1s}} \bigg] -  n_H\bigg[ B_1 ( \delta_b^{(1)}\delta_b^{(1)} + \delta_b^{(2)}) + B_2 \delta x_{e}^{(2)} + B_3\delta_b^{(1)}\delta x_{e}^{(1)} + B_4\delta x_{e}^{(1)}\delta x_{e}^{(1)}  \bigg] \bigg\}\\
&- \frac{\delta_b^{(1)}}{n_H} \bigg\{ A\delta N^{(1)}_0 - n_H \bigg[B_1\delta_b^{(1)} + B_2\delta x_e^{(1)} \bigg]  \bigg\}\\
&= \frac{1}{4\pi a n_H} \bigg\{ \delta_m^{(1)}\int d\hat{\bd{n}} \; 
\hat{\bd{n}} \cdot \nabla \delta N^{(1)} - \int d\hat{\bd{n}} \; 
\hat{\bd{n}} \cdot \nabla \delta N^{(2)}] \bigg\}.\\
}
Taking a spatial average, we find that the second term vanishes due to the gradient. Using the first-order result from Eq.~\eqref{eq:dN1_cont}, we obtain the continuum contribution to the background recombination rate
\eq{
\langle \delta \dot{x}_e^{(2)}\vert_{\text{cont}} \rangle &=  \int d\ln k \frac{\Delta_B^2(k)}{2B_0^2} \bigg\{ 1 - \frac{Aa}{k}\arctan\bigg(\frac{k}{Aa}\bigg) \bigg\} \bigg[ B_1 \overline{T}_{\delta_b\delta_b}(k)+ B_2\overline{T}_{\delta_b\delta x_e}(k)\bigg]. \\
}
This contribution is added to the $n = 2$ contribution from Eq.~\eqref{eq:dxe2_RT} in computing the shift to the background recombination rate.

\pagebreak

\bibliographystyle{apsrev4-2}
\bibliography{refs} % Entries are in the refs.bib file

\end{document}